\documentclass[a4paper,11pt]{article}
\pdfoutput=1
\usepackage{jheppub}
\usepackage{amsmath}
\usepackage{slashed}
\usepackage{siunitx}
\usepackage{xcolor}
\usepackage{graphicx}
\usepackage{multirow}
\usepackage{amsmath}    
\usepackage{amssymb}    %
\usepackage{mathtools}
\usepackage{subcaption}

\catcode`\$=\active
\gdef$#1${\texorpdfstring{\(#1\)}{\detokenize{#1}}}
\newcommand\sss{\scriptscriptstyle}

\newcommand{\lfact}[1]{\frac{#1}{16\pi^2}}

\newcommand{\divfact}[1]{\frac{#1\Delta}{16\pi^2}}

\newcommand{\inmath}[1]{\relax\ifmmode#1\else$#1$\fi}
\def\ee/{\inmath{e^{+}e^{-}}}
\def\zh/{\inmath{ZH}}
\def\zhh/{\inmath{ZHH}}
\def\zhhh/{\inmath{ZHHH}}
\def\vvh/{\inmath{\nu_e\bar{\nu}_eH}}
\def\vvhh/{\inmath{\nu_e\bar{\nu}_eHH}}
\def\vvhhh/{\inmath{\nu_e\bar{\nu}_eHHH}}
\def\eezh/{\ee/\inmath{\to}\zh/}
\def\eezhh/{\ee/\inmath{\to}\zhh/}
\def\eezhhh/{\ee/\inmath{\to}\zhhh/}
\def\eevvh/{\ee/\inmath{\to}\vvh/}
\def\eevvhh/{\ee/\inmath{\to}\vvhh/}
\def\eevvhhh/{\ee/\inmath{\to}\vvhhh/}
\newcommand{\mh}{m_{ \sss H}}
\newcommand{\mw}{m_{ \sss W}}
\newcommand{\mz}{m_{ \sss Z}}
\newcommand{\mv}{m_{ \sss V}}
\newcommand{\mt}{m_{t}}

\newcommand{\tril}{\lambda_{3}}
\newcommand{\trilsm}{\tril^{\rm SM}}
\newcommand{\qual}{\lambda_{4}}
\newcommand{\qualsm}{\qual^{\rm SM}}
\newcommand{\cbs}{\bar{c}_6}
\newcommand{\cbe}{\bar{c}_8}

\newcommand{\ktre}{\kappa_{3}}
\newcommand{\kqual}{\kappa_{4}}
\newcommand{\kpenta}{\kappa_{5}}

\def\beq{\begin{equation}}
\def\beqn{\begin{eqnarray}}
\def\eeq{\end{equation}}
\def\eeqn{\end{eqnarray}}
\def\beal{\begin{align}}
\def\endal{\end{align}}

\DeclarePairedDelimiter\bra{\langle}{\rvert}
\DeclarePairedDelimiter\ket{\lvert}{\rangle}
\newcommand{\MSbar}{{\rm \overline{MS}}}

\begin{document}

\title{Constraining~the~Higgs~self-couplings~at~e$^+$e$^-$~colliders}

\author[a]{Fabio Maltoni,}
\author[b]{Davide Pagani,}
\author[a]{Xiaoran Zhao}
\affiliation[a]{Centre for Cosmology, Particle Physics and Phenomenology (CP3),
Universit\'e catholique de Louvain, B-1348 Louvain-la-Neuve, Belgium}
\affiliation[b]{Technische Universit\"at M\"unchen, James-Franck-Str.~1, D-85748 Garching, Germany}
\emailAdd{fabio.maltoni@uclouvain.be}
\emailAdd{davide.pagani@tum.de}
\emailAdd{xiaoran.zhao@uclouvain.be}

\note{Preprint: MCnet-18-02, CP3-18-17, TUM-HEP-1132/18.}

\abstract{
    We study the sensitivity to the shape of the Higgs potential of single, double, and triple Higgs production at future $e^+e^-$ colliders. Physics beyond the Standard Model is parameterised through the inclusion of higher-dimensional operators $(\Phi^\dagger \Phi- v^2/2)^n/\Lambda^{(2n-4)}$ with $n=3,4$, which allows a consistent treatment of  independent deviations of the cubic and quartic self-couplings beyond the tree level. We calculate the effects induced by a modified potential up to one loop in single and double Higgs production and at the tree level in triple Higgs production, for both $Z$ boson associated and $W$ boson fusion  production mechanisms. We consider two different scenarios. First, the dimension six operator provides the dominant contribution (as expected, for instance, in a linear effective-field-theory (EFT)); we find in this case that the corresponding Wilson coefficient can be determined at $\mathcal{O}(10\%)$ accuracy by just combining accurate measurements of single Higgs cross sections at $\sqrt{\hat s}=$240-250 GeV and double Higgs production in $W$ boson fusion at higher energies. Second,  both operators of dimension six and eight can give effects of similar order, {\it i.e.},  independent quartic self-coupling deviations are present. Constraints on Wilson coefficients can be best tested  by combining measurements from single, double and triple Higgs production. Given that the sensitivity of single Higgs production to  the dimension eight operator is presently unknown,  we consider double and triple Higgs production and show that combining their information colliders at higher energies will provide  first coarse constraints  on the corresponding Wilson coefficient.  
    }

\maketitle
\flushbottom

\section{Introduction}
\label{sec:introduction}
In the Standard Model (SM), the breaking of the electroweak symmetry, $SU(2)_L \times U(1)_Y \to U(1)_{\rm QED}$, is induced by the potential:
\begin{equation}
V^{\rm SM}(\Phi)=-\mu^2 (\Phi^\dagger \Phi)+\lambda (\Phi^\dagger \Phi)^2\, ,
\label{VSM}
\end{equation}
where $\Phi$ is the Higgs doublet and the parameters $\mu$ and $\lambda$ depend on 
the vacuum expectation value of the Higgs field $v$ (or equivalently, the Fermi constant $G_F$) and the Higgs boson mass $\mh$, {\it i.e.}, $\mu^2=\mh^2/2$ and $\lambda=\mh^2/(2 v^2)$. The form of eq.~(\ref{VSM}) is dictated by the symmetries of the SM and the requirement of renormalisability. It is therefore a firm prediction of the SM that, once $m_H$ is known, the Higgs boson ($H$) self interactions are uniquely determined; $\trilsm=\qualsm=\lambda$,
 where $\lambda^{\rm SM}_{3}(\lambda^{\rm SM}_{4})$ is the factor in front of the $v H^3$ ($H^4/4$) interaction  in the SM Lagrangian after ElectroWeak Symmetry Breaking (EWSB).
Since its discovery in 2012 \cite{Chatrchyan:2012xdj, Aad:2012tfa} a wealth of information has been accumulated on the scalar particle at 125 GeV of mass.  Its couplings to vector bosons and third generation fermions~\cite{Khachatryan:2016vau} are so far all compatible with the SM expectations. However, no confirmation on the SM form of the Higgs potential is yet available from collider experiments.  The reason of this lies in the intrinsic difficulty of accessing the relevant information experimentally. 

Determining the form of the Higgs potential necessarily implies measuring the strength of the Higgs three-  and four- (and possibly higher) point self-couplings. This is a challenging task at colliders for several reasons.  As mentioned above, the self-couplings are proportional to $\lambda$, which in the SM is about 1/8 and therefore rather weak, {\it i.e.} of the same order of the Higgs couplings to the vector bosons and significantly smaller than the Yukawa coupling to the top quark. In addition, for a direct sensitivity on $\tril$($\qual$), processes featuring at least three(four) Higgs bosons  need to be considered, namely double(triple) Higgs production  As a result, effects associated to the self-couplings in the range of the SM values are in general very small. This simple fact has two immediate implications. First, one will need considerable statistics to be collected at the LHC Run II and III  and in future electron-positron colliders before reaching the sensitivity to values close the SM predictions. Second, the precision of the experimental determinations of the self-couplings will critically depend on that of the other Higgs boson couplings entering the same process (or more in general the observable) under consideration.  For example, at the LHC the largest production rate is due to gluon fusion processes via a top-quark loop. While the leading contribution from the top-quark Yukawa coupling $y_t$ scales as $y_t^4$,  the leading contribution from the Higgs self-coupling scales as $(\tril)^2$ and is kinematically suppressed at large $m(HH)$ invariant-mass values.

Many studies have been performed for the LHC at $\sqrt{\hat s}=13$ TeV aiming to directly access $\tril$
from double Higgs production measurements \cite{Baur:2003gp,  Dolan:2012rv, Papaefstathiou:2012qe, Baglio:2012np,Yao:2013ika, Barger:2013jfa, deLima:2014dta, Englert:2014uqa, Wardrope:2014kya, Liu:2014rva, Azatov:2015oxa, Li:2015yia, Lu:2015jza,  Cao:2015oaa, Cao:2015oxx, Behr:2015oqq,  Cao:2016zob, Adhikary:2017jtu}.
However, due to the complexity of the corresponding realistic experimental setups, it is still
unclear what is the final precision that could be achieved on the determination of $\tril$. 
At the moment the strongest experimental bounds on $\tril$ (from non-resonant double-Higgs production) have been obtained in the CMS analysis for the $b\bar{b}\gamma \gamma$ signature \cite{CMS:2017ihs}. Exclusion limits on $\tril$ have been found to strongly depend on the value of the top Yukawa coupling $y_t$. In particular, in the case of an SM $y_t$ value,  order $ \tril <-9 ~\tril^{\rm SM}$ and $\tril > 15 ~\tril^{\rm SM}$ limits are obtained. According to optimist {\it experimental} projection studies \cite{ATL-PHYS-PUB-2014-019}, even with the high luminosity (HL) option of 3000 fb$^{-1}$ it may be possible to exclude values only in the range $\tril<-1.3 ~\tril^{\rm SM}$ and  $\tril>8.7 ~ \tril^{\rm SM}$.  

Concerning the quadrilinear coupling $\qual$, it is instead incontrovertibly clear that the possibility of constraining  $\qual$ via the measurement of triple Higgs production at the LHC is quite bleak \cite{Plehn:2005nk, Binoth:2006ym, Maltoni:2014eza, Baglio:2015wcg}. Even at a future 100 TeV proton--proton collider only loose bounds may be obtained with a considerable amount of integrated luminosity \cite{Chen:2015gva,Kilian:2017nio,Fuks:2017zkg}.

\medskip

Additional and complementary strategies for the determination of $\tril$ and $\qual$ are therefore desirable not only at the moment but also for the (near) future. To this purpose, a lot of new results have recently appeared aiming to access $\tril$ via indirect loop-induced effects. This idea has been pioneered by McCullough in the context of $e^+e^-$ colliders \cite{McCullough:2013rea}, where loop-induced effects in single-Higgs production have been investigated for  $ZH$ associated production~\cite{Ellis:1975ap, Lee:1977eg, Ioffe:1976sd}. A first evaluation of analogous loop effects at the LHC   has been presented in ref.~\cite{Gorbahn:2016uoy} for  $gg \to H \to \gamma \gamma$. At the same time, the complete set of (one- and two-) loop computations for all relevant  single-Higgs observables at the LHC together with the proposal of combining inclusive and differential observables, has been put forward in \cite{Degrassi:2016wml}. Since then several studies have appeared: the computation of 
the factorisable QCD corrections to the single-Higgs EW production at the LHC~\cite{Bizon:2016wgr}, two-loop effects in precision EW observables~\cite{Degrassi:2017ucl,Kribs:2017znd} and, more recently, 
further investigations on the impact of the differential information and the relevance of SM electroweak corrections \cite{Maltoni:2017ims}.
Furthermore,  global analyses in the context of an SMEFT (SM-EFT) have also been presented for present and future measurements at the LHC \cite{DiVita:2017eyz} and even for the case of future $e^+e^-$ colliders \cite{Barklow:2017awn, DiVita:2017vrr}. On the other hand, in these works,  effects of $\qual$ have been either ignored, being irrelevant for the calculation considered, or assumed to be determined in turn by the $\tril$ value.

In the present work  we investigate for the first time the (combined) sensitivity to both the $\tril$  and $\qual$ self-couplings in (multi-)Higgs production at future $e^+e^-$ colliders. We consider  $H$, $HH$, and $HHH$ production both in association with a $Z$ boson or via $W$-boson fusion (WBF) \cite{Jones:1979bq}.
These processes are listed in Tab.~\ref{tableprocesses},
\begin{table}[!h]
\begin{center}
\begin{tabular}{ccc}
 Process & $\tril$ &$\qual$ \\
 \hline
 $ZH$, $\nu_e \bar \nu_e H$  (WBF)  &  one-loop & two-loop \\
 $ZHH$,  $\nu_e \bar \nu_e HH$ (WBF) & tree & one-loop \\
 $ZHHH$, $\nu_e \bar \nu_e HHH$ (WBF) & tree & tree
\end{tabular}
\end{center}
\caption{Processes considered in this work and the order at which the $\tril$ and $\qual$ dependence appears. We do not calculate two-loop effects, but we do calculate one-loop effects for both single and double Higgs production. 
\label{tableprocesses}}
\end{table}
where we have also specified at which level in perturbation theory the $\tril$  and $\qual$ dependence appears (we do not calculate two-loop effects in this work).
In particular, we perform the computation of one-loop effects in single and (for the first time) double Higgs production. The former pose no theoretical challenge, confirm the results of ~\cite{McCullough:2013rea, DiVita:2017vrr} (and {\it mutatis mutandis}, of \cite{Degrassi:2016wml, Bizon:2016wgr}); they are presented here for completeness and are also used in our analysis.  On the other hand, one-loop effects in double Higgs production can be computed only within a complete and consistent EFT approach, where UV renormalisation can be performed. To this purpose, we work in a theoretical and computational framework where the cubic and quartic couplings can independently deviate from the SM predictions and loop computations can be consistently performed. Specifically, we add the two higher-dimensional operators $c_{2n}(\Phi^\dagger \Phi- v^2/2)^n/\Lambda^{(2n-4)}$ with $n=3,4$ to the SM Lagrangian, where the presence of the ``$-v^2$'' term considerably simplifies the technical steps of the one-loop calculation in double Higgs production. On the other hand, Wilson coefficients in this basis or in the standard $c'_{2n}(\Phi^\dagger \Phi)^n/\Lambda^{(2n-4)}$ parameterisation can be easily related at any perturbative order and also after the running to a different scale. While the $c'_{2n}$ coefficient are more suitable for the matching to a UV-complete model, the $c_{2n}$ ones feature simple relations to Higgs self-couplings, and are more convenient for phenomenological predictions such as those performed here for (multi-)Higgs production. 
Independently from the choice of the basis, it will be clear in the text that the Wilson coefficients $c'_{6}$ and $c'_{8}$, or $c_{6}$ and $c_{8}$, are the relevant parameter to be considered and not directly the $\tril$ and $\qual$ couplings.  

By comparing and combining the direct and indirect sensitivities on the Higgs self-couplings that could be obtained at future $e^+e^-$ colliders (CEPC\cite{CEPC-SPPCStudyGroup:2015csa},
FCC-ee\cite{Gomez-Ceballos:2013zzn},
ILC\cite{Baer:2013cma} and
CLIC\cite{CLIC:2016zwp,Abramowicz:2016zbo}) running at different energies and luminosities, we explore the final reach of such colliders to constrain the Higgs potential. In general, we assume that BSM effects due to the Higgs interactions with the other SM particles are negligible w.r.t. those induced by  self interactions. In practice, we work under the same assumptions of the first calculations of one-loop $\tril$ effects in single Higgs production at $e^+e^-$ \cite{McCullough:2013rea} or proton--proton collisions \cite{Gorbahn:2016uoy, Degrassi:2016wml}, which represented an unavoidable input for the analyses considering a more general class of BSM scenarios \cite{DiVita:2017eyz, Barklow:2017awn, DiVita:2017vrr}. On the other hand, precisely one of these recent global analyses, ref.~\cite{DiVita:2017vrr}, has shown that in high-energy $e^+e^-$ collisions, where $ZHH$ and WBF$~HH$ production are kinematically available, working with our assumption or allowing for additional BSM effects does not affect the constraints that can be obtained for the trilinear Higgs self-coupling, thus justifying our working assumption. In this work we will investigate the precision that can be achieved on the trilinear Higgs self-coupling, not only when it is close to its SM value, {\it i.e.}, we will explore also regimes where its effects entering via loop corrections can be relevant also in double-Higgs production. Moreover, these loop corrections, similarly to $ZHHH$ and WBF$~HHH$ production at the Born level,  involve effects due to the quadrilinear Higgs self-coupling. We will investigate the constraints that can be set on this coupling under two different assumptions. In the first, we consider the case of a well behaving EFT expansion, where dimension-eight operators induce effects smaller than dimension-six ones. In other words, the value of the trilinear coupling automatically sets also the value of  the quadrilinear coupling. In the second, we lift this assumptions and we allow for independent trilinear and quadrilinear couplings, namely, we allow for similar effects from $(\Phi^\dagger \Phi)^3$ and $(\Phi^\dagger \Phi)^4$ operators.

The paper is organised a follows. 
In section \ref{sec:theory} we introduce the notation and we discuss the EFT framework used in our calculation. The details concerning the definition of the renormalisation scheme at one loop and all the necessary counterterms for the calculation performed here are given in Appendix \ref{appendix:ren}. In section
\ref{sec:results} we provide the predictions for cross sections of single, double and triple Higgs production at different energies, discussing their dependence on the $c_6$ and $c_8$ parameters. The one-loop calculations in single and double Higgs production are performed via one-loop form factors, the explicit results for which are provided in Appendix \ref{sec:formfactors}. 
In section~\ref{sec:bounds} we determine the reach of several experimental setups at future $e^+e^-$ colliders for constraining the cubic and quartic couplings; both individual and combined results from single, double and triple Higgs production are scrutinised. The maximum $c_6$ and $c_8$ values beyond which perturbative convergence cannot be trusted are derived in Appendix \ref{validity}. We draw our conclusions in section \ref{sec:conclusions}. 

\vfill

\section{Theoretical setup}
\label{sec:theory}
\subsection{Notation and parametrisation of New Physics effects}
\label{notation}
In this work we are interested to the effect induced by the modification $V^{\rm SM}(\Phi) \rightarrow V(\Phi)$ defined as
\begin{equation}
V(\Phi)=V^{\rm SM}(\Phi)+ V^{\rm NP}(\Phi)\, ,\qquad \Phi =  \begin{pmatrix}G^+ \cr  \frac{1}{\sqrt{2}}(  v+H+i G^0)\, 
 \end{pmatrix}\  \, ,
\label{V}
\end{equation}
where the New Physics (NP) modifications of the potential are all included in $V^{\rm NP}$ and the symbol $\Phi$ denotes the Higgs doublet. The term $V^{\rm SM}$ has already been defined in eq.~\eqref{VSM}.

Following the convention of  ref.~\cite{Boudjema:1995cb}, the most general form of $V^{\rm NP}$ that is invariant under $SU(2)$ symmetry can be written as 
\begin{equation}
V^{\rm NP}(\Phi)\equiv  \sum_{n=3}^{\infty}\frac{c_{2n}}{\Lambda^{2n-4}}\left(\Phi^\dag\Phi -\frac{1}{2}v^2 \right)^n \, .
\label{VNP}
\end{equation}

It is important to specify from the beginning why for our calculation it is convenient to parametrise the NP contributions as done in eq.~\eqref{VNP} and not using the standard EFT parameterisation
\begin{equation}
    V^{\rm NP}_{\rm~std}(\Phi)\equiv  \sum_{n=3}^{\infty}\frac{c_{2n}^{\prime}}{\Lambda^{2n-4}}\left(\Phi^\dag\Phi \right)^n\, .
\label{VNP_Stnotation}
\end{equation}

The advantages of the parametrisation in eq.~\eqref{VNP} w.r.t the one in eq.~\eqref{VNP_Stnotation} are due to the fact that after EWSB any $\left(\Phi^\dag\Phi \right)^n$ originates $H^i$ terms with $1\le i \le 2n$, while any $\left(\Phi^\dag\Phi -\frac{1}{2}v^2 \right)^n$ originates $H^i$ terms only with $n\le i \le 2n$.
In other words, at tree-level, the trilinear Higgs self-coupling receives modifications only from $c_6$ and the quadrilinear only from $c_6$ and $c_8$.
Needless to say, when they are  summed to $V^{\rm SM}$, equations \eqref{VNP} and  \eqref{VNP_Stnotation}  not only refer to the same quantity parametrised in a different way ($V^{\rm SM}+V^{\rm NP}_{\rm~std}=V^{\rm SM}+V^{\rm NP}$), but they are also fully equivalent for any truncation of the series at a given order  $n$.

Writing $V^{\rm SM}(\Phi)+ V^{\rm NP}(\Phi)$ after EWSB as
\begin{equation}
 V(H) = \frac{1}{2} m_H^2 H^2 + \tril v H^3 + \frac{1}{4}\qual H^4 +  {\lambda_5} \frac{H^5}{v} + O(H^6)
\end{equation}
allows to define the self-couplings $\lambda_n$, which can be parametrised by 
the quantities\footnote{Note that $\ktre$ and $\kqual$ are defined differently than $\kappa_5$. The former are the ratios of the trilinear and quadrilinear couplings with their SM values. The latter is the value normalised to $\lambda$, being a tree-level $H^5$ interaction not present in the SM.} 
\begin{eqnarray}
\label{k3}
\ktre \equiv\frac{\tril}{\trilsm} = 1+ \frac{c_6 v^2}{\lambda \Lambda^2} &\equiv& 1+\bar c_6,\\
\label{k4}
\kqual  \equiv\frac{\qual}{\qualsm}  = 
1+ \frac{ 6 c_6 v^2}{\lambda \Lambda^2} +
\frac{ 4 c_8 v^4}{\lambda \Lambda^4}
&\equiv& 1+ 6 \bar c_6 + \bar c_8 \, ,\\
\label{k5}
\kpenta \equiv\frac{\lambda_5}{\lambda} = 
\frac{ 3 c_6 v^2}{4 \lambda \Lambda^2} +
\frac{ 2 c_8 v^4}{\lambda \Lambda^4}+
\frac{  c_{10} v^6}{\lambda \Lambda^6}
&\equiv& \frac34 \bar c_6 +  \frac12 \bar c_8 +  \bar c_{10}  \, . 
\end{eqnarray}   
where $\trilsm(\qualsm)$ is the value of $\tril(\qual)$ in the SM and reads
\begin{equation}\label{l34}
\trilsm=\qualsm=\lambda=\frac{\mh^2}{2 v^2}\,.
\end{equation}
In other words, $\bar c_{6},\bar c_{8}$ and $\bar c_{10}$ are  $c_6,c_8$ and $c_{10}$ normalised in such a way that  can be easily related to $\ktre, \kqual$ and $\kappa_5$. In particular
\begin{eqnarray}
    \bar c_6&\equiv &\frac{c_6 v^2}{\lambda \Lambda^2}=\ktre-1\,, \\
    \quad\bar c_8&\equiv& \frac{4c_8v^4}{\lambda\Lambda^4}=\kqual-1-6(\ktre-1)\,. \label{k4inv}
\end{eqnarray}
Eqs.~\eqref{k3} and \eqref{k4} make two important  points manifest. First, while the trilinear coupling only depends on $c_6$, the quadrilinear depends on both $c_6$ and $c_8$. Thus, in a well-behaved EFT, where the effects of higher dimensional operators are systematically suppressed by a large scale,  one expects deviations both in $\ktre$ and $\kqual$ and such that $(\kqual -1) \simeq 6  (\ktre -1)$, see also eq.~\eqref{k4inv}.  Second,  $\ktre$ and $\kqual$ do not depend on any $c_{2n}$ coefficient with $n>4$, without any assumption on the  $c_{2n}$ values with $n>4$. In other words, for the study of trilinear and quadrilinear Higgs self-couplings {\it at the tree level,} only the $c_{6}$ and $c_{8}$ Wilson coefficients are relevant. On the other hand,  {\it at one-loop level} also $c_{10}$ is in principle needed.

As already mentioned in the introduction, in this paper we will calculate one-loop corrections to double Higgs production, taking into account both $c_6$ and $c_8$ effects. In general, when loop corrections are calculated and $c_{2n}$ coefficients themselves are renormalised, the parametrisation of  eq.~\eqref{VNP} is convenient also for other reasons that are particularly relevant when the Wilson coefficients are renormalised. At variance with  eq.~\eqref{VNP_Stnotation}, the values of the coefficients $c_{2n}$ influence only the value of the Higgs self-couplings; they do not alter the SM relations among $\mh$, $v$, $\mu$ and $\lambda$ 
\begin{equation}\label{muSM}
\mu^2=\frac{\mh^2}{2} \, ,
\end{equation}
\begin{equation}\label{lamSM}
\lambda=\frac{\mh^2}{2 v^2} \, .
\end{equation}  
This is convenient because the physical quantities are $\mh$ and $v$ and not $\mu$ and $\lambda$, while using eq.~\eqref{VNP_Stnotation} one has to determine before $\mu$ and $\lambda$ and then the self-couplings (see Appendix A of \cite{Degrassi:2016wml}). Especially, and this is the main motivation for this work, thanks to eqs.~\eqref{muSM} and \eqref{lamSM}  the modification $V^{\rm SM}(\Phi) \rightarrow V^{\rm SM}(\Phi) + V^{\rm NP}(\Phi)$   allows to keep the SM relations between the renormalisation constants and the definition of the renormalisation counterterms as done in ref.~\cite{Denner:1991kt}. On the other hand, the explicit insertion of $v^2$ in  eq.~\eqref{VNP} deserves a particular attention for the renormalisation procedure and leads to additional counterterms. All the necessary ingredients for the calculations performed here are provided in Appendix \ref{appendix:ren}.  

It is important to note that the coefficients $c_{2n}$ in eq.~\eqref{VNP} and $c'_{2n}$ in eq.~\eqref{VNP_Stnotation} are connected by very simple relations and can be converted ones into the others at any step of the calculation. Thus, our calculation is fully equivalent to using the $c'_{2n}$ coefficients and renormalising them in the $\MSbar$ scheme (see Appendix \ref{appendix:ren}). Via the simple tree-level relations among $c_{2n}$ and $c'_{2n}$ coefficients one can convert results at any order from one convention to the other, including the renormalisation-group equations. As already said, while $c'_{2n}$ coefficients are more suitable for the matching to a UV-complete model, the $c_{2n}$ coefficients are more convenient for one-loop calculations of hard-scattering matrix elements such as those performed here for double Higgs production. Independently from the basis choice, the NP effects should be parametrised via $c'_{6}$ and $c'_{8}$, or $c_{6}$ and $c_{8}$, rather than directly through the $\tril$ and $\qual$ couplings. The reason is that the individual Wilson coefficients are the  quantities entering the renormalisation procedure, and $c'_{6}$, or equivalently $c_{6}$, is affecting both $\tril$ and $\qual$ values. In the following we will mainly parametrise NP via the $\bar{c}_{2n} $ coefficients, which are simply related to both $c_{2n}$ and $\kappa_n$ (see eqs.\eqref{k3}-\eqref{k4inv}).

\section{Single, double and triple Higgs production: $\cbs$ and $\cbe$ dependence}
\label{sec:results}
In this section we describe the calculations for
 single, double and triple Higgs production via WBF, the $\eevvh/(H(H))$ processes, or in association with a $Z$ boson, the $\eezh/(H(H))$ processes.  We will denote the latter also as $Z H^n $.  Besides the $Z H^n $ and WBF production modes, at $e^+e^-$ colliders   there are other possibly relevant processes such as $t \bar t H^n$, $Z$-boson-fusion (ZBF) or loop-induced $H^n$ production via photon fusion~\cite{Boudjema:1995cb} from the initial-state radiation. However, these processes have considerably smaller cross sections than WBF or $Z H^n $ production modes, so we do not consider them in our analysis. Part of them have been considered in ref.~\cite{DiVita:2017vrr} and their impact has been indeed found to be negligible.

While triple Higgs production processes are calculated at the Born level, for both single and double Higgs production we take into account also one-loop corrections involving the additional $\bar{c}_6$ and $\bar{c}_8$ dependence. The sensitivity on $\bar{c}_6$ and $\bar{c}_8$, and in turn on $\ktre$ and $\kqual$,  depends on the multiplicity of Higgs bosons in the final state and the numbers of loop corrections considered, as summarised in Tab.~\ref{tableprocesses}. 
We expect complementary information from  $Z H^n $ and WBF, when the different collider energies of the possible future $\ee/$ colliders are considered. While the cross section of $Z H^n $ is maximal for energies slightly larger than its production threshold, the WBF cross section grows with the energy. Moreover, based on results in refs.~\cite{McCullough:2013rea, Degrassi:2016wml, Bizon:2016wgr, Maltoni:2017ims, DiVita:2017vrr}, in $ZH^n$ production we expect a strong dependence of the Higgs self-coupling effects from loop corrections, with larger effects at lower energies. On the contrary, in WBF this energy-dependence is expected to be much smaller. 
 
It important to note that $ZH^n$ and WBF cross sections, and in turn their sensitivity on $\bar{c}_6$ and $\bar{c}_8$, depend on beam polarisations,  which can be tuned at linear colliders.\footnote{Due to the Sokolov Ternov effect \cite{Sokolov:1963zn}, the tuning of beam polarisations is much more difficult in circular colliders. }
First of all, WBF  contributes only via the $LR$ polarisations,\footnote{Here, as in the following, we are neglecting power-suppressed terms that depend on the mass of the electron.} since $W$-boson couples only to the left-chirality fermions. 
Conversely, the $ZH^n$ processes can also originate from $RL$ polarisations (right-handed $e^{-}$, left-handed $e^{+}$), also denoted as $P(e^{-},e^{+})=(1.0,-1.0)$. On the other hand, results for $RL$ polarisations  can be easily obtained from those with $LR$ via the relation
\begin{align}
    \sigma_{RL}=\sigma_{LR}\left(\frac{2\sin^2\theta_W}{1-2\sin^2\theta_W}\right)^2 \approx ~ 0.65 \sigma_{LR}\, , \label{RLLR}
\end{align}

In all our calculations we use following input parameters \cite{Patrignani:2016xqp}
\begin{eqnarray}
    G_{\mu}&=&\SI{1.1663787e-5}~\si{GeV^{-2}}\,,\quad \mw=~80.385~\si{GeV}\,,\quad \mz=91.1876~\si{GeV}\, \nonumber \\
   \mh&=&125~\si{GeV}\, , \quad m_t=173.21~\si{GeV}\, ,
\end{eqnarray}
We assume $\bar{c}_6$ and $\bar{c}_8$ measured at the scale $\mu_r= 2 \mh$, which therefore we will also use as $\MSbar$ renormalisation scale for  $\bar{c}_6$ in the double Higgs computation.

The WBF production modes feature the same final states of $ZH^n$ production with $Z \to \nu_e \bar{\nu}_e$ decays. The latter are not considered as part of the  WBF contribution in our calculation. Although NLO EW corrections in the SM would jeopardise the gauge invariance of this classification at the amplitude level \cite{Denner:2003iy}, this is not the case for the one-loop corrections induced by additional $c_{2n}$ interactions, which is the kind of effects we consider in this work on top of LO effects, as better specified later in eqs.~\eqref{sHNLO} and \eqref{sHHNLOci}.  Moreover, the interference of $ZH^n$-type diagrams with ``genuine'' WBF configurations is negligible

\subsection{Single Higgs production}
\label{singleH}
\begin{figure}[t]
    \begin{center}\vspace*{-1.0cm}
        \includegraphics[width=0.25\textwidth]{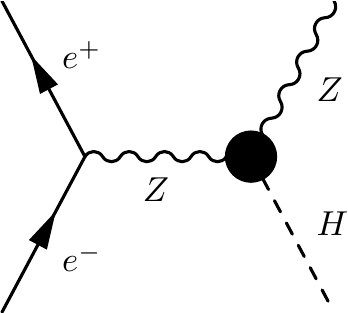}
        \hspace*{1.5cm}
        \includegraphics[width=0.25\textwidth]{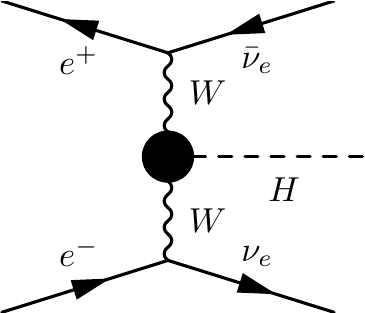}
        \caption{Feynman diagrams for single Higgs production. The black blob corresponds to the one-loop $HVV$ form factors.}
        \label{single-H-figure}
    \end{center}
\end{figure}

In this section we briefly (re)-describe the calculation of loop-induced effects from $\cbs$($\ktre$) in $ZH$ and single-Higgs WBF production at $e^+e^-$ colliders (representative diagrams are shown in Fig.~\ref{single-H-figure}). We introduce the notation that will be generalised to the case of double Higgs and triple Higgs  production and we show how it is related to the previous calculations \cite{McCullough:2013rea, Degrassi:2016wml, Maltoni:2017ims, DiVita:2017vrr}.  

For both WBF and $ZH$ production channels no Higgs self-coupling contributes at the tree level. On the other hand, one-loop corrections
depends on the trilinear Higgs  self-coupling, but not on the quartic.
Thus, while the LO cross section $\sigma_{\rm LO}(H)$ is only of SM origin, NLO predictions includes also effects from $\cbs$:
\begin{eqnarray}
    \sigma_{\rm NLO} (H)&=& \sigma_{\rm LO}(H) + \sigma_{\rm 1-loop}(H)\, ,\\
    \sigma_{\rm LO}(H)&=&\sigma_{\rm LO}^{\rm SM}(H)\, ,\\
     \sigma_{\rm 1-loop}(H)&=& \sigma_{0}    +\sigma_{1} \bar c_6    +\sigma_{2} \bar c_6^2 \, , \label{sHNLO}
\end{eqnarray}
where $\sigma_{\rm 1-loop}$ involves one-loop corrections of both SM origin and induced by $c_6$. The quantity $\sigma_{0}$ consists of the  NLO EW corrections from the SM, $\sigma_1$ represents the leading contribution in the EFT expansion (order $(v/\Lambda)^2$), while
$\sigma_2$ is of order  $(v/\Lambda)^4$, also arising from one-loop corrections.\footnote{Whenever we refer to NLO EW corrections of SM origin, those include also real emissions of photons. On the contrary, one-loop effects induced by  $c_{2n}$ are infrared safe and involve just virtual corrections. Needless to say, at one-loop, NLO QCD corrections are vanishing for the processes considered in this paper.}  Note that within our choice of operators there is no contribution proportional to $\cbe$ or any other $c_{2n}$ coefficient in this expansion, meaning that eq.~\eqref{sHNLO} is actually exact; no other terms can enter at all even for higher orders in the $(v/\Lambda)$ expansion. 
Furthermore, we remind that, at variance with the case of double Higgs production, in single Higgs production at one-loop the anomalous coupling approach ($\ktre$) is fully equivalent to the calculation in the EFT ($\cbs$).

For our phenomenological study we ignore the SM NLO EW corrections \cite{Denner:2003yg, Denner:2003iy}. Our main focus is not the precise determination of $\cbs$, but the study of its impact via its leading contributions. As discussed in detail in ref.~\cite{Maltoni:2017ims},  SM NLO EW corrections have a tiny impact on the extraction of the value of $\cbs$ and do not affect the accuracy of the determination of $\cbs$.    
Therefore, we consider $\bar{c}_6$ effects at one loop via the following approximation
\begin{equation}
    \sigma_{\rm NLO}^{\rm pheno}(H)=\sigma_{\rm LO}+\sigma_1\bar{c}_6+\sigma_2\bar{c}_6^2\, .
\end{equation}
With this approximation,
the sensitivity to the trilinear coupling can be expressed via the ratio
\begin{equation}
    \delta \sigma(H)\equiv\frac{\sigma_{\textrm{NLO}}^{\rm pheno}-\sigma_{\rm LO}}{\sigma_{\rm LO}}=\frac{\sigma_{1} \bar c_6    +\sigma_{2} \bar c_6^2}{\sigma_{\rm LO}}=(\kappa_3-1)C_1+(\kappa_3^2-1) C_2 \label{1h-c1}\,,
\end{equation}
\begin{equation}
    C_2=
    \delta Z_H^{{\rm SM}, \lambda}\label{c2here}\,,
 \end{equation}
 where we have expressed the $\sigma_{i}/ \sigma_{\rm LO}$ ratios directly\footnote{Note that $\ktre^2-1=(\ktre-1)^2+2(\ktre-1)$, so $ \sigma_{2}=C_2\sigma_{\rm LO}$ and $ \sigma_{1}=(C_1+2C_2)\sigma_{\rm LO}$} using the symbols $C_1$ and $C_2$ introduced in ref. \cite{Degrassi:2016wml}. $C_1$ denotes the one-loop virtual contribution  involving one triple Higgs vertex, while $C_2$ originates from the 
Higgs wave-function renormalisation constant (see eqs.~\eqref{dZHSMNP},\eqref{dZHNP} and \eqref{dZHHHH}), which is the only source of $\bar c_6^2$ and thus $\ktre^2$ dependence at one loop level. 
Both $C_1$ and $C_2$ are independently UV-finite and, for simplicity,  we choose not to resum higher-orders contributions to the wave function, at variance with ref.~\cite{Degrassi:2016wml}.  Indeed, given the results already presented in ref.~\cite{McCullough:2013rea}, we expect to bound $\ktre$ close to the SM ($\ktre=1$) and in this scenario  such a resummation would not make a noticeable difference anyway. Moreover,
 even considering $\ktre$ in the range $| \ktre |<6$ from ref.~\cite{DiLuzio:2017tfn}, the difference between the formula in eq.~\eqref{1h-c1} and including the resummed higher-order contributions to $Z_H$ is below $1\%$ (see also ref.~\cite{Maltoni:2017ims}).  Considering $C_2$ in eq.~\eqref{c2here}, the difference w.r.t. the definition in ref.~\cite{Degrassi:2016wml} is only due to this choice, however, in the limit $\bar c_6 \to 0$($\ktre \to 1$) the two different definitions are equivalent as can be seen from the value of $C_2$: 
\begin{align}
    C_2=\delta Z_H^{{\rm SM}, \lambda}\equiv-\frac{9}{16}\frac{G_{\mu}\mh^2}{\sqrt{2}\pi^2}\left(\frac{2\pi}{3\sqrt{3}}-1\right)\approx -0.00154 \,,
\end{align}
Moreover, in the limit $\bar c_6 \to 0$, a linear expansion of eq.~\eqref{1h-c1} for $ZH$ would lead to the result in ref.~\cite{McCullough:2013rea}.
As explained in ref.~\cite{Degrassi:2016wml} for hadronic processes, $C_1$ parametrises contributions that are process and kinematic dependent.

\begin{figure}[t]
    \begin{center}
    \includegraphics[width=0.45\textwidth]{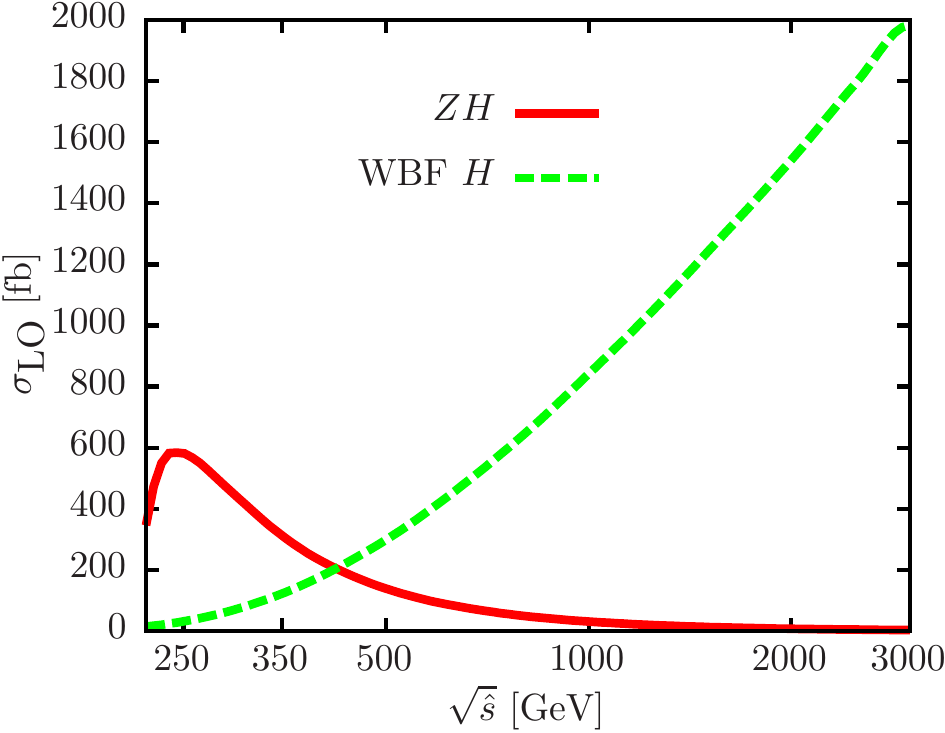}
    \hfill
    \includegraphics[width=0.45\textwidth]{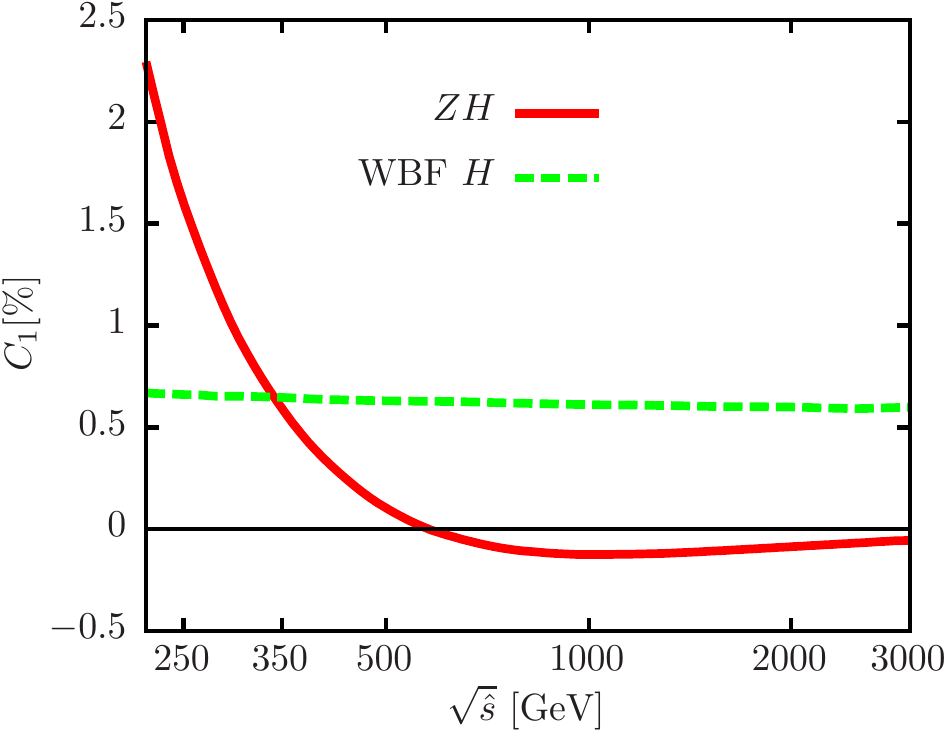}
    \caption{LO cross section (left) and $C_1$ (right) as function of the  center of mass energy $\sqrt{\hat s}$ for $P(e^{-},e^{+})=(-1.0,1.0)$. }
    \label{xs-h}
    \end{center}
\end{figure}

In Fig.~\ref{xs-h}, we show $\sigma_{\rm LO}$ (left plot) and $C_1$ (right plot)  for $ZH$ (red) and WBF (green) production as function of the energy of the collider $\sqrt{\hat s}$.  
As expected, while $C_1$ strongly depends on $\sqrt{\hat s}$ for $ZH$, it does very mildly for WBF$~H$. In particular, for $ZH$, when increasing the energy, $C_1$ decreases at the beginning, then changes its sign around $\sqrt{\hat s}=550~\si{GeV}$ and remains small. On the other hand, the total cross section for $ZH$ production peaks at around $\sqrt{\hat s}=240~\si{GeV}$ and decreases as $\sqrt{\hat s}$ increases, while the cross section for WBF$~H$ production increases with $\sqrt{\hat s}$. Thus, while for the range $200-500~{\rm GeV}$ the $ZH$ production is expected to be more sensitive than WBF on $\cbs$($\ktre$), at higher energies the situation is reversed. The information from collisions at different energies, or even at different colliders, increases the sensitivity on $\kappa_3$, as it has been discussed in ref.~\cite{ DiVita:2017vrr}. We will show analogous results in sec.~\ref{sec:bounds}.
We have also looked at the differential distribution for the transverse momentum of the Higgs boson, but we have not seen any strong dependence on $C_1$. Hence, for single Higgs production at $\ee/$ colliders differential distribution cannot increase the sensitivity on $\ktre$, at variance with the case of hadron colliders \cite{Degrassi:2016wml, Bizon:2016wgr, DiVita:2017eyz, Maltoni:2017ims} and  of double-Higgs production \cite{Tian:2013qmi}.

The range of validity of this calculation in $\ktre$ and in turn $\cbs$ is mainly dictated by the effects from $\delta Z_H^{\rm NP}$, as discussed in ref.~\cite{Degrassi:2016wml}, from which the bound $|\ktre|<20$ can be also straightforwardly applied here. A more cautious and conservative condition can be derived  by requiring perturbative unitarity for the $HH\to HH$ scattering amplitude and/or perturbativity for the loop corrections to the $HHH$ vertex in any kinematic configuration. This bound has been derived in ref.~\cite{DiLuzio:2017tfn} and leads to the requirement $|\ktre|<6$, independently from the value of $\kqual$. However, the kinematic configuration leading to this bounds are those involving two Higgses on-shell and the virtuality of the third Higgs close to $2 \mh$, which is not relevant for the trilinear interaction entering in single-Higgs production. We independently re-investigated this bound on $\cbs$ (and analogous ones on $\cbe$) in Appendix \ref{validity}, where its derivation is discussed in detail.
%
%
\subsection{Double Higgs production}
\label{sec:calcHH}
\begin{figure}[t]
    \begin{center}
        \includegraphics[width=0.22\textwidth]{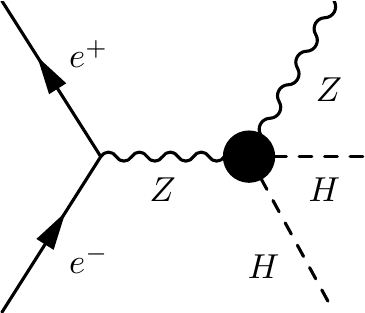}
        \includegraphics[width=0.22\textwidth]{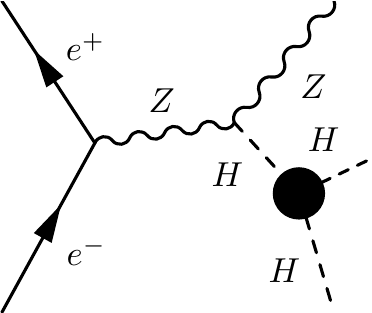}
        \includegraphics[width=0.22\textwidth]{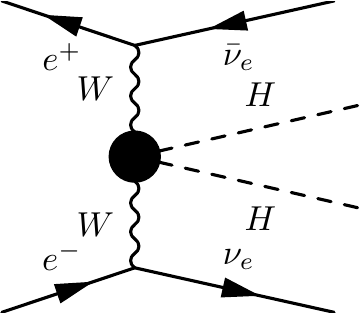}
        \includegraphics[width=0.22\textwidth]{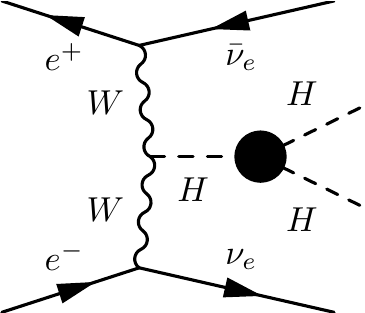}
        \caption{Representative Feynman diagrams for double Higgs production. The black blobs correspond to the one-loop $HHVV$ and $HHH$ form factors.}
        \label{double-H-figure}
    \end{center}
\end{figure}
We now consider double Higgs production. The cross sections for the production of two Higgs bosons in association with a $Z$ bosons (\eezhh/)   and  via WBF (\eevvhh/) do depend on the trilinear Higgs  self-coupling already at the tree level (see diagrams in Fig.~\ref{double-H-figure}). Moreover, for both processes, one-loop corrections  depend on both the trilinear and quartic Higgs self-couplings.
At leading order $ZHH$ and WBF$~HH$ cross sections can be written as~\footnote{The $\sigma_{i}$ terms entering eq.~\eqref{sHHLO} are not the same quantities appearing in eq.~\eqref{sHNLO}. }:
\begin{eqnarray}
 \sigma_{\rm LO}(HH) &=& \sigma_{0} +\sigma_{1} \bar c_6 + \sigma_{2} \bar c_6^2\, , \label{sHHLO} 
\end{eqnarray}
where $\sigma_{0}$ is the SM result, $\sigma_1$ represents the leading contribution in the EFT expansion (order $(v/\Lambda)^2$), while
$\sigma_2$ is the squared EFT term of order  $(v/\Lambda)^4$. Note that within our choice of operators there is no contribution proportional to $\cbe$ in this expansion. Actually, no $c_{2n}$ coefficient with $n>3$ enters at the tree level. 

The NLO corrections involve several different contributions. First we classify all of them and then we specify those  relevant for our study. Using a  notation that is analogous to  eq.~\eqref{sHHLO}, the cross section at NLO accuracy can be parametrised as 
\begin{eqnarray}
    \sigma_{\rm NLO} (HH)&=& \sigma_{\rm LO}(HH) + \sigma_{\rm 1-loop}(HH) \,\label{sHHNLOci} \, ,\\
    \sigma_{\rm 1-loop}(HH)&=& \sigma_{00}
     + \sigma_{10} \bar c_6    
     + \sigma_{20} \bar c_6^2 \label{sHHNLO1} \\
     &+&    \sigma_{30}\bar c_6^3    +\sigma_{40} \bar c_6^4 \label{sHHNLO2} \\
    &+&\bar c_8 \Big[\sigma_{01} +\sigma_{11} \bar c_6 +\sigma_{21}\bar c_6^2\Big]  \label{sHHNLO3}\\
     &+& \bar c_{10} \Big[ { \sigma_{001} + \sigma_{101} \bar c_{6} } \Big] \label{sHHNLO4}\,, 
\end{eqnarray}
where the $\sigma_{ij}$ quantities refer to the one-loop terms that factorise $\cbs^i \cbe^j$ contributions and the $\sigma_{i0j}$ to those proportional to $\cbs^i \bar c_{10}^j$. 
Some comments on the terms in \eqref{sHHNLO1}, ~\eqref{sHHNLO2},~\eqref{sHHNLO3} and \eqref{sHHNLO4} are in order. 

The terms in \eqref{sHHNLO1} are the NLO EW corrections to the contributions  that appear already at LO. The quantity $\sigma_{00}$, for instance, corresponds to the NLO EW corrections in the SM and has been calculated in ref.~\cite{Belanger:2003ya}.  The terms $\sigma_{10}$ and $\sigma_{20}$ are $\mathcal{O}(\alpha)$ corrections to $\sigma_{1}$ and $\sigma_{2}$, respectively, and thus they are always subdominant. They should be included for precise determination of $\cbs$ values, yet being subdominant,  we neglect them together with $\sigma_{00}$ in this first analysis, similarly to the case of single Higgs production.

\begin{figure}[t]
\begin{center}\vspace*{-1.0cm}
\includegraphics[width=0.9\textwidth]{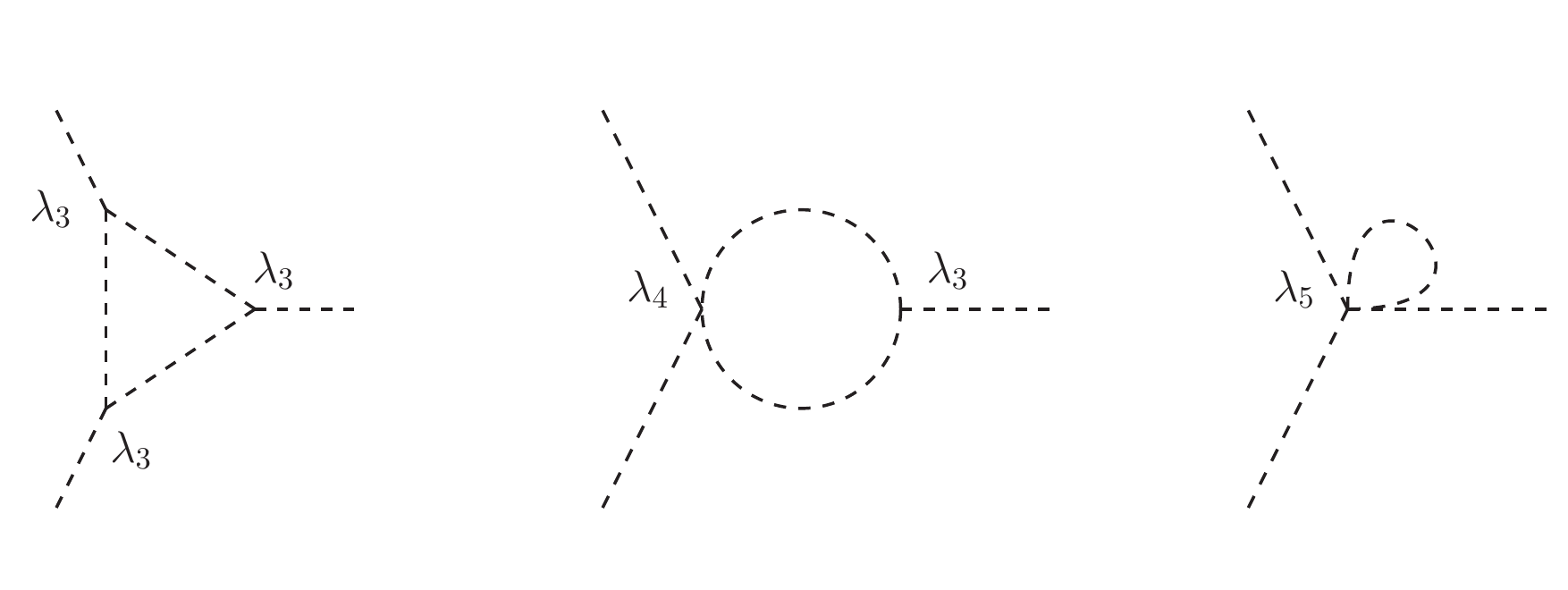}
\caption{Feynman Diagrams contributing to the $HHH$ form factor at one loop. }
\label{tril_loop}
\end{center}
\end{figure}

The terms in \eqref{sHHNLO2} collect contributions that appear at NLO for the first time.
For small values $\bar c_6 \ll1 $, these terms are suppressed w.r.t.~$\sigma_{1}$ and $\sigma_{2}$ in \eqref{sHHLO},  and may be neglected. However, at variance with $\sigma_{10}$ and $\sigma_{20}$, for large values of $\bar c_6$, they are not subdominant. Thus, we keep them in order to study the $\cbs$ and in turn $\ktre$ dependence beyond the linear approximation, which as explained is not sufficient for large values of $\cbs$. Moreover, this allows also to better quantify the range of validity of our perturbative calculation (see Appendix \ref{validity}). These contributions originate from the left diagram of Fig.~\ref{tril_loop}, which shows also the other possible one-loop corrections to the $HHH$ vertex.
This diagram induces a $(\cbs)^3$ dependence in the amplitude and in turn a $(\cbs)^4$ dependence in $\sigma_{\rm 1-loop}(HH)$ via the interference with Born diagrams. As it has been discussed in ref.~\cite{DiLuzio:2017tfn}, its contribution can be large. Also, the presence of $(\cbs)^3$ effects indicates that terms up to the order $(v/\Lambda)^6$  have to be taken into account in the one-loop amplitudes and thus in the renormalisation constants. Schematically, each order in the $(v/\Lambda)$ expansion implies that the following terms can be in principle present
\begin{eqnarray}
    (v/\Lambda)^2\rightarrow& \{\bar c_6\} &\rightarrow \{(\tril-\lambda_3^{\rm SM})\}\, ,   \label{ordinil2}\\
(v/\Lambda)^4\rightarrow& \{(\bar c_6)^2\,,~\bar c_8\} &\rightarrow \{(\tril-\lambda_3^{\rm SM})^2\, , ~(\qual-\lambda_4^{\rm SM})\}\, , \label{ordinil4}\\
(v/\Lambda)^6\rightarrow& \{(\bar c_6)^3\,,~\bar c_8 \bar c_6\,,~\bar c_{10}\} &\rightarrow \{(\tril-\lambda_3^{\rm SM})^3\,, ~(\qual-\lambda_4^{\rm SM})( \tril-\lambda_3^{\rm SM})\,, ~ \lambda_5 \}\, . \label{ordinil6}
\end{eqnarray}
Thus, the full dependence on $\tril$ and $\qual$ of the diagrams appearing in Fig.~\ref{tril_loop} is taken into account.\footnote{This classification is general enough to include also the effects present at one loop in the $HHVV$ and $HVV$ vertexes, as can be seen in Appendix \ref{sec:formfactors}.} 
On the other hand, $(v/\Lambda)^6$ terms include $c_{10}$ contributions, which we reparametrised in term of $ \bar c_{10}\equiv (c_{10} v^6)/(\lambda \Lambda^6)$; they lead to an independent value also for $\lambda_5$, the factor in front of the $H^5 / v$ term appearing in $V^{\rm NP}(\Phi)$ after EWSB. 
The origin of the terms in \eqref{sHHNLO3} and \eqref{sHHNLO4} can be now understood on the base of  Fig.~\ref{tril_loop} and eqs.~\eqref{ordinil2}-\eqref{ordinil6} and are commented in the following. 
The terms in \eqref{sHHNLO3} are the contributions that depend on $\bar c_8$. Thus, $\sigma_{01}$, $\sigma_{11}$ and $\sigma_{21}$ are the most relevant quantities in our one-loop study of double Higgs production, as they provide the sensitivity to $\cbe$ and therefore to the deviation from the quadrilinear that one expects on top of the one determined by $\bar c_6$ only.  Although $\sigma_{11}$ and $\sigma_{21}$ would be suppressed for small $c_6$ we keep them to study the validity of our calculation in the $(\bar c_6,\bar c_8)$ plane, or equivalently  $(\ktre,\kqual)$ plane. 

Finally, the last term \eqref{sHHNLO4}, is related to $\bar c_{10}$-dependent contributions. These contributions arise from the diagram  with the $H^5$ interactions in Fig.~\ref{tril_loop} and the corresponding term in the renormalisation constant $\delta c_{6}$ (see eq.~\eqref{dc6} for the explicit  $\delta c_{6}$ formula), and can be expressed as 
\begin{equation}
\bar c_{10} \Big[ { \sigma_{001} + \sigma_{101} \bar c_{6} } \Big]=\left(\sigma_{1} + 2 \sigma_{2} \bar c_6 \right) \frac{5\lambda\bar c_{10}}{4\pi^2}\left(1-\log\frac{\mh^2}{\mu_r^2}\right)\, . \label{Dk3}
\end{equation} 
At one-loop in $ZHH$ or WBF production their sum can be written as a kinematically independent shift to $\cbs$,
\begin{equation}
    \bar c_6\to \bar c_6+\frac{5\lambda \bar c_{10}}{4\pi^2}\Big(1-\log\frac{\mh^2}{\mu_r^2}\Big) \sim \bar c_6 +0.016\bar c_{10} \Big(1-\log\frac{\mh^2}{\mu_r^2}\Big)\,.
     \label{c_6_shift}
\end{equation}
In practice we can only constrain a linear combination of $\bar c_6$ and $\bar c_{10}$  that is in eq.~\eqref{c_6_shift}.  In the following we work in the assumptions that $\bar c_{10}$ effects are negligible and we set $\bar c_{10}=0$, however, for not too large values of $\bar c_{10}$, {\it i.e.}, where the linear expansion in $\bar c_{10}$ is reliable, results of $\bar c_6$ can be translated into a linear combination of $\bar c_6$ and $\bar c_{10}$ via eq.~\eqref{c_6_shift}.\footnote{If $\bar c_{10}$ is so large that the shift induced by eq.~\eqref{c_6_shift} is even larger than $\cbs$ itself, then squared loop-diagrams involving the $H^5$ vertex would be larger than their interferences with Born diagrams. Thus, one-loop contributions, and consequently the level of accuracy of our calculation, would not be sufficient.}  In order to be directly sensitive to $\bar c_{10}$ one would need to consider one-loop effects in triple Higgs production, or evaluate quadruple Higgs production at the tree level. 
  
In conclusion, in our phenomenological analysis, we evaluate $\bar c_6$ and $\bar c_8$ effects at one loop via the following approximation
\begin{eqnarray}
    \sigma^{\rm pheno}_{\rm NLO} (HH) &=& \sigma_{\rm LO}(HH) + \Delta\sigma_{\bar c_6}(HH)+ \Delta\sigma_{\bar c_8}(HH)\, ,\nonumber\\
    \Delta\sigma_{\bar c_6}(HH)&=&  \bar c_6^3   \Big[ \sigma_{30}   +\sigma_{40} \bar c_6 \Big]\, , \nonumber\\
    \Delta\sigma_{\bar c_8}(HH)&=&\bar c_8 \Big[\sigma_{01} +\sigma_{11} \bar c_6 +\sigma_{21}\bar c_6^2\Big]  \,.\label{sHHNLOpheno}
\end{eqnarray}
The analytical results for the form factors used for the calculation of $\Delta\sigma_{\bar c_6}(HH)$ and $\Delta\sigma_{\bar c_8}(HH)$ are given in Appendix \ref{sec:formfactors}. We show now the impact of $\cbs$ and $\cbe$ in the $\sigma^{\rm pheno}_{\rm NLO}$ predictions at different energies.
\begin{figure}[!t]
    \begin{center}
        \includegraphics[width=0.45\textwidth]{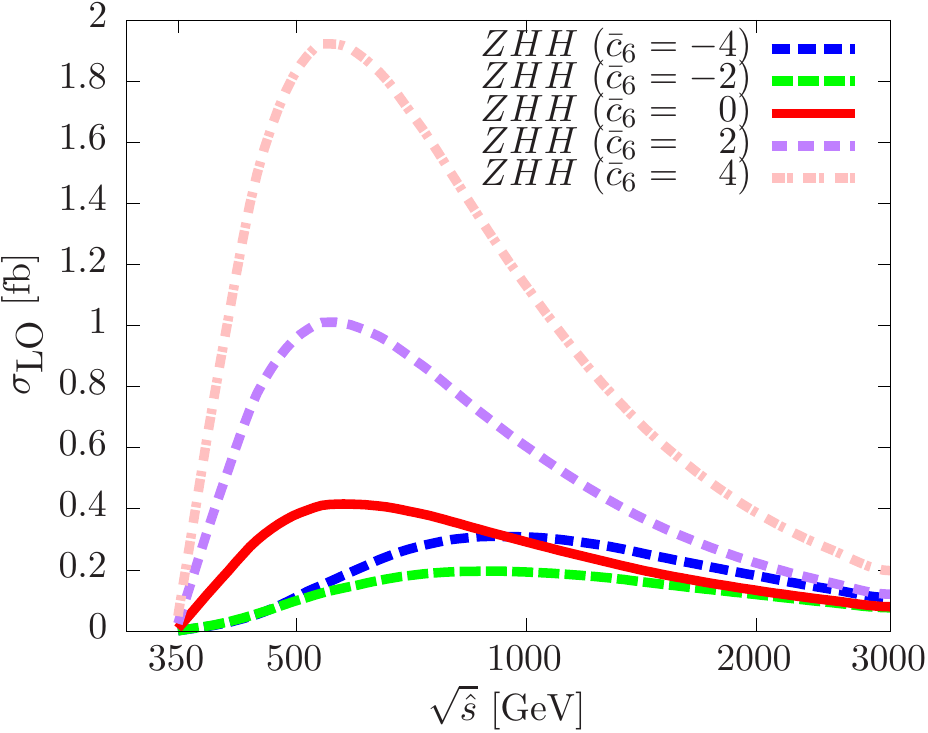}
        \includegraphics[width=0.45\textwidth]{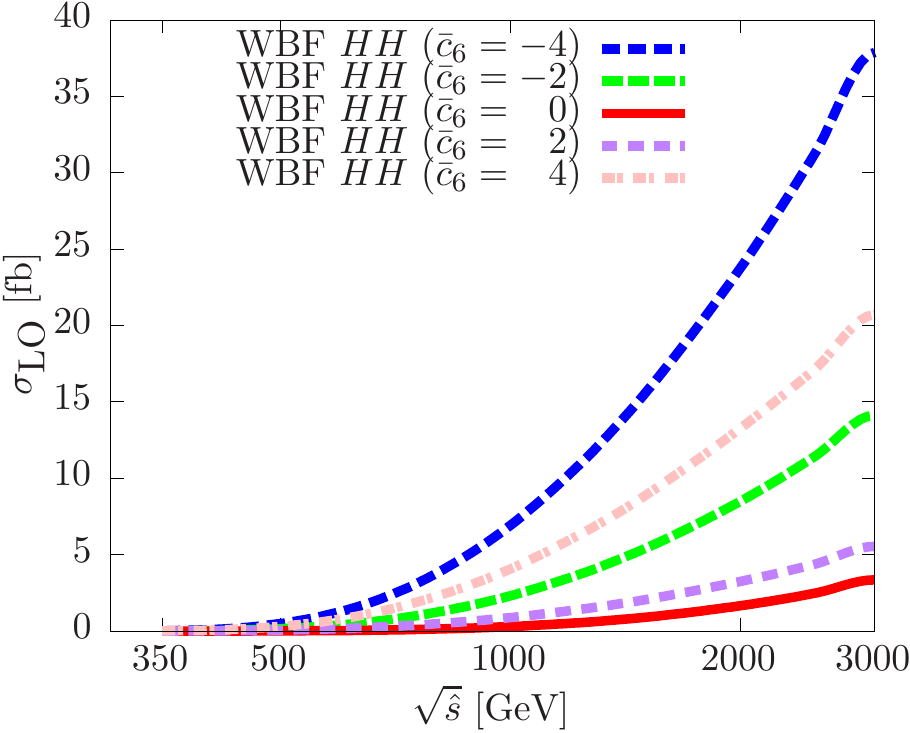}
        \caption{LO cross section of  $ZHH$ and WBF$~HH$ as function of $\sqrt{\hat s}$ for different values of $\cbs$. Results refer to $P(e^{-},e^{+})=(-1.0,1.0)$. }
        \label{xs-hh}
    \end{center}
\end{figure}
\begin{figure}[!h]
    \begin{center}
        \includegraphics[width=0.45\textwidth]{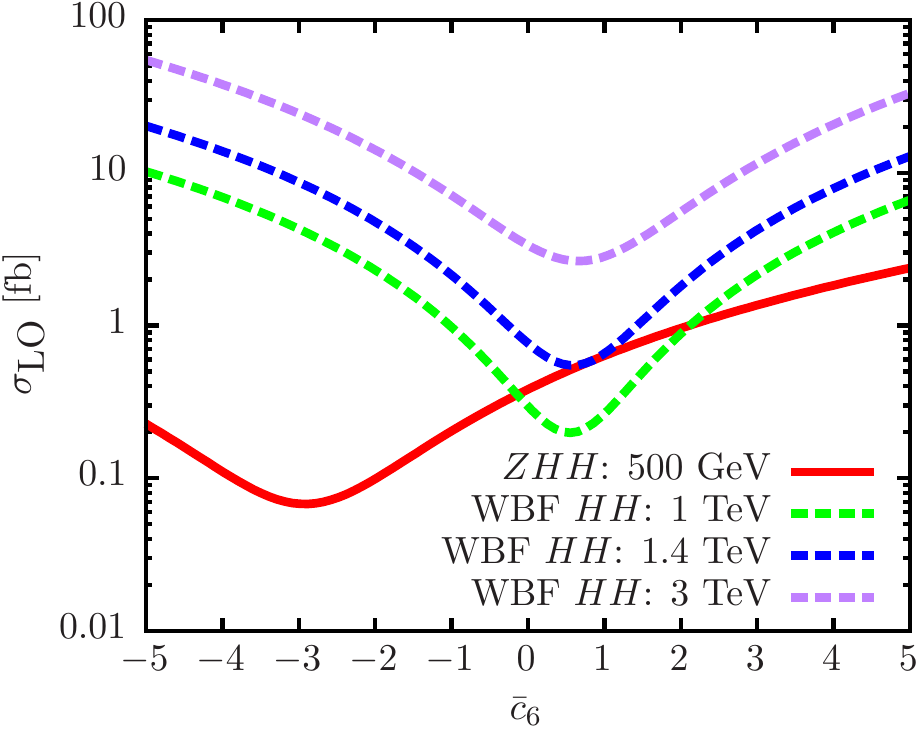}
                \includegraphics[width=0.45\textwidth]{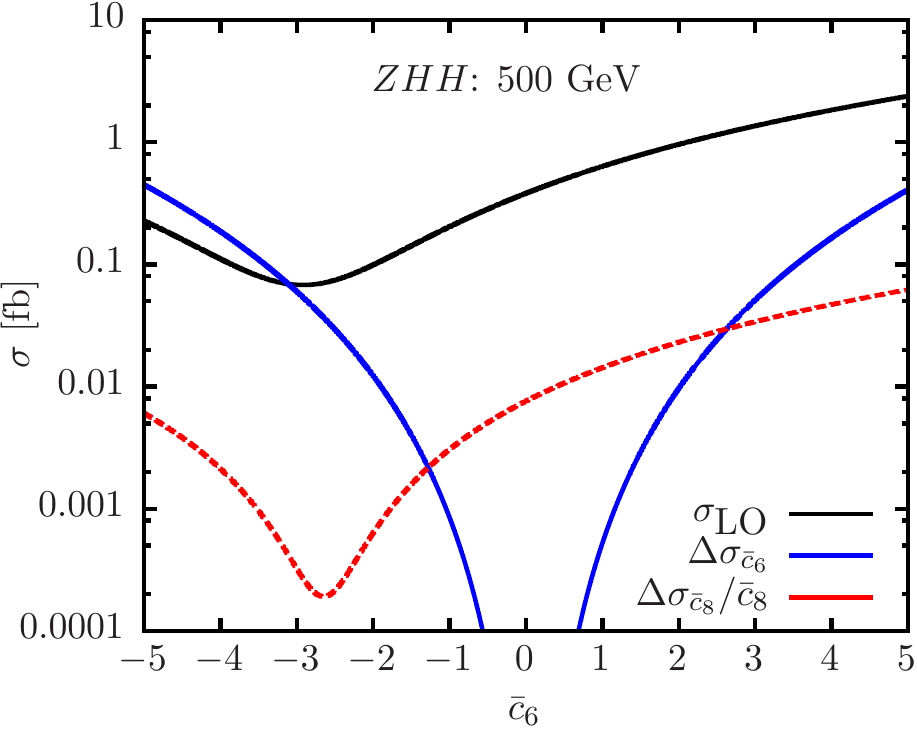}
        \hfill
                \includegraphics[width=0.45\textwidth]{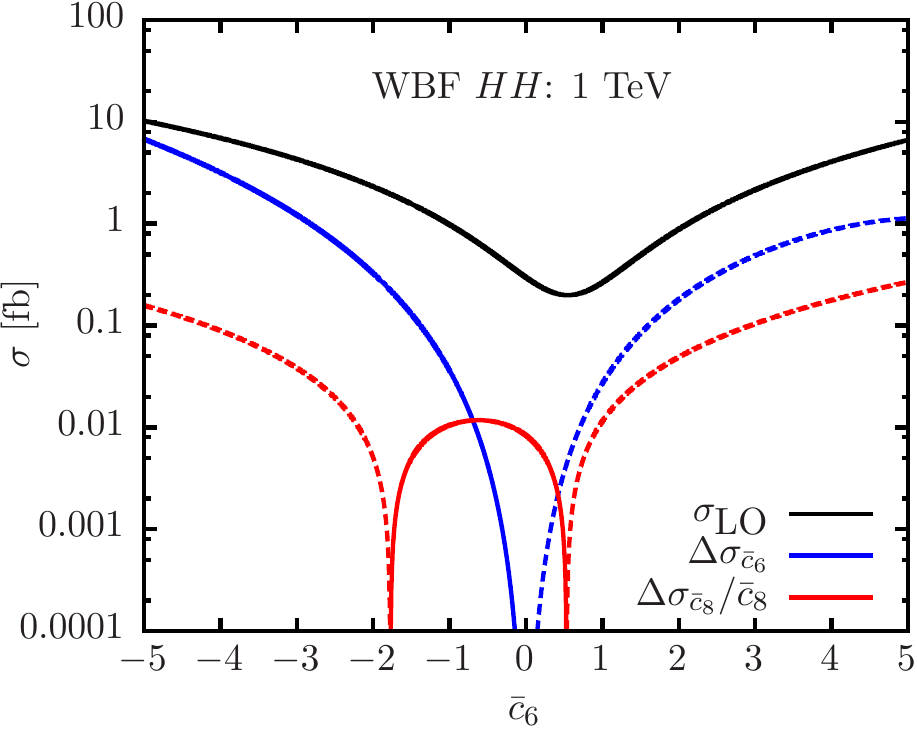}
        \includegraphics[width=0.45\textwidth]{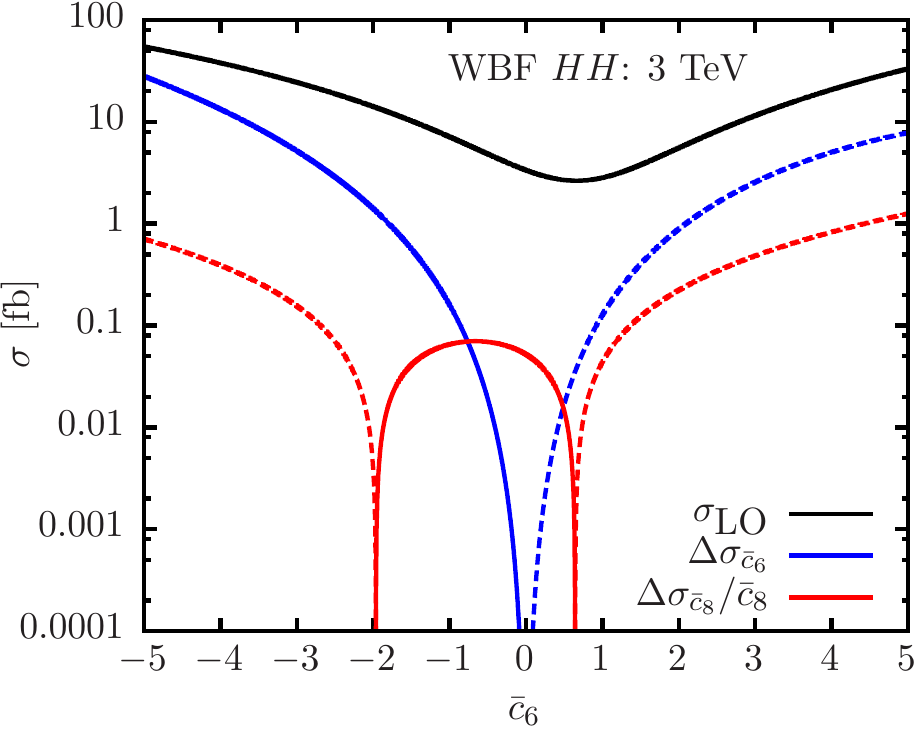}
                \hfill
        \caption{Top-left plot: $\cbs$ dependence of the LO cross section for $ZHH$ and WBF$~HH$ at different energies. The three other plots show the $\cbs$ dependence of $ \Delta\sigma_{\bar c_6}(HH)$,  $\Delta\sigma_{\bar c_8}(HH)/\cbe$ and again $\sigma_{\rm LO}$ for $ZHH$ production at 500 GeV (top-right), and WBF$~HH$ at 1000 GeV (bottom-left) and 3000 GeV  (bottom-right). Negative values are displayed as short-dashed lines.
   }
        \label{xs-hh-large}
    \end{center}
\end{figure}

First of all, in Fig.~\ref{xs-hh} we show the LO cross section $\sigma_{\rm LO}$  of $ZHH$ (left) and WBF (right) production as function of $\sqrt{\hat s}$ for different values of $\cbs$.
In $ZHH$ production, the LO cross section peaks around $\sqrt{\hat s}=500~\si{GeV}$,
which is the optimal energy for measuring this processes, while WBF$~HH$ cross section grows with energy. As can be seen by comparing the left and right plot, the dependence on $\cbs$ is different in $ZHH$ and WBF$~HH$ production. Especially, at variance with $ZHH$, WBF$~HH$ cross sections in general increase when $\cbs\neq 0$.
This feature is even more clear in the top-left plot of Fig.~\ref{xs-hh-large}, where we show the dependence of $\sigma_{\rm LO}$ on $\cbs$ for the different phenomenologically relevant configurations that will be analysed in sec.~\ref{sec:bounds}, namely, $ZHH$ at 500 GeV collisions and  WBF$~HH$ at 1, 1.4 and 3 TeV collisions. 

Using a similar layout, in Fig.~\ref{xs-hh-large} we display three other plots, which show the dependence of $\sigma_{\rm LO}$, $ \Delta\sigma_{\bar c_6}(HH)$ and  $\Delta\sigma_{\bar c_8}(HH)/\cbe$  on $\cbs$ for different processes and energies. Specifically, in the upper-right plot we show the case of $ZHH$ at 500 GeV, while in the lower plots we show WBF$~HH$ at 1 TeV (left) and 3 TeV (right). In these three plots we display $\sigma_{\rm LO}$, which has also been shown in the top-left plot, as a black line and $ \Delta\sigma_{\bar c_6}(HH)$ and  $\Delta\sigma_{\bar c_8}(HH)/\cbe$  as a blue and red line, respectively. Thus the blue line directly shows the $\cbe$-independent part of $\sigma^{\rm pheno}_{\rm NLO}$, while the red one corresponds to the coefficient in front of the $\cbe$-dependent part $\Delta\sigma_{\bar c_8}(HH)$, which in turn depends on $\cbs$. For both cases, a short-dashed line is used when $ \Delta\sigma_{\bar c_6}(HH)$ or  $\Delta\sigma_{\bar c_8}(HH)/\cbe$ are  negative.
From Fig.~\ref{xs-hh-large} we can see that not only for the LO prediction (top-left plot) but also for one-loop effects the $\cbs$ (as well $\cbe$) dependence is very different in $ZHH$ (top-right plot) and WBF$~HH$ (lower plots) production. On the other hand, as can be seen in the lower plots, besides a global rescaling factor, WBF$~HH$ results are not strongly affected by the energy of $e^+e^-$ collisions.\footnote{In the case of $ZHH$ there are larger differences with the energy, but  in our analysis we consider it only at 500 GeV. } In the case of $ZHH$ production at 500 GeV, the minimum of the LO cross section is at $\cbs\sim -3$, while for  WBF$~HH$ it is at $\cbs\sim 0.5$. This minimum is given by cancellations induced by the interference of diagrams featuring or not the $HHH$ vertex. Such pattern of cancellations is different in the  $ \Delta\sigma_{\bar c_6}(HH)$ one-loop contribution, which in absolute value is instead minimal at $\cbs=0$ and very large at large values of $\cbs$. For this reason, {\it e.g.}, for $\cbs < -3$ the  $ \Delta\sigma_{\bar c_6}(HH)$ one-loop contribution is larger than the LO cross section. This does not signal the breaking of the perturbative convergence, rather it is due to the large cancellations that are present in this region only in the LO cross section; as already said, the perturbative limits, which are derived in Appendix \ref{validity}, require $|\cbs|<5$ and correspond to the range of the plot. In the case of WBF$~HH$ production $ \Delta\sigma_{\bar c_6}(HH)$ is always smaller than $\sigma_{\rm LO}$, being negative for $\cbs>0$ and positive for $\cbs<0$. 

Regarding the $ \Delta\sigma_{\bar c_8}(HH)$ contribution, which we display in the red lines normalised with $1/\cbe$, the effect is very different in $ZHH$ and WBF$~HH$ production. In the case of $ZHH$ production $ \Delta\sigma_{\bar c_8}(HH)$ is always negative and the minimum in absolute value is very close to the minimum of the LO prediction. In the case of WBF$~HH$ production $ \Delta\sigma_{\bar c_8}(HH)$ change sign at $\cbs\sim-2$ and $\cbs\sim 0.5$, being positive between these two values and negative outside them. In general, in absolute value, the ratio $ \Delta\sigma_{\bar c_8}(HH)/\sigma_{\rm LO}$ is always below $\cbe \cdot 2 \%$ value. Still, given the allowed perturbative range $|\cbe|<31$ (see Appendix \ref{validity}), effects from  large values of $\bar c_8$ can be in principle probed.
\begin{figure}[t]
    \begin{center}\vspace*{-1.0cm}
        \includegraphics[width=0.22\textwidth]{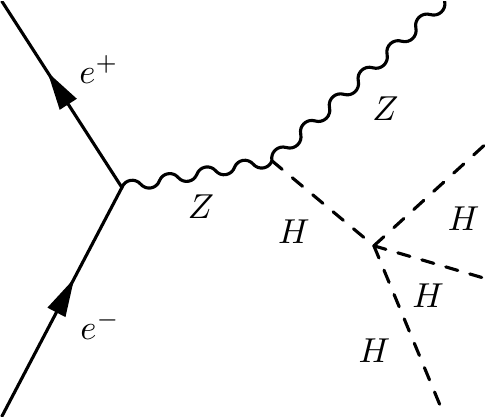}
        \includegraphics[width=0.22\textwidth]{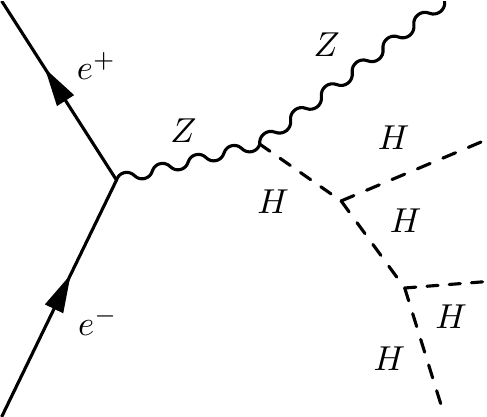}
        \includegraphics[width=0.22\textwidth]{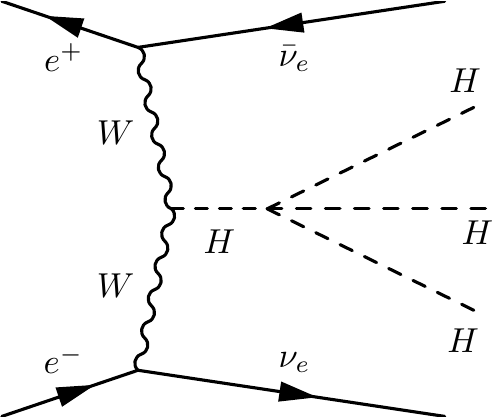}
        \includegraphics[width=0.22\textwidth]{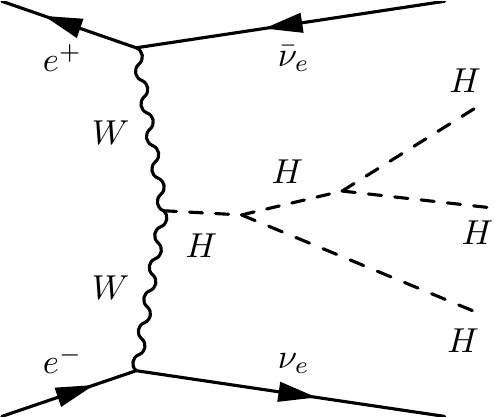}
        \caption{Representative tree-level Feynman diagrams for triple Higgs production.}
        \label{triple-H-figure}
    \end{center}
\end{figure}
\subsection{Triple Higgs production}
\begin{figure}[t]
    \begin{center}
        \includegraphics[width=0.45\textwidth]{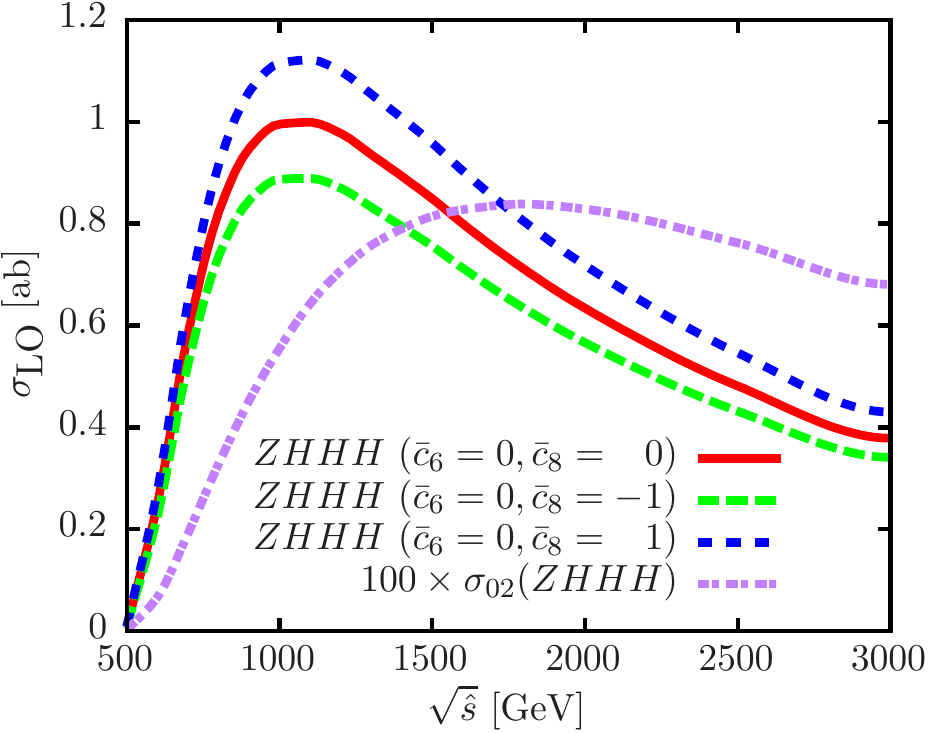}
        \hfill
        \includegraphics[width=0.45\textwidth]{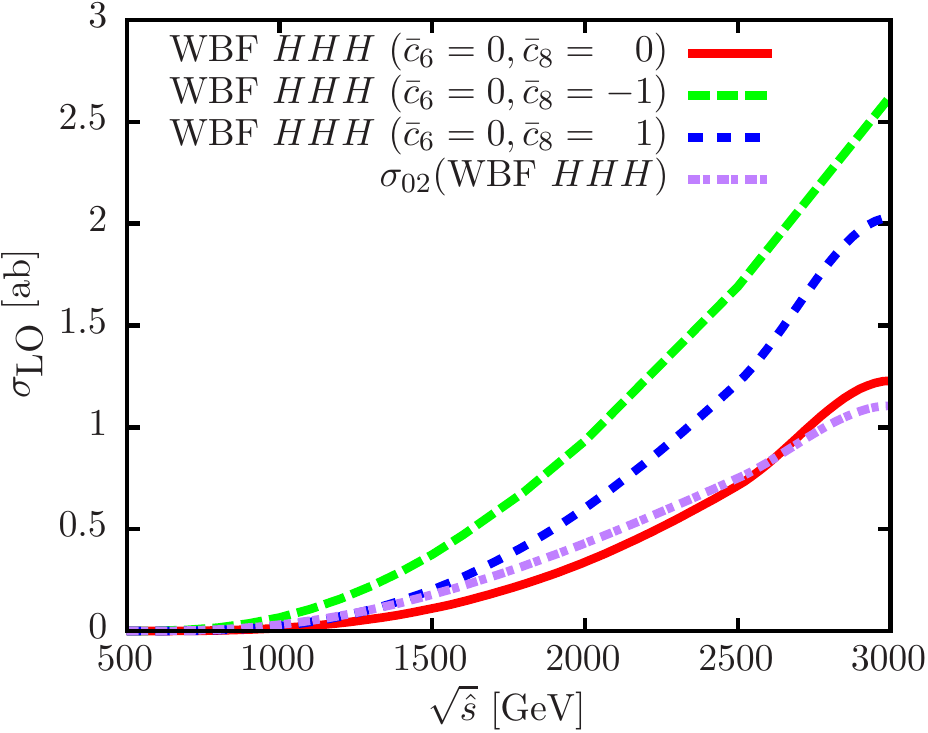}
        \caption{LO cross section of  $ZHHH$ and WBF$~HHH$ as function of $\sqrt{\hat s}$ for representative values of $\cbs$ and $\cbe$. The $\sigma_{02}$ component is also explicitly shown. Results refer to $P(e^{-},e^{+})=(-1.0,1.0)$. 
}
        \label{xs-hhh}
    \end{center}
\end{figure}

In triple Higgs production cubic and quartic self-couplings are present already at the tree-level and therefore both the leading dependences on $\cbs$ and $\cbe$ are already present at LO (see diagrams in Fig.~\ref{triple-H-figure}). Following the same notation used for double Higgs production, the cross section used for our phenomenological predictions can be written as
\begin{eqnarray}
    \sigma_{\textrm{LO}}(HHH)=\sigma_{00} + \sum_{1 \le i+2j\le 4} \sigma_{ij} \bar c_6^i \bar c_8^j \,,
    \label{sHHHLO}
\end{eqnarray}
where the $\sigma_{00}$ term corresponds to the LO SM prediction.  Similarly to the case  of double Higgs production at one loop, terms up to the eighth power in the $(v/\Lambda)$ expansion are present at the cross section level, although in this case only the fourth power is present at the amplitude level.
The upper bounds on $\cbs$ and $\cbe$ mentioned in the previous section and discussed in Appendix \ref{validity} have to be considered also in this case. It is important to note that although for large values of $\cbs$ and $\cbe$ loop corrections may be sizeable, at variance with double Higgs production, $\cbs$ and $\cbe$ are both entering at LO. Thus, when limits on $\cbs$ and $\cbe$ are extracted, loop corrections may slightly affect them, but only for large  $\cbs$ and $\cbe$ values.
\begin{table}[t]
\center
    \begin{tabular}{|c|c|c|c|c|}
        \hline
        ratio over  $\sigma_{00}$ &$\sigma_{10}$ & $\sigma_{20}$ & $\sigma_{30}$ & $\sigma_{40}$  \\
         \hline
         500 GeV &$(2.2,-9.0)$&$(1.4,8.5)$&$(0.3,34)$&$(0.02,19)$  \\
         \hline
        1 TeV &$(2.2,-3.7)$ &$(1.5,16)$&$(0.2,17)$&$(0.01,6)$   \\
                 \hline
        1.4 TeV &$(2.2,-3.4)$ &$(1.6,16)$&$(0.2,12)$&$(0.01,3.8)$   \\
         \hline
        3 TeV &$(2.2,-2.1)$ &$(1.9,7.6)$&$(0.2,3.8)$&$(0.01,1.0)$   \\
        \hline
                         \hline
        ratio over  $\sigma_{00}$ & $\sigma_{01}$ & $\sigma_{11}$ & $\sigma_{21}$ & $\sigma_{02}$ \\
                         \hline
         500 GeV &$(0.1,-4.0)$&$(0.1,-14)$&$(0.01,16)$& $(0.002,3.3)$  \\
         \hline
        1 TeV &$(0.1,-1.5)$&$(0.2,10)$&$(0.02,7.1)$&$(0.006,2.3)$   \\
                 \hline
        1.4 TeV &$(0.1,-1.0)$&$(0.2,9.2)$&$(0.02,5.2)$&$(0.009,2.0)$   \\
         \hline
        3 TeV &$(0.1,-0.3)$&$(0.3,4.1)$&$(0.03,1.6)$&$(0.02,0.9)$   \\
        \hline
 \end{tabular}
    \caption{$\sigma_{ij}/\sigma_{00}$ ratios for ($ZHHH$,~WBF$~HHH$). $\sigma_{ij}$ are defined in eq.~\eqref{sHHHLO}.\label{tablesij}}
\end{table}
In Tab.~\ref{tablesij} we give all the $\sigma_{ij}/\sigma_{00}$ ratios, so that the size of all the relative effects from the different NP contributions can be easily inferred.\footnote{There are large cancellations among the different contributions; more digits than those shown here have to be taken into account in order to obtain a reliable result.} 
In Fig.~\ref{xs-hhh}, we show $\sigma_{\textrm{LO}}$ at different energies for representative values of $\cbs$ and $\cbe$, including the SM case $(\cbs=0,\cbe=0)$ where  $\sigma_{\textrm{LO}}= \sigma_{00}$. 
There,  we also explicitly show the value of the $\sigma_{02}$ component, which factorises the $(\cbe)^2$ dependence. We can see that for $ZHHH$ production (left) 
the sensitivity to $\cbe$ is rather weak. The $\sigma_{02}$ component is just around 1\% of  $\sigma_{00}$, which means that even for large values of $\cbe$ the total cross section would not be large enough to be measurable at the future colliders considered in this study (see discussion in sec.~\ref{sec:bounds}). On the other hand, the total cross section of WBF$~HHH$ increases with the energy, as for single and double Higgs production. Especially, the $\sigma_{02}$ component is much larger; it is of the same order of  the SM $\sigma_{00}$ component. As an example, assuming $\cbe=1$($\cbe=-1$) and $\cbs=0$, $\sigma_{\textrm{LO}}$ at 3 TeV is 1.7 (2.2) times larger than $\sigma_{00}$. For large $\cbe$ values, $\sigma_{\rm LO}\approx \cbe ^2  \sigma_{02}\approx \cbe ^2 \sigma_{00}$.  As can be seen in Tab.~\ref{tablesij}, WBF is also very sensitive on $\cbs$; for large values of  $\cbs$ indeed $\sigma_{\rm LO}\approx \cbs ^4  \sigma_{40}$ and in particular  $ \cbs ^4  \sigma_{40} \approx \cbs ^4  \sigma_{00}$ at 3 TeV. All these effects are even larger at lower energies. 
%
\section{Bounds on the Higgs self-couplings}
\label{sec:bounds}
In this section we study how the $\cbs$ and $\cbe$ parameters can be constrained at future lepton colliders via the analysis of single, double, and triple Higgs production.
We consider four future \ee/ colliders, 
CEPC~\cite{CEPC-SPPCStudyGroup:2015csa},
FCC-ee~\cite{Gomez-Ceballos:2013zzn},
ILC~\cite{Baer:2013cma},
and CLIC~\cite{CLIC:2016zwp,Abramowicz:2016zbo},
with different operations modes\footnote{At the ILC also an operation mode at $\sqrt{\hat{s}}\sim350~\si{GeV}$ is expected, but studies  mainly focused on the scan of the $t\bar{t}$ production
threshold,  ignoring Higgs physics. At CLIC also a slightly different  scenario at $380~\si{GeV}$ instead of $350~\si{GeV}$ may be possible.}
that are summarised in Tab.~\ref{runmode}.
In the following, we will refer to the different scenarios as ``collider-$\sqrt{\hat{s}}$'' like, {\it e.g.}, CLIC-3000.
\begin{table}[tbp]
\center
    \begin{tabular}{|c|c|c|c|c|}
        \hline
        & $\sqrt{\hat{s}}$ [GeV] & $P(e^{-},e^{+})$ & Luminosity [ab$^{-1}$] & Relevant final states\\
        \hline
        CEPC & 250 & (0.0,0.0) & 5.0 & $ZH$, WBF$~H$\\
        \hline
        \multirow{2}{*}{FCC-ee} & 240 & (0.0,0.0) & 10.0 & $ZH$, WBF$~H$\\
                                & 350 & (0.0,0.0) & 2.6 & $ZH$, WBF$~H$\\
        \hline
        \multirow{3}{*}{ILC} & 250 & (-0.8,0.3) & 2.0 & $ZH$, WBF$~H$\\
                             & 500 & (-0.8,0.3) & 4.0 & $ZHH$, WBF$~H$\\
                             & 1000 & (-0.8,0.2) & 2.0 & $ZHHH$, WBF$~H(H(H))$\\
        \hline
        \multirow{3}{*}{CLIC} & 350 & (-0.8,0.0) & 0.5 & $ZH$, WBF$~H$\\
                              & 1400 & (-0.8,0.0) & 1.5 & $ZHHH$, WBF$~H(H(H))$\\
                              & 3000 & (-0.8,0.0) & 2.0 &  WBF$~H(H(H))$\\
        \hline
    \end{tabular}
    \caption{The different operation modes for \ee/ colliders considered in this work.\label{runmode}}
\end{table}
Although higher integrated luminosities can 
be attained at the CEPC and FCC-ee, energies 
as high as at the ILC and CLIC cannot be reached, since they are circular colliders.
As a result, only single Higgs production can be measured at the CEPC and FCC-ee,
and therefore only indirect constraints via loop corrections  can be set on $\cbs$.
Instead, at  the ILC and CLIC double Higgs production can be measured. With this process, both $\cbs$   and $\cbe$ can be constrained, the former via the direct dependence at the Born level and the latter via the indirect dependence through loop corrections.
Moreover, even triple Higgs production is kinematically allowed at the ILC and CLIC, allowing to set direct constraints on  $\cbe$.

In our analysis we consider the following two scenarios\footnote{One may be tempted to explore the regime $\cbs=0$ and $\cbe\ne 0$, too. However, this condition is neither motivated by an EFT expansion nor protected by any symmetry. As can be seen from eq.~\eqref{dc6}, a large $\bar c_8$ automatically generates a $\bar c_6$ component via loop corrections. }:
\begin{enumerate}
    \item  As expected from a well-behaving EFT expansion, the contribution from $\bar c_8$ is suppressed and we can safely set $\bar c_8=0$.
        We explore how well we can measure $\bar c_6$, not only assuming $\cbs\sim0$, {\it i.e.}, an SM-like configuration, but also allowing for large BSM effects via $\cbs\ne 0$. 
        
    \item The value of $\bar c_8$ can be different from zero and leads to non-negligible effects. We explore how well we can constrain $\bar c_8$ and how much $\bar c_8$ can affect the measurement of $\bar c_6$.
\end{enumerate}

First, we study the sensitivity of $ZH^n$ and WBF processes at the various colliders considered. Then we show combined results for the  ILC and CLIC. It is important to note, however, that single Higgs production depends on $\cbe$ only via two-loop effects, which we did not calculate in this work (see Tab.~\ref{tableprocesses}). Thus, we cannot directly combine single Higgs with double Higgs and triple Higgs in the case of Scenario 2. Nevertheless, we discuss the limit that can be obtained in single Higgs production under the assumption that the $\cbe$-dependent two-loop effects are negligible.
\subsection{Single Higgs production}
\begin{table}
\footnotesize
    \begin{tabular}{|c|c|c|c|c|c|c|}
        \hline
        & $\sqrt{\hat{s}}~$[GeV] & process & $\epsilon~[\%]$ & $C_1~[\%]$ &  $\bar c_6(\pm1\sigma)$ & $\bar c_6(\pm2\sigma)$\\
        \hline
        CEPC & 250 & $ZH$ & $0.51$ & 1.6 & $(-0.38,0.42)\cup(8.0,8.8)$ & $(-0.73,0.88)\cup(7.5,9.1)$\\
        \hline
        \multirow{3}{*}{FCC-ee} & 240 & $ZH$ & $0.4$ & 1.8 & $(-0.26, 0.28)\cup(9.4,9.9)$ & $(-0.51.0.57)\cup(9.1,10.2)$\\
                                & 240 & WBF$~H$ & $2.2$ & 0.66 & $(-2.81,5.1)$ & $(-4.3,6.6)$\\
                                & 350 & WBF$~H$ & $0.6$ & 0.65 & $(-1.15,3.4)$ & $(-1.89,4.1)$\\
        \hline
        \multirow{2}{*}{ILC}
        & 250 & $ZH$ & $0.71$ & 1.6 & $(-0.52,0.59)\cup(7.8,8.9)$ & $(-0.98,1.3)\cup(7.1,9.4)$\\
        & 500 & WBF$~H$ & $0.23$ & 0.63 & $(-0.56, 2.7)$ & $(-0.97,3.1)$ \\
        & 1000 & WBF$~H$ & $0.33$ & 0.61 & $(-0.78, 2.7)$ & $(-1.3,3.3)$ \\
        \hline
        \multirow{3}{*}{CLIC} & 350 & $ZH$ & $1.65$ & 0.59 & $(-2.48,4.3)$ & $(-3.80,5.6)$\\
                              & 1400 & WBF$~H$ & $0.4$ & 0.61 & $(-0.91,2.9)$ & $(-1.50,3.5)$ \\
                              & 3000 & WBF$~H$ & $0.3$ & 0.59 & $(-0.75,2.6)$ & $(-1.26,3.1)$ \\
        \hline
    \end{tabular}
    \caption{Expected precision $\epsilon$ for the measurements of single Higgs production modes and the expected
        $1\sigma$ and $2\sigma$ constraints on $\cbs$, assuming an SM measurement, are listed. The value of $\epsilon$ for the CEPC has been taken or obtained via a luminosity rescaling from ref.~\cite{CEPC-SPPCStudyGroup:2015csa},
        for the FCC-ee from ref.~\cite{Gomez-Ceballos:2013zzn},
        for the  ILC from refs.~\cite{Baer:2013cma, Barklow:2017suo} and
    for the CLIC from ref.~\cite{Abramowicz:2016zbo}.}
    \label{1h-prec}
\end{table}
In this section we discuss the constraints that can be obtained on $\cbs$ via the single-Higgs production modes.
As said, since the effects of $\bar c_8$ are unknown, we restrict our study to the case where it can be ignored, {\it i.e.}, Scenario 1.
We start by considering the case in which we assume that the Higgs potential is like in the SM ($\cbs=0$) and then we consider the BSM case with  $\cbs\ne0$.

In Tab.~\ref{1h-prec} we show $1\sigma$ and $2\sigma$ constraints on $\cbs$ that can be obtained via $ZH$ and WBF$~H$ at different energies and colliders, using eq.~\eqref{1h-c1}.
We show also the value of $C_1$  and the accuracy $\epsilon$ that can be achieved in any experimental setup, as provided in \cite{CEPC-SPPCStudyGroup:2015csa,Gomez-Ceballos:2013zzn, Baer:2013cma, Abramowicz:2016zbo, Barklow:2017suo} or obtained from them via a luminosity rescaling.\footnote{In the case of WBF$~H$ at ILC, {\it e.g.}, only the $H\to b \bar b$ has been considered for obtaining the value of $\epsilon$. Thus, smaller values of $\epsilon$ may be also achieved.} In general in this work, unless differently specified, we assume Gaussian distributions for the errors and no correlations among them, and the errors are rescaled according to cross section in BSM cases. In the results of Tab.~\ref{1h-prec} we did not take into account effects due to $\cbs$ in the Higgs decay, since, at variance with the LHC case, they can be in principle neglected at $e^+e^-$ colliders. Indeed, the total cross section of $e^+e^- \to ZH$ production can be measured via the recoiling mass method~\cite{Baer:2013cma}, without selecting a particular $H$ decay channel. Using the same method, the branching ratio of any (visible) decay channel can be precisely measured and used as input in the WBF$~H$ analysis, so that also in this case effects due to $\cbs$ in the Higgs decay can be neglected. Nevertheless, we explicitly checked that taking into account $\cbs$ effects in the decay for the $H\to b \bar b$ channel, which will be the one most precisely measured, results in Tab.~\ref{1h-prec} are almost unchanged.

As can be seen in eq.~\eqref{1h-c1}, not only a linearly $\cbs$ dependent term is present, but  also a $\cbs^2$ one. Since $C_2$ is negative and $C_1$ is positive for both  $ZH$ and WBF$~H$, the SM cross section value is degenerate in $\cbs$; besides the SM case $\cbs=0$ also a second different $\cbs\ne 0$ condition is giving the same value of the cross section. While for the WBF$~H$ this second solution is close to $\cbs=2$, in $ZH$ at 240-250 GeV this is around  $\cbs=9$, depending on the energy. As a result, the two solutions being close to each other, in WBF$~H$ the 1$\sigma$ and 2$\sigma$ intervals are always broad, while in $ZH$  at 240-250 GeV we see two narrow intervals: one around $\cbs=0$ and one around $\cbs=9$. Note that for CLIC-350 also $ZH$ is yielding a broad interval as a constraint, since $\epsilon$  is larger and $C_1$ is smaller. Via the combined measurement of $ZH$ and WBF$~H$ processes, or including LHC results in a global fit, the $\cbs$ region around   $\cbs=9$ can be excluded. In conclusion, assuming no other BSM effects, the best constraints on $\cbs$ via {\it single Higgs production} can be obtained at low energy and high luminosity.
        \begin{figure}[t]
        \center
        \includegraphics[width=0.5\textwidth]{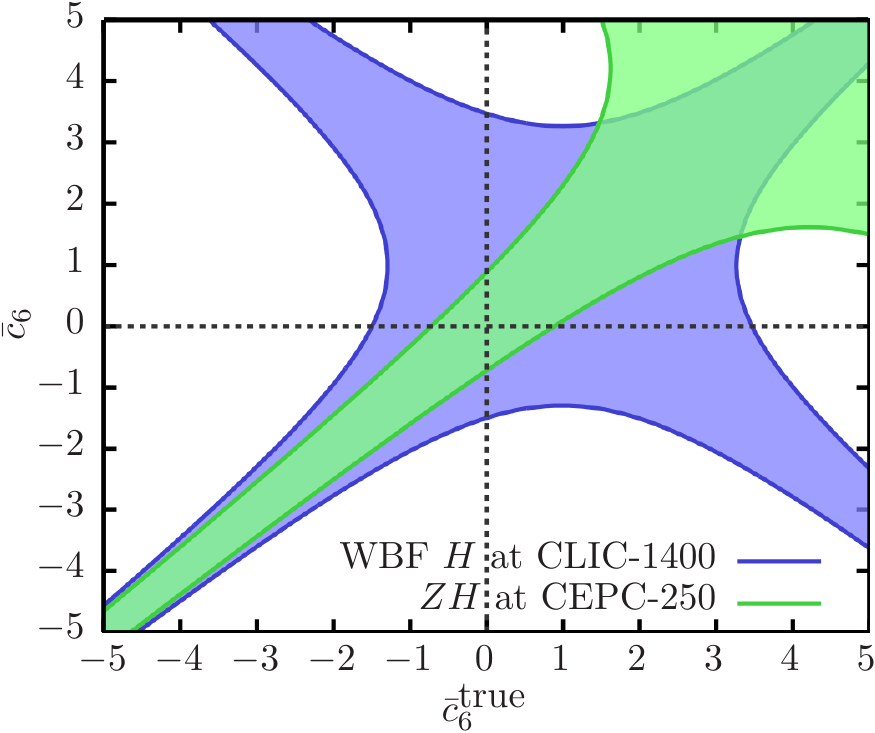}
    \caption{The $2\sigma$ bounds on $\bar c_6$ as a function of $\cbs^{\rm true}$ in single Higgs production in the Scenario 1 described in the text. Two representative cases are considered:  $ZH$ at CEPC-250 and WBF$~H$ at CLIC-1400.\label{fig:1h-c6-bounds}}
        \end{figure}

We now consider the situation in which $\cbs$ has a value different from zero, which will denote as $\cbs^{\rm true}$, and we explore the constraints that can be set on $\cbs$, by varying the value of $\cbs^{\rm true}$. In Fig.~\ref{fig:1h-c6-bounds} we consider $ZH$ at CEPC-250 and WBF$~H$ at CLIC-1400 as examples.\footnote{The case of CLIC-350 can be directly seen in Fig.~\ref{fig:combined-c6}, which is described later in the text. In this case, since the value of $\epsilon$ is larger, weaker constraints can be set w.r.t. CEPC-250. } The bands in the plot show which constraints on $\cbs$ (y-axis) can be set, depending on the value of $\cbs^{\rm true}$ (x-axis). We considered only the $-5<\cbs, \cbs^{\rm true}<5$ range, so that results can be directly compared with the analogous analysis performed in the next section for double Higgs production, where this range cannot be extended without violating perturbativity (see Appendix \ref{validity}). The ``X'' shapes of the $ZH$ and WBF$~H$ bands can be understood as follows. In the limit of zero uncertainties two solutions can be obtained from the equation      $ \sigma^{\rm pheno}_{\rm NLO}(\cbs)=\sigma^{\rm pheno}_{\rm NLO}(\cbs^{\rm true})$:
\begin{eqnarray}
\cbs&=&\cbs^{\rm true}\, ,\label{sol1}\\
\cbs&=&-\frac{C_1}{\delta Z_H^{{\rm SM}, \lambda}}-2-\cbs^{\rm true}\, , \label{sol2}
\end{eqnarray}
which intersect each other at the point $P=\big(-\frac{C_1}{2\delta Z_H^{{\rm SM}, \lambda}}-1, -\frac{C_1}{2\delta Z_H^{{\rm SM}, \lambda}}-1\big)$. For $ZH$ at CEPC-250 $P=(4.2,4.2)$ and for WBF$~H$ at CLIC-1400  $P=(1.0,1.0)$.\footnote{If we consider $ZH$ at FCC-ee-240 we obtain $P=(4.9,4.9)$.}
The uncertainties  $\epsilon$, however,  are not negligible and determine the width of the branches, which are centred on the solutions in eqs.~\eqref{sol1} and \eqref{sol2}.

For $ZH$ production, due to the large value of $C_1$, only one branch is present in the $-5<\cbs, \cbs^{\rm true}<5$  region. Instead, for WBF$~H$,
        since $C_1$ is small, SM-like scenarios $\cbs^{\rm true}\sim0$ lies in the
        intersection region of the branches. Thus, as already previously discussed, $ZH$ provides stronger constraints for $\cbs^{\rm true}\sim0$. On the contrary, for  $\cbs^{\rm true}\sim 4$, WBF$~H$ constraints are stronger. We remind the reader that it is not obvious that the LHC, even after accumulating 3000 fb$^{-1}$ of luminosity, will be able to exclude a value $\cbs \sim 4$. Still, with a single measurement for $\cbs^{\rm true}\sim 4$ both the intervals around $\cbs \sim 4$ and
$\cbs \sim -3$ are allowed, but the latter may be probed also at the LHC. As shown in Tab.~\ref{1h-prec}, also for $ZH$ and $\cbs^{\rm true}\sim0$ there is a second interval in the constraints, but it is outside the range of the plot.
        \subsection{Double Higgs production}

We now turn to the case of double Higgs production. The expected precisions $\epsilon$ for the measurements considered in our analysis\footnote{ Note that the value of $\epsilon$ listed in ref.~\cite{Behnke:2013lya, Abramowicz:2016zbo} are for a different luminosities than those considered in Tab.~\ref{runmode}. Since the statistical uncertainty is the dominant one, the values of $\epsilon$ in Tab.~\ref{2h-prec} have been obtained by rescaling those of ref.~\cite{Behnke:2013lya, Abramowicz:2016zbo} proportionally to the square root of the luminosity.} 
 are listed in Tab.~\ref{2h-prec}. Although double Higgs production cannot be measured as precise as single Higgs production, it depends on $\bar c_6$ at LO and therefore the sensitivity on this parameter is much higher. 
        \begin{table}[t]
        \center
            \begin{tabular}{|c|c|c|c|}
                \hline
                & $\sqrt{\hat{s}}$ [GeV] & process & $\epsilon$ \\
                \hline
                \multirow{2}{*}{ILC \cite{Behnke:2013lya}} & 500 & $ZHH$ & $19\%$  \\
                                                          & 1000 & WBF$~HH$ & $23\%$ \\
                \hline
                \multirow{2}{*}{CLIC \cite{Abramowicz:2016zbo}} & 1400 & WBF$~HH$ & $33\%$ \\
                                                               & 3000 & WBF$~HH$ & $15\%$ \\
                \hline
            \end{tabular}
            \caption{Expected precision $\epsilon$ for the measurements of double Higgs production processes. For the ILC the values of $\epsilon$ have been obtained rescaling the values in \cite{Behnke:2013lya} to the luminosity of Tab.~\ref{1h-prec}. In the case of CLIC, we have derived the values of $\epsilon$ via the relations in section 9 of ref.~\cite{Abramowicz:2016zbo}. }
            \label{2h-prec}
        \end{table}

\begin{figure}[t]
                \center
    \includegraphics[width=0.6\textwidth]{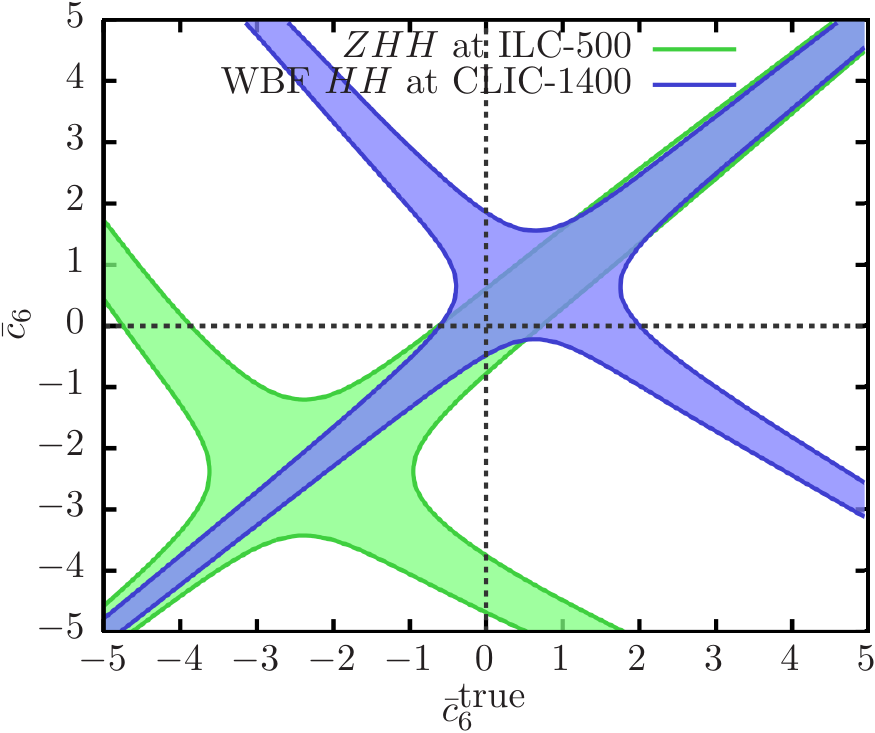}
    \caption{The $2\sigma$ bounds on $\bar c_6$ as a function of $\cbs^{\rm true}$ in double Higgs production in the Scenario 1 described in the text. Two representative cases are considered:  $ZHH$ at ILC-500 and WBF$~HH$ at CLIC-1400.\label{fig:2h-c6-bounds}}
        \end{figure}

We start our analysis considering Scenario 1, where we set $\bar c_8=0$.
  As can be seen in sec.~\ref{sec:calcHH}, the WBF$~HH$ dependence on $\bar c_6$ is similar for different energies. For this reason, for Scenario 1, we show  WBF$~HH$ only for CLIC-1400,  together with $ZHH$ at ILC-500.
Similarly to Fig.~\ref{fig:1h-c6-bounds}, which concerns the case of single Higgs production, in Fig.~\ref{fig:2h-c6-bounds} we plot the constraints that can be set on $\cbs$, by varying the value of $\cbs^{\rm true}$.   Also in $\sigma_{\rm LO}(HH)$ both a linear and quadratic dependence on $\bar c_6$ are present, leading to ``X''-shape bands. The ``X''-shape is slightly asymmetric due to the one-loop $\sigma_{30}$ and $\sigma_{40}$ contributions that are present in $\sigma^{\rm pheno}_{\rm NLO} (HH)$, see eq.~\eqref{sHHNLOpheno}, which we always use in our study. The central points of the ``X'' bands are around $(\cbs^{\rm true},\cbs)=(-2.5,-2.5)$ for $ZHH$ at ILC-500, and around $(\cbs^{\rm true},\cbs)=(0.5,0.5)$ for WBF$~HH$ at CLIC-1400. For this reason, although the WBF$~HH$ band is narrower due to a larger $\bar c_6$ dependence, for values $\cbs^{\rm true}\sim0$,  $ZHH$ at ILC-500 is giving better constraints. On the other hand, for values $\cbs^{\rm true}\ne0$ and especially $\cbs^{\rm true}\sim-2.5$, WBF$~HH$ at CLIC-1400 is leading to better constraints.
It interesting to note that the central points of the ``X'' bands in WBF $H$ and WBF $HH$ are very close, while for $ZH$ and $ZHH$ they are different.
This implies that the combination of the information from WBF single and double Higgs production would not exclude any of the branches of the ``X'' shape. Thus, the information from $ZH$ or $ZHH$ is necessary for this purpose. We will comment again this point in sec.~\ref{sec:combined}.

We now  consider Scenario 2. Specifically, we assume that the true value for $\cbs$ is $\cbs^{\rm true}$ and that the measured cross section for double Higgs production is $ \sigma^{\rm measured}=\sigma^{\rm pheno}_{\rm NLO}(\cbs=\cbs^{\rm true},\cbe=0)$ and we show which value of $(\cbs, \cbe)$ can be constrained via the prediction of $\sigma^{\rm pheno}_{\rm NLO}(\cbs,\cbe)$. 
Starting with the SM case, we show results for $ \sigma^{\rm measured}=\sigma^{\rm pheno}_{\rm NLO}(\cbs=0,\cbe=0)$  in Fig.~\ref{s2-0-0}. We consider the range $|\cbs| < 5$ and $|\cbe| < 31$, because as explained in Appendix \ref{validity} for larger values the perturbative calculations cannot be trusted. The plot on the left shows the constraints for $ZHH$ and WBF$~HH$ at the ILC-500, while the one on the right those for WBF$~HH$ at CLIC-1400 and CLIC-3000.  First of all we can notice that the constraints on $\cbs$ are weaker than in Scenario 1. Also, no constraints on $\cbe$ independently from $\cbs$ can be set. On the other hand, the largest part of the $(\cbs,\cbe)$ plane can be excluded and the shape of the band depends on the process. It is important to note that this results depend on the choice of the renormalisation scale $\mu_r$ and therefore the scale at which  $\cbs(\mu_r)$ and $\cbe(\mu_r)$ are measured. Our results refers to $\mu_r=2 \mh$, which corresponds to the production threshold for the $HH$ pair and therefore to the phase-space region associated to the bulk of the cross section. While the region close to the SM $(\cbs\sim 0,\cbe\sim 0)$ is very mildly affected by this choice, we warn the reader that the border of the plane $|\cbs| \sim 5$ and $|\cbe| \sim 31$ can be strongly affected.

\begin{figure}[t
        ]
            \includegraphics[width=0.45\textwidth]{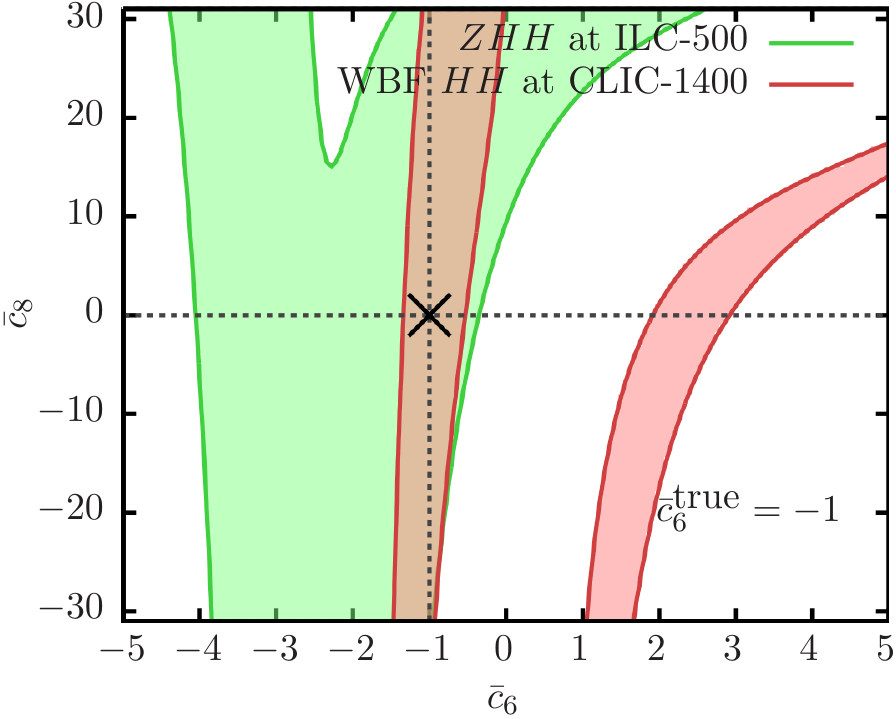}
            \hfill
            \includegraphics[width=0.45\textwidth]{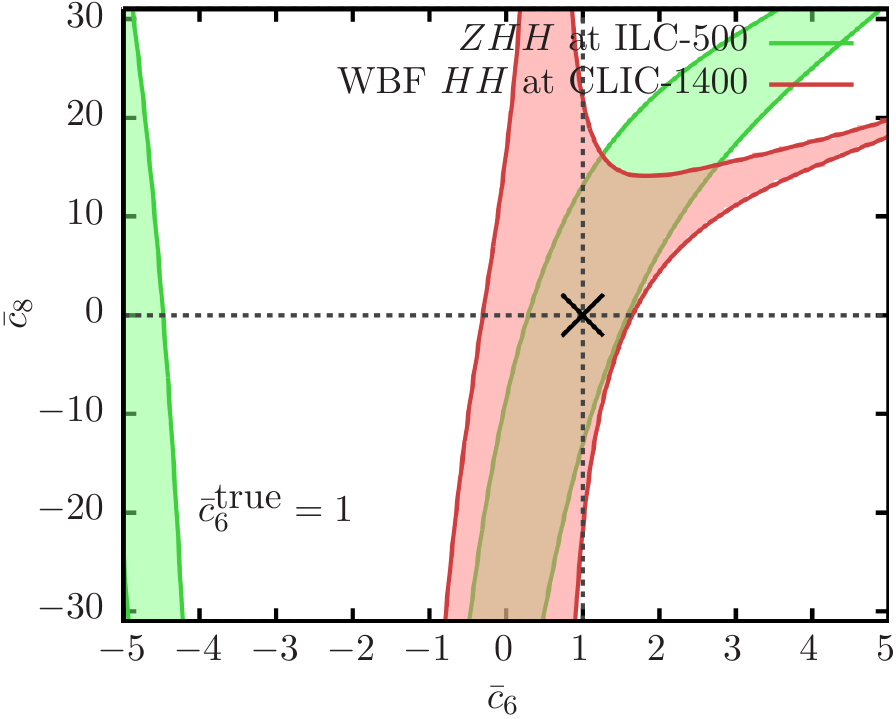}

            \includegraphics[width=0.45\textwidth]{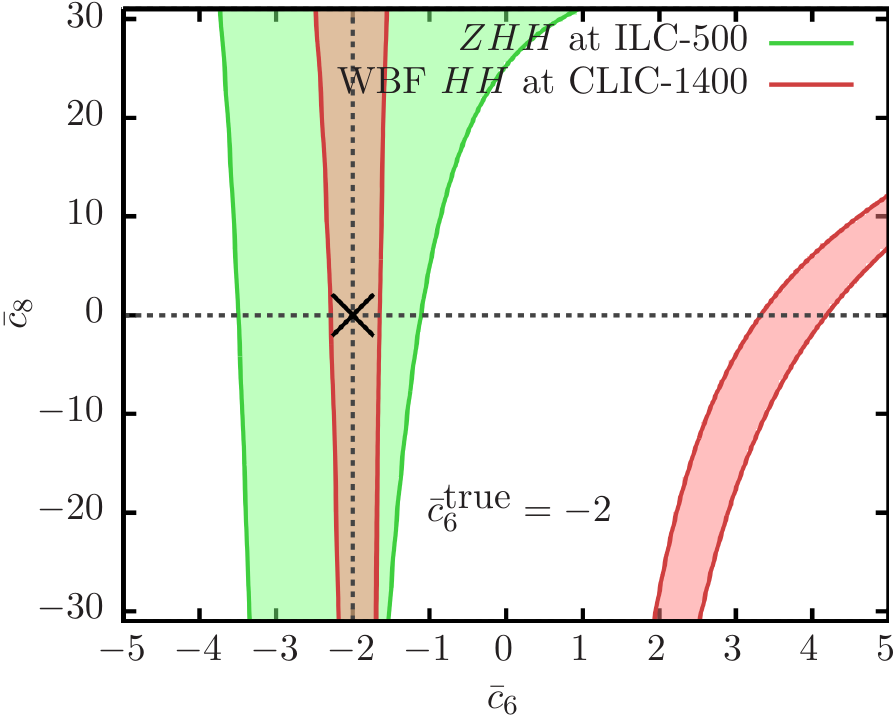}
            \hfill
            \includegraphics[width=0.45\textwidth]{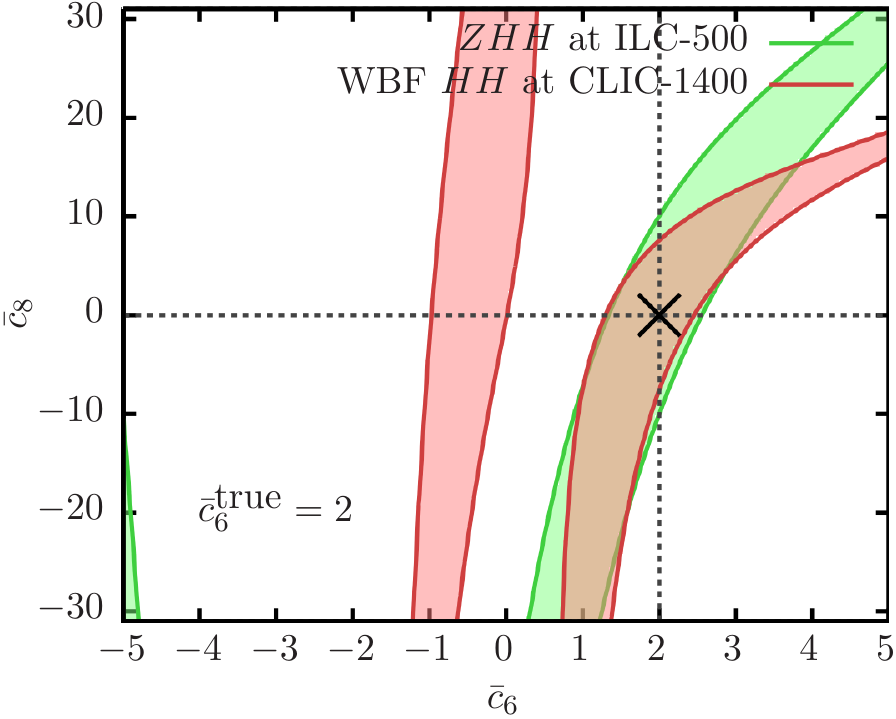}

            \includegraphics[width=0.45\textwidth]{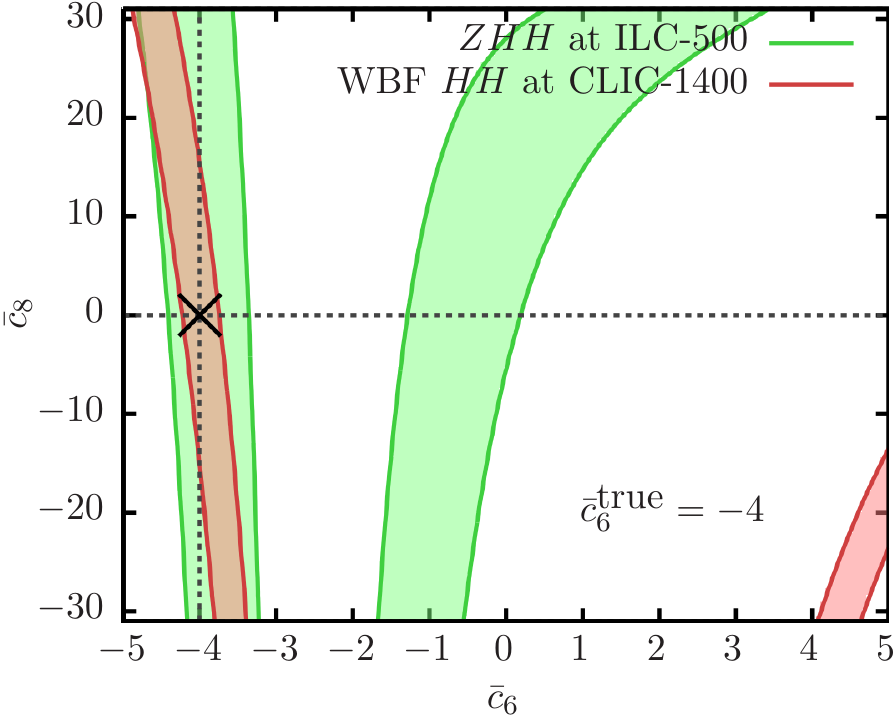}
            \hfill
            \includegraphics[width=0.45\textwidth]{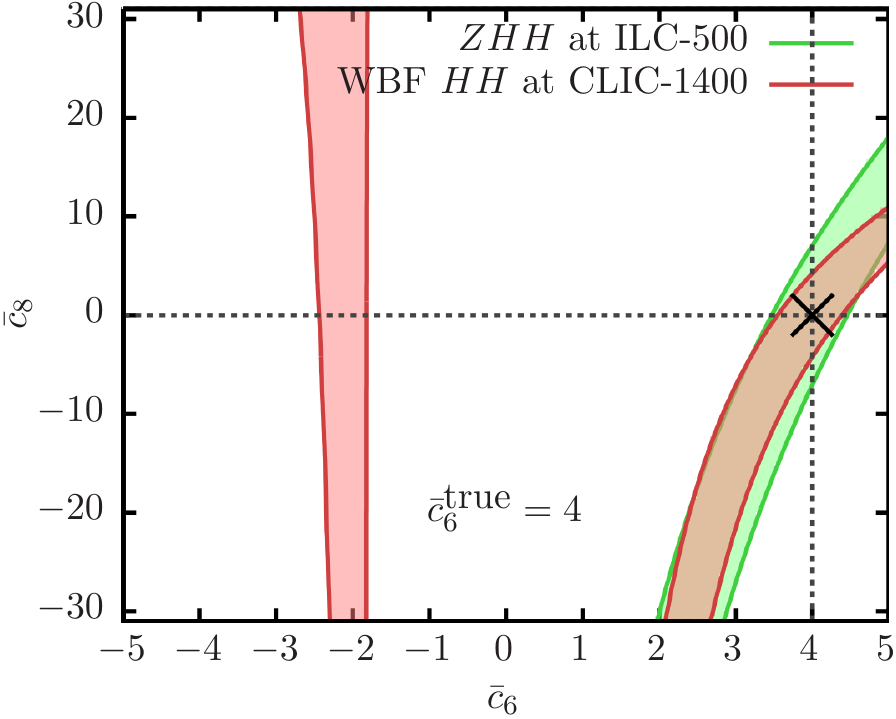}
            \caption{$2\sigma$ bounds in the $(\bar c_6, \bar c_8)$ plane assuming BSM cross sections in double Higgs production corresponding to $(\bar c_6^{\rm true}, \bar c_8^{\rm true} =0)$ in the Scenario 2 described in the text, with  $\bar c_6^{\rm true}=-4,-2,-1,1,2,4$ marked in the plots with a cross. All plots show results for  $ZHH$  at ILC-500 and WBF$~HH$ at CLIC-1400. \label{s2-nonsm}}
        \end{figure}

We then consider how the constraints in the $(\bar c_6,\bar c_8)$ plane depend on the value of $ \sigma^{\rm measured}$. We consider BSM configurations $ \sigma^{\rm measured}=\sigma^{\rm pheno}_{\rm NLO}(\cbs=\cbs^{\rm true},\cbe=0)$ with $\cbs^{\rm true} \ne 0$.\footnote{As the total cross section depends on $\cbe$ mildly, we do not expect that the constraints depend on $\cbe^{\rm true}$} In Fig.~\ref{s2-nonsm} we show the plots for the values of $\cbs^{\rm true}=-4,-2,-1,1,2,4$; in each plot the point $(\cbs^{\rm true},\cbe^{\rm true}=0)$ is displayed with a cross and the value of $\cbs^{\rm true}$ is given. For these plots, only results for $ZHH$ at ILC-500 and WBF$~HH$ at ILC-1000 are displayed. Similarly to the SM case, given a value of $\cbs^{\rm true}$, the constraints on $\cbs$ independent from $\cbe$ are weaker than those in Scenario 1. However, also in these cases, the largest part of the $(\cbs, \cbe)$ plane can be excluded and the shapes of the bands strongly depend both on the process and the value of $\cbs^{\rm true}$. In all cases, $ZHH$ and WBF$~HH$ sensitivities are complementary; as we will see in sec.~\ref{sec:combined}, their combination improves the constraints in the $(\cbs, \cbe)$ plane.
This is a clear advantage for the ILC, where both $ZHH$ and WBF$~HH$ can be precisely measured.

The shapes of the green and red bands can be qualitatively explained as follow. Without $\cbe$ effects the green and red bands would simply consist of either two separate (narrow) bands or a single large band, consistently with the results that could be obtained by vertically slicing the bands in Fig.~\ref{fig:2h-c6-bounds}. The $\cbe$ effects bend the bands, leading to the shapes that can be observed in Fig.~\ref{s2-nonsm}. It is interesting to note that the improvement from CLIC-1400 to CLIC-3000 is rather mild. The main reason is that the increment of the WBF $HH$ cross section is compensated by the decrement of its dependence on $\cbs$, which can be directly observed in the top-left plot of Fig.~\ref{xs-hh-large}. 
        \begin{figure}[tbp]
            \includegraphics[width=0.45\textwidth]{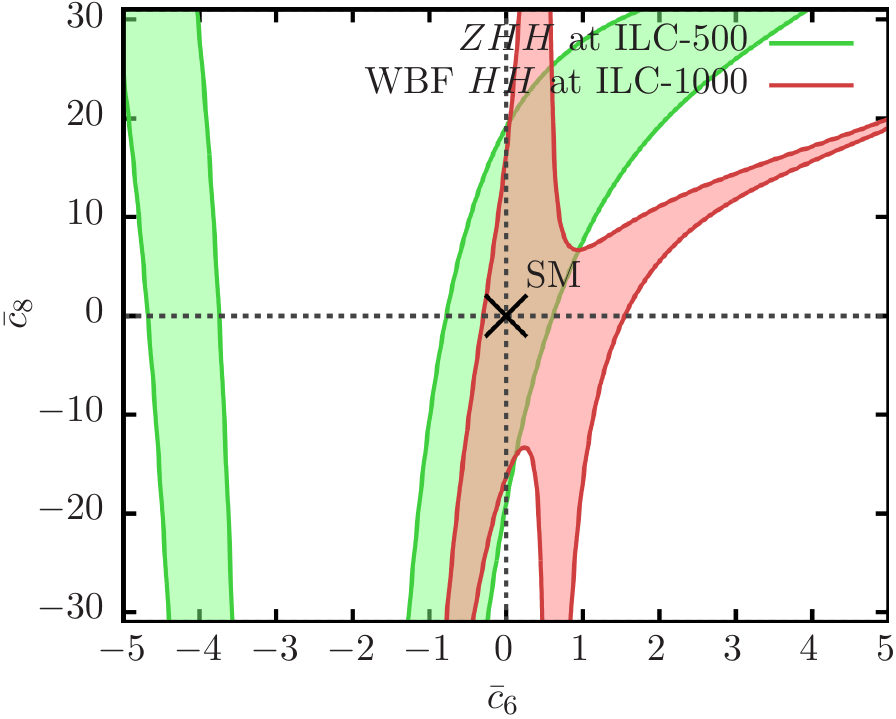}
            \includegraphics[width=0.45\textwidth]{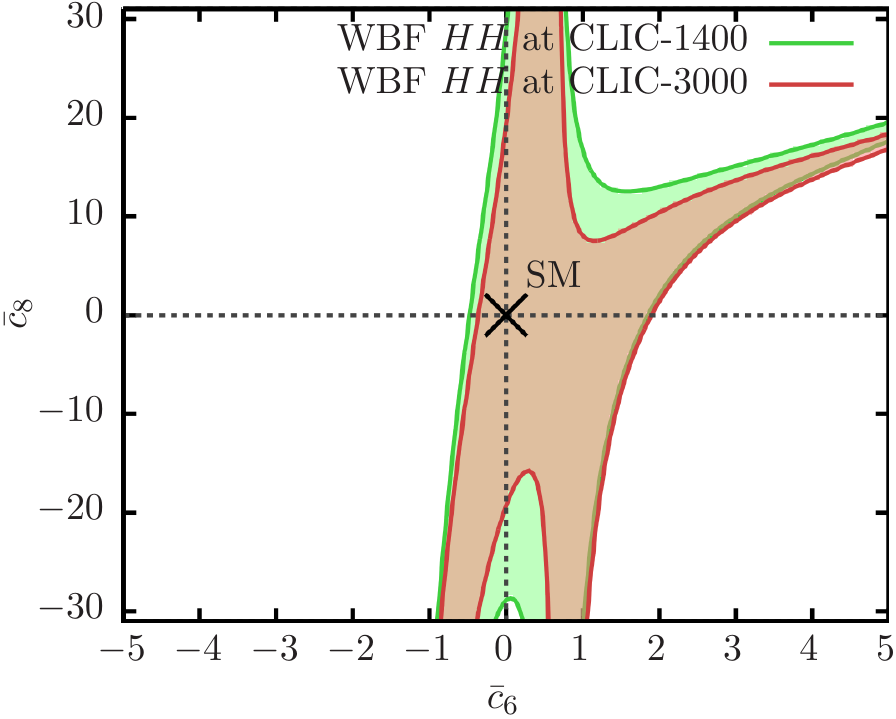}
            \caption{$2\sigma$ bounds in the $(\bar c_6, \bar c_8)$ plane assuming SM cross sections for double Higgs production in the Scenario 2 described in the text. Left:  $ZHH$ at ILC-500 and WBF$~HH$ at ILC-1000. Right: WBF$~HH$ at CLIC-1400 and CLIC-3000.\label{s2-0-0}}
        \end{figure}
        \subsection{Triple Higgs production}
        \label{triple}
We now consider the case of triple Higgs production. In the SM $ZHHH$ and WBF$~HHH$ production processes have a too small cross section for being observed.
As an example, if we consider $LR$-polarised beams at 1 TeV and the dominant  decay into a $b \bar b$ pair for the three Higgs bosons  and into jets for the $Z$ boson, about 6 $\si{ab^{-1}}$ of integrated luminosity would be necessary for one signal event in the SM. As can be seen in Fig.~\ref{xs-hhh}, with WBF$~HHH$ the cross section is even smaller in the SM,  on the other hand this process has a strong sensitivity on $\cbe$, due to the large value of $\sigma_{02}$ factorising the $\cbe^2$ dependence. Thus, limits on $\cbs$ and $\cbe$ can be set, but only considering Scenario 2 where $\cbe$ can be different from zero.

At variance with double Higgs production, given the very small number of events, we cannot set limits on the $(\bar c_6, \bar c_8)$ plane by assuming $ \sigma^{\rm measured}(HHH)=\sigma_{\rm LO}(\cbs=\cbs^{\rm true},\cbe=0)$. Indeed, the number of events expected is close to zero and a Gaussian fit cannot be performed. Rather, we have to assume events are zero and compare them with the expected value of events for a given  $(\bar c_6^{\rm true}, \bar c_8^{\rm true})$ performing a Poissonian analysis.\footnote{In fact, for the case of CLIC-3000, large $\cbs$ values would lead to $\sim 5$ expected events. We will consider this effect in the combined analysis in sec.~\ref{sec:combined}. }         We assume that the other SM backgrounds are giving zero events and we estimate the signal efficiency  $\varepsilon_{HHH}$ by rescaling the one known for  WBF$~HH$ production $\varepsilon_{HH}$. In practice, for both WBF$~HHH$   and $ZHHH$ production we estimate the signal efficiency to be $\varepsilon_{HHH}=\varepsilon_{HH}^{\frac{3}{2}}=4.7\%$, where $\varepsilon_{HH}$ has been taken from ref. \cite{Tian:2013qmi}.

\begin{figure}[t]
        \includegraphics[width=0.45\textwidth]{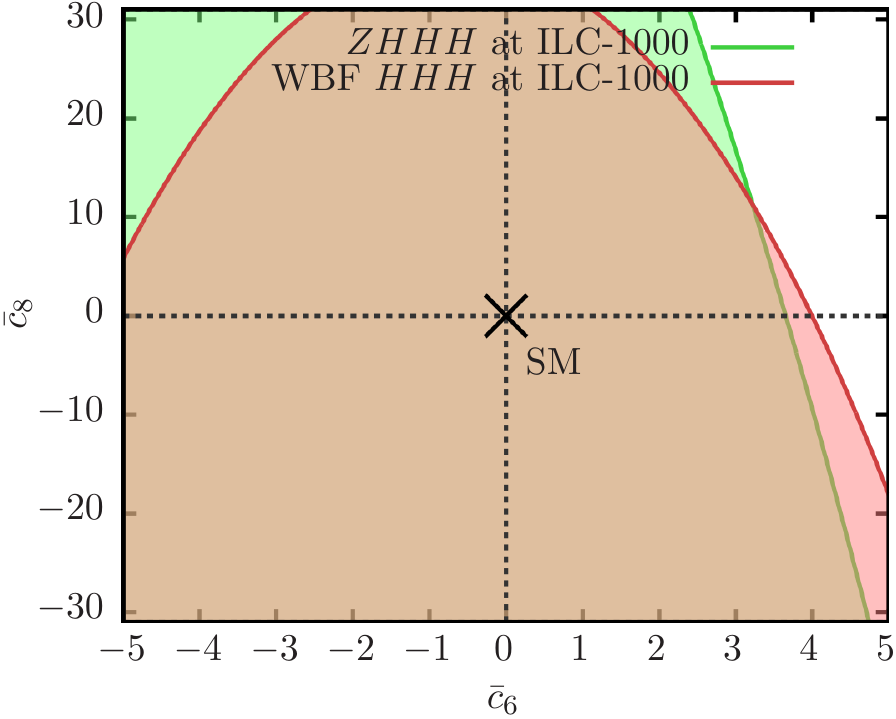}
            \hfill
        \includegraphics[width=0.45\textwidth]{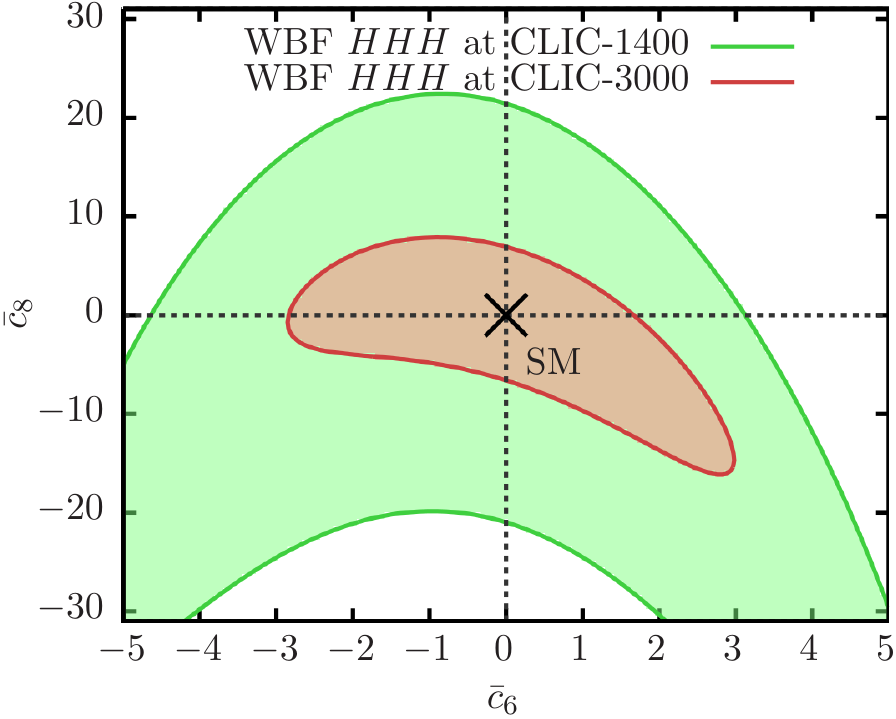}
        \caption{$2\sigma$ bounds in the $(\bar c_6, \bar c_8)$ plane assuming SM cross sections in triple Higgs production in the Scenario 2 described in the text. Left:  $ZHHH$ and WBF$~HHH$ at ILC-1000. Right: WBF$~HHH$ at CLIC-1000 and CLIC-1400. }
            \label{mle-3h}
        \end{figure}

In Fig.~\ref{mle-3h}, we show the  $2\sigma$ bounds in the  $(\bar c_6,\bar c_8)$ plane.  The plot on the left shows the constraints for $ZHHH$ and WBF$~HHH$ at ILC-1000, while the one on the right those for WBF$~HHH$ at CLIC-1400 and CLIC-3000. 
As can be seen, at ILC-1000 almost all the $(\bar c_6,\bar c_8)$ plane is compatible with a zero event condition, both for $ZHHH$ and WBF$~HHH$  production. On the other hand, at CLIC-1400 and  especially at CLIC-3000 a vast area of the plane can be excluded via the study of WBF$~HHH$ production. In particular, at CLIC-3000, the constraint on $\bar c_8$ are comparable to those obtainable at a future 100 TeV hadron collider~\cite{Chen:2015gva,Kilian:2017nio}. The constraints on $\bar c_6$ are instead worse than in the double Higgs production case.
\subsection{Combined bounds}
\label{sec:combined}
We now investigate the constraints that can be obtained via the combination of the information from single, double and triple Higgs production. We consider both Scenarios 1 and  2 and, as already mentioned,
in the  case of Scenario 2 we combine only results from double and triple Higgs production. We show in parallel the limits on $\cbs$ from single Higgs production by assuming that the $\cbe$-dependent two-loop effects are small.

We start discussing the Scenario-1 analysis, separately considering the ILC and CLIC. For both colliders we progressively include results at higher energies in three stages. In the case of the ILC, we start with $ZH$ at ILC-250, in a second step we include $ZHH$ and WBF$~H$ results from ILC-500 and finally $ZHHH$ and WBF$~H(H(H))$ from ILC-1000. Instead, in the case of the CLIC, we start with $ZH$ at CLIC-350, in a second step we include WBF$~H(H(H))$ and $ZHHH$ results from CLIC-1400 and finally  WBF$~H(H(H))$ results from CLIC-3000. In the case of triple Higgs production we assume that we observe as many events as predicted by $\sigma_{\textrm{LO}}(HHH)$ in eq.~\eqref{sHHHLO}, with $\cbe=0$.

\begin{figure}[t]
    \includegraphics[width=0.45\textwidth]{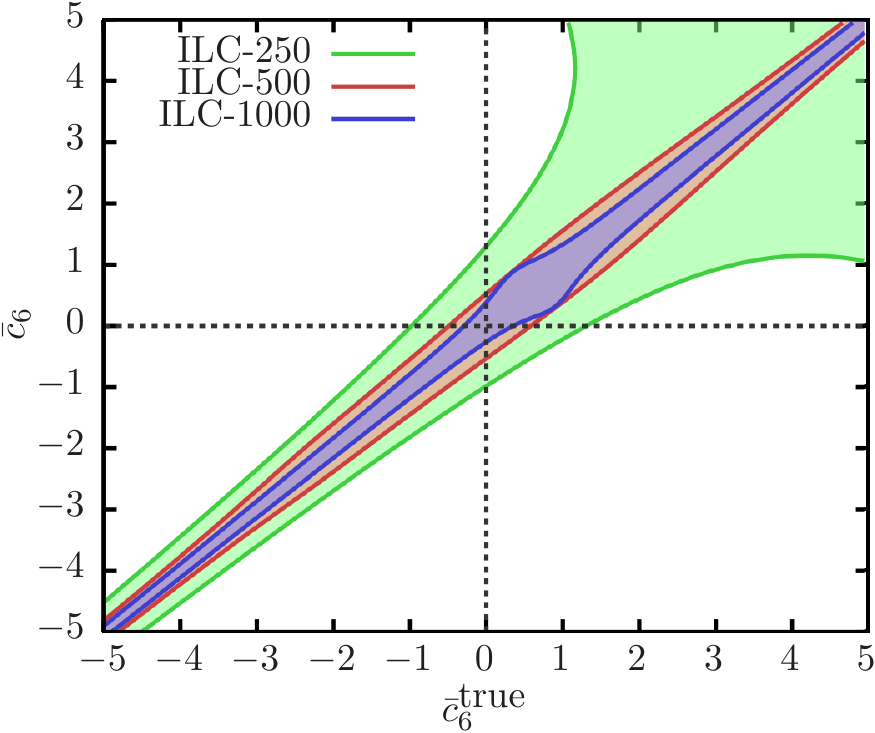}
    \hfill
    \includegraphics[width=0.45\textwidth]{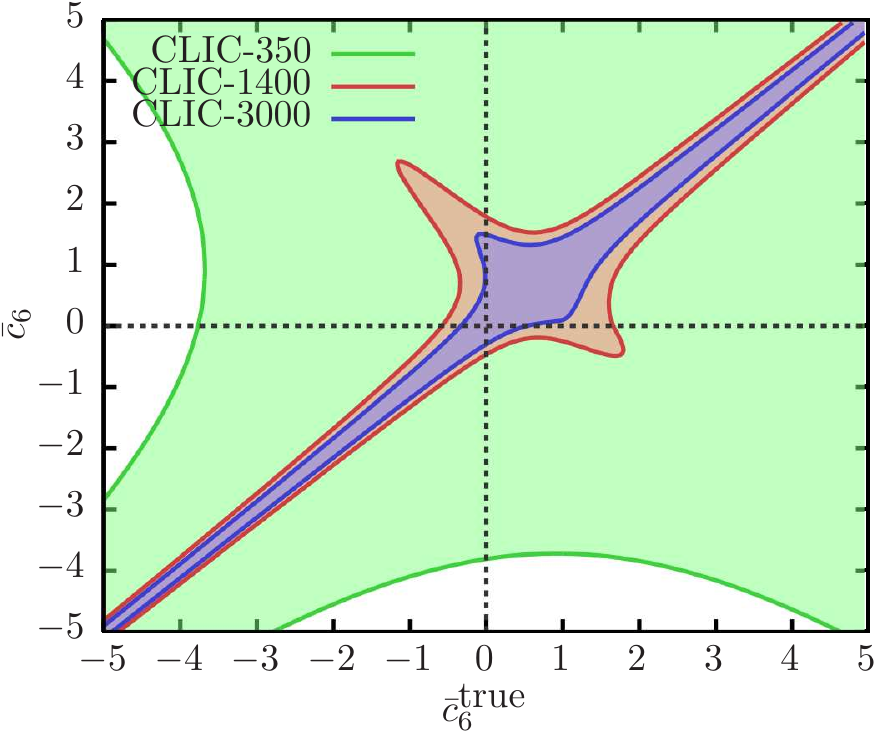}
    \caption{ Combined $2\sigma$ constraints on $\bar c_6$ as a function of $\cbs^{\rm true}$ for the ILC (left) and the CLIC (right) in the Scenario 1 described in the text.\label{fig:combined-c6} }
\end{figure}

In Fig.~\ref{fig:combined-c6}, we show the combined results for the ILC (left) and CLIC (right) assuming Scenario 1.
In the first stage, both ILC-250 and CLIC-350 constraints are worse than those of CEPC-250 shown in Fig.~\ref{fig:1h-c6-bounds}. This is due to a lower precision in the measurements ($\epsilon$) and for CLIC-350 also a smaller value of $C_1$. However, in the second stage, including results at higher energies, for both colliders constraints are much stronger,  since double Higgs production becomes available. Especially, combining single and double Higgs production the ``X'' shape disappears and only the band around the line $\cbs=\cbs^{\rm true}$ remains.\footnote{In the case of CLIC, where the $ZHH$ information is not entering the combination, also the information from triple Higgs production is necessary for this purpose.} In the case of the CLIC, bumps are still present at $\cbs\sim 1$, which originate from the centre of the ``X''-shape band for WBF$~H(H)$ at CLIC-1400, see Fig.~\ref{fig:1h-c6-bounds} and Fig.~\ref{fig:2h-c6-bounds}.  For the same reason, also for the ILC the band is slightly larger around $\cbs \sim 1$. In the third stage, constraints are improved  both for the ILC and CLIC. Still, the weaker bounds can be set for  $ \sim 0 < \cbs^{\rm true} < 1$, where  the center of the ``X''-shape band for WBF$~HH$ is located. In this region, constraints are better at the ILC thanks to the $ZHH$ contribution at 500 GeV, which helps to resolve this region.

We now consider Scenario 2. As done in the case of double Higgs production we assume that the true value for $\cbs$ is $\cbs^{\rm true}$ and  that $ \sigma^{\rm measured}=\sigma^{\rm pheno}_{\rm NLO}(\cbs=\cbs^{\rm true},\cbe^{\rm true}=0)$ while that for triple Higgs production we observe as many events as predicted by $\sigma_{\textrm{LO}}(HHH)$ in eq.~\eqref{sHHHLO}, with $\cbe=0$.
 In the case of the ILC, we consider $ZHH$ at ILC-500 and its combination with ILC-1000 results from $ZHHH$ and WBF$~HH(H)$ production. In the case of CLIC, we consider $ZHHH$ and WBF$~HH(H)$ production at CLIC-1400 and its combination with WBF$~HH(H)$ at CLIC-3000. Thus, while ILC-500 is not a combined result, being simply obtained for $ZHH$ production, all the others include information from both double and triple Higgs production. As already said, single Higgs production cannot be directly included in the combination, since its $\cbe$ dependence starts at two-loop level.

\begin{figure}[t]
\center
\includegraphics[width=0.45\textwidth]{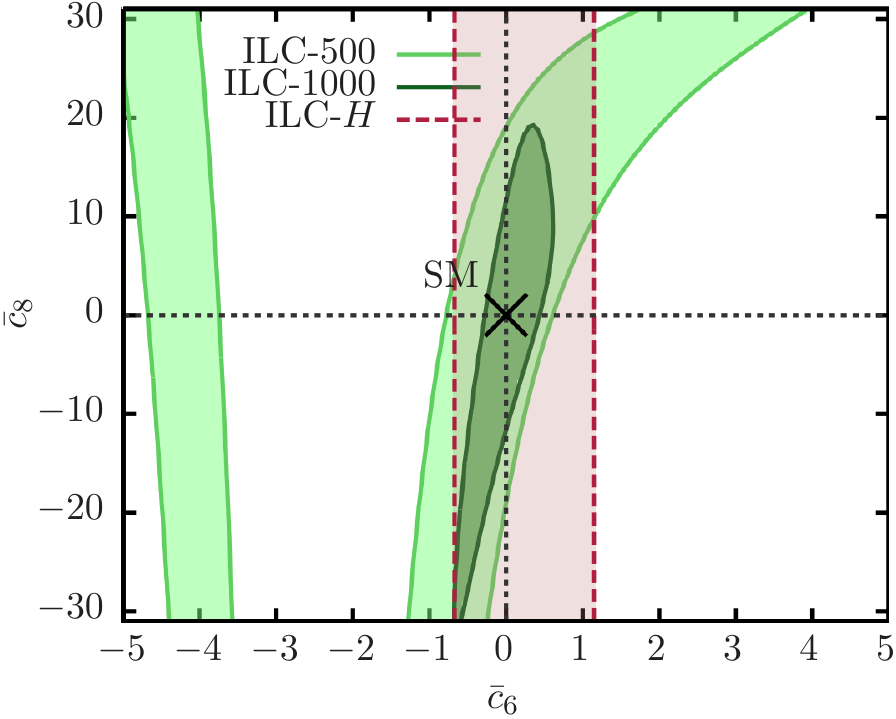}
\includegraphics[width=0.45\textwidth]{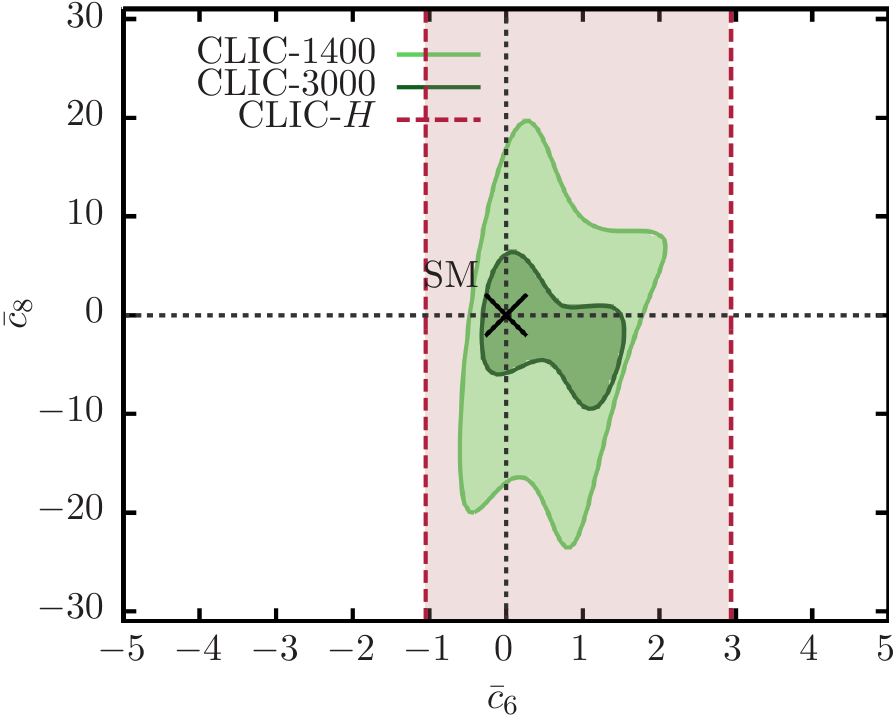}
    \caption{Combined $2\sigma$ constraints in the  $(\bar c_6,\bar c_8)$ assuming SM cross sections, at the ILC (left) and CLIC (right), in the Scenario 2 described in the text. ILC-$H$ and CLIC-$H$ refer to a combination of all single Higgs measurements at all energy stages for each collider under study.
\label{fig:combinedSM-c6c8}}
\end{figure}

 In Fig.~\ref{fig:combinedSM-c6c8} we show results for the SM case $(\cbs^{\rm true}=0,\cbe^{\rm true}=0)$ as green bands. There 
 we also show as red bands the limits on $\cbs$ extracted from single Higgs measurements at the ILC and CLIC\footnote{ More specifically,
 for the ILC, the single Higgs limit are combined results from $ZH$ at ILC-250,
 WBF $H$ at ILC-500, and WBF $H$ at ILC-1000,
 while for the CLIC, the single Higgs limit are combined results from $ZH$ at CLIC-350, WBF $H$ at CLIC-1400, and WBF $H$ at CLIC-3000.}, 
 assuming that the two-loop $\cbe$ dependence is negligible.
 Due to the available higher energies,  combined double and triple Higgs constraints at the CLIC are better than at the ILC. Indeed the WBF $HH(H)$ production cross section increases with the energy. On the other hand, single Higgs production can be better measured at the ILC and therefore the corresponding constraints on $\cbs$ are better than at the CLIC. 
 We notice that the only case where single Higgs results may be relevant in a further combination with those from double and triple Higgs production is the case of ILC-500, which is actually coming from only $ZHH$ production. Indeed, the combination of $ZH$ at ILC-250 and WBF $H$ at ILC-500 would help in removing the band around $\cbs=-4$, and shrinking the possible region for the band around SM value. On the contrary, at higher energies the 
 WBF $HH$  production is more relevant in constraining $\cbs$. 
Thus, with the exception of ILC-500, single Higgs production could be helpful in constraining the $(\cbs,\cbe)$ plane only if the dependence on $\cbe$ at two-loop is larger than what we assumed or if low-energy runs at higher luminosity, such as those at circular colliders, are considered.

\begin{figure}[t]
\center
\includegraphics[width=0.45\textwidth]{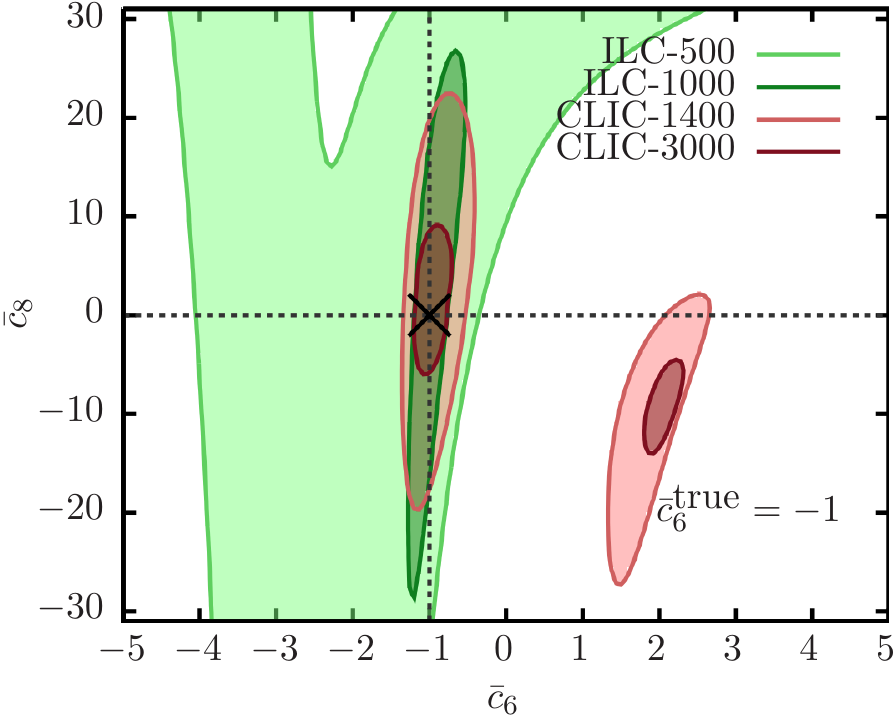}
\hfill
\includegraphics[width=0.45\textwidth]{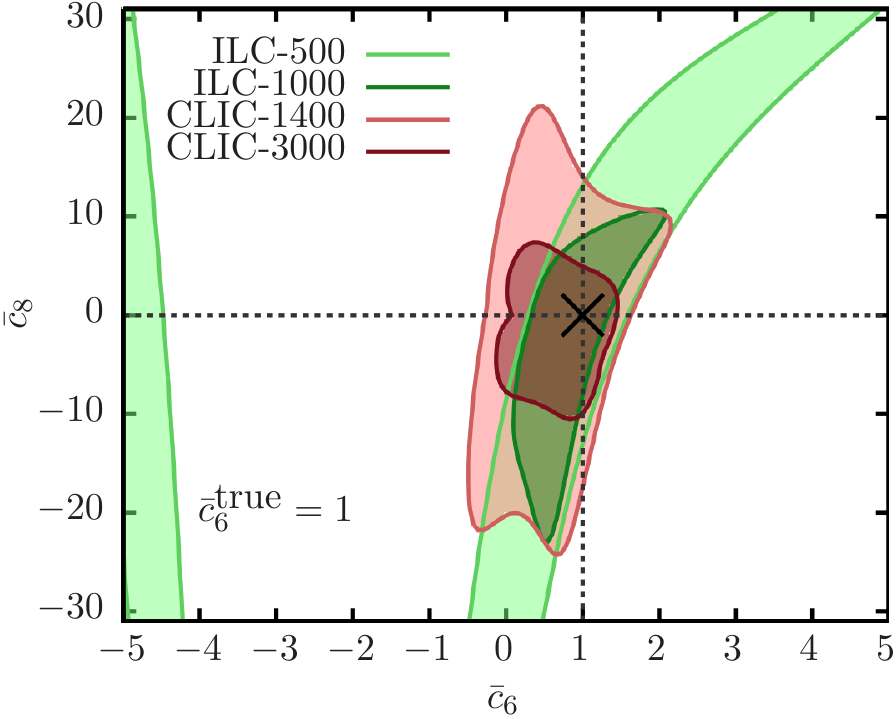}
\includegraphics[width=0.45\textwidth]{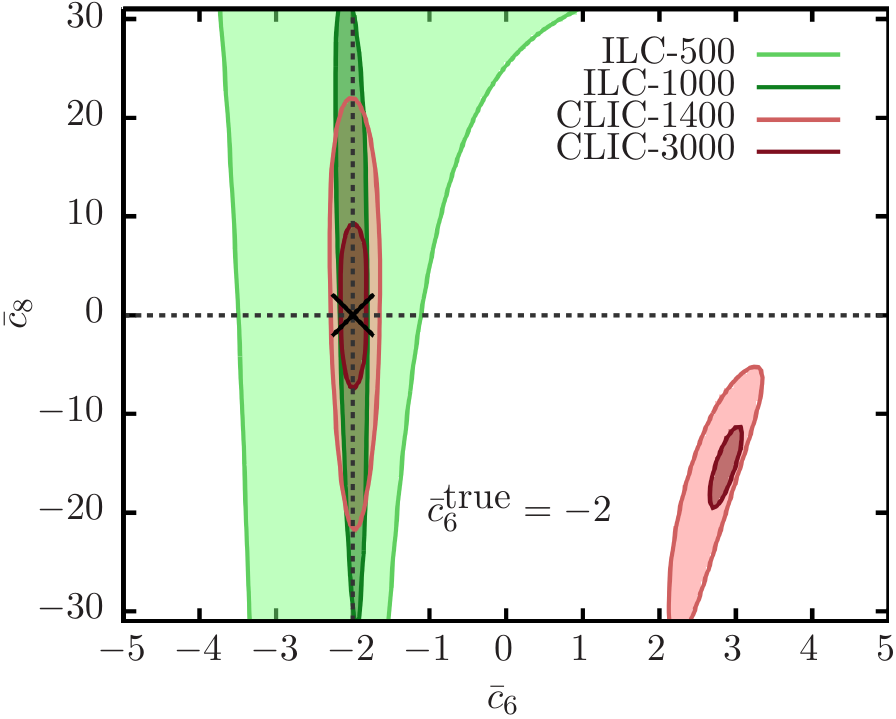}
\hfill
\includegraphics[width=0.45\textwidth]{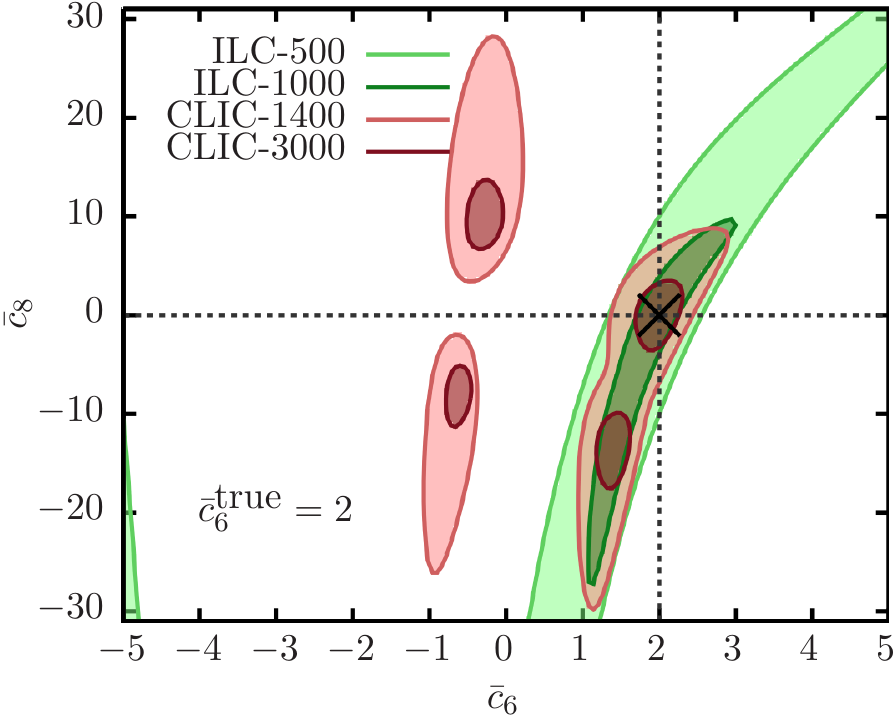}
\includegraphics[width=0.45\textwidth]{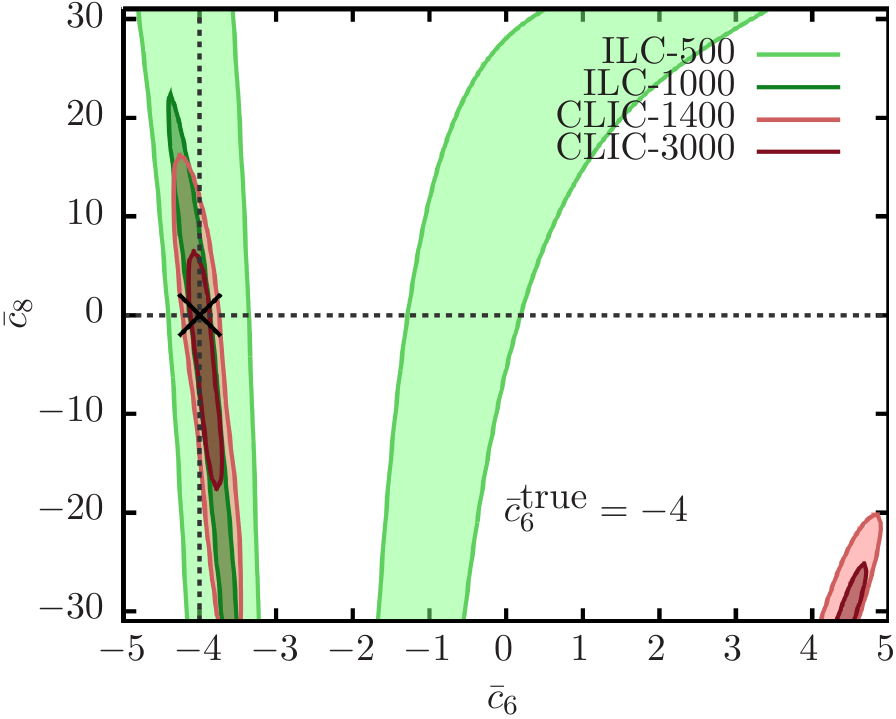}
\hfill
\includegraphics[width=0.45\textwidth]{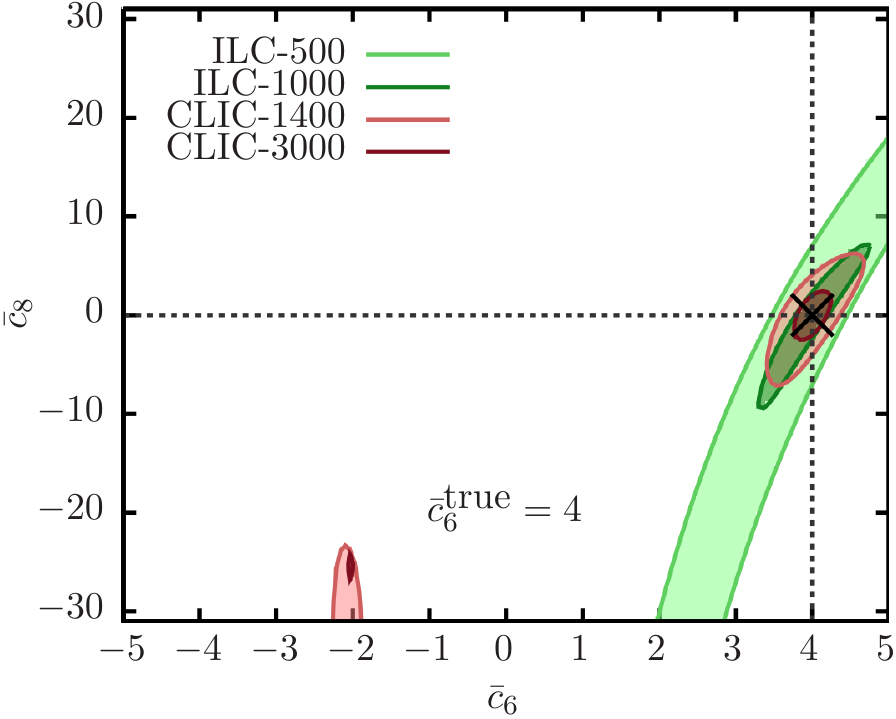}
    \caption{
    Combined $2\sigma$ bounds in the $(\bar c_6, \bar c_8)$ plane assuming BSM cross sections corresponding to $(\bar c_6^{\rm true}, \bar c_8^{\rm true} =0)$ in the Scenario 2 described in the text, with  $\bar c_6^{\rm true}=-4,-2,-1,1,2,4$ marked in the plots with a cross. 
\label{fig:combined-c6c8}}
\end{figure}

 In Fig.~\ref{fig:combined-c6c8} we show the constraints from the combination of double and triple Higgs for BSM cases $\cbs^{\rm true}=-4,-2,-1,1,2,4$. As already discussed for the SM case, constraints from single Higgs production are negligible for high energy $e^+ e^-$ colliders in this scenario under our assumptions and for this reason they are not shown. We display in each plot both CLIC and ILC bounds.
As we can see, both in the SM and in all BSM cases considered, the combination of results from double and triple Higgs production is always strongly improving the bounds. Also, with higher energies, stronger constraints can be set; the best results can be obtained combining results at CLIC-1400 with those at CLIC-3000, especially for $\cbs^{\rm true}\ne0$ since a non-zero number of events can be observed. It is interesting to note that CLIC bounds around $(\cbs^{\rm true}, \cbe^{\rm true})$ are less sensitive than at the ILC
on the value of $\cbs^{\rm true}$, featuring vertical elongated contours in the $(\cbs, \cbe)$ plane. The reason is that at CLIC bounds mainly comes from WBF$~HHH$, while at the ILC  mainly  from double Higgs production, both $ZHH$ and WBF$~HH$.

In conclusion we observed that low- and high-energy runs are useful for constraining the shape of the Higgs potential. Under the assumption of Scenario 1,  we have shown the complementarity of $ZH$ production at low energy with WBF$~HH$ information at higher energies. Under the Scenario 2, we have shown that the combination of the information from double and triple Higgs production, which is possible only at high energy, improves the constraints in the $(\cbs, \cbe)$ plane ({\it cf.} fig.~\ref{s2-nonsm} with fig.~\ref{fig:combined-c6c8}). 
\section{Conclusions}
\label{sec:conclusions}
Determining whether the scalar potential for the Higgs boson is the minimal one predicted by the SM is among the main targets of the current and future colliders. In this work, we have investigated the possibility of setting constraints on the shape of the Higgs potential via the measurements of single, double and triple Higgs production at future $e^+e^-$ colliders, considering the two dominant channels, {\it i.e.}, $Z$ boson associate production ($ZH^n$) and $W$ boson fusion WBF. In order to leave the possibility for the trilinear and quadrilinear couplings to vary independently, we have added to the SM potential two EFT operators $\frac{c_{6}}{\Lambda^{2}}\left(\Phi^\dag\Phi -\frac{1}{2}v^2 \right)^3$ and $\frac{c_{8}}{\Lambda^{4}}\left(\Phi^\dag\Phi -\frac{1}{2}v^2 \right)^4$ and calculated the tree-level and one-loop dependence on $c_6$ and $c_8$ for single and double Higgs production as well as tree-level results for triple Higgs production (see also Tab.~\ref{tableprocesses} in sec.\ref{sec:introduction}). 

One-loop corrections to single Higgs production, which depends only on $\lambda_3$ and thus $c_6$, have already been calculated and studied in the literature and we have confirmed previous results. On the other hand, the one-loop dependence on $\qual$ and therefore on $c_6$ and $c_8$ of double Higgs production has been calculated for the first time here. At variance with the case of single Higgs production, the EFT parametrisation is in this case compulsory and an anomalous coupling approach cannot be consistently used; the $c_6$ parameter is itself renormalised and receives corrections from both $c_6$ and $c_8$. We have provided all the necessary renormalisation constants and counterterms and expressed the finite one-loop results via analytical form factors that can be directly used in phenomenological applications. We have also motivated the inclusion of the ``$-\frac{1}{2}v^2 $'' term in the EFT parametrisation, which simplifies the renormalisation procedure by preserving the relations among the SM counterterms.  Nevertheless, results can always be easily translated to the $\frac{c'_{6}}{\Lambda^{2}}\left(\Phi^\dag\Phi \right)^3$ and $\frac{c'_{8}}{\Lambda^{4}}\left(\Phi^\dag\Phi \right)^4$ basis.    

In our phenomenological analyses we have considered several experimental setups at future $e^+ e^-$ colliders (CEPC, FCC-ee, ILC and CLIC) and have analysed the constraints that can be set on  $   \bar c_6 \equiv \frac{c_6 v^2}{\lambda \Lambda^2}$ and $\bar c_8\equiv \frac{4c_8v^4}{\lambda\Lambda^4}$. To this purpose we have considered two scenarios:
\begin{itemize}
    \item{Scenario 1:} the effects of $\cbe$ are negligible and we analyse the constraints that can be set on $\cbs$, both for the SM potential and in the case $\cbs^{\rm true}\ne 0$.
    \item{Scenario 2:} the effects of $\cbe$ are not assumed to be negligible and we analyse the constraints that can be set on the ($\cbs$, $\cbe$) plane, both for the SM potential and in the case $\cbs^{\rm true}\ne 0$.
\end{itemize}
In Scenario 1 the value of $\qual$ directly depends on $\tril$, while in Scenario 2 they are independent. We verified that requiring perturbative convergence sets upper bounds on the absolute values of $\cbs$ and $\cbe$, {\it i.e.},  $|\cbs | < 5$ and $|\cbs | < 31$. Thus, we have analysed the constraints that can be set in this region of the ($\cbs$, $\cbe$) plane.

In Scenario 1, the best constraints on $\cbs$ can be obtained from the combination of $ZH$ results from low-energy high-luminosity runs and  results from high-energy runs for $ZHH$ and WBF $(HH, HHH)$ production. On the other hand, in BSM cases $\cbs^{\rm true}\ne 0$,  WBF $H$ gives stronger constraints than $ZH$ production and similarly  WBF $HH$ production can be more sensitive than $ZHH$ production. 

In Scenario 2, since two-loop $\cbe$ effects for single Higgs production are not available, we combine only double and triple Higgs production,
and show the single Higgs bounds under the assumption that two-loop effects are negligible.
The combination of high-energy results from double and triple Higgs production gives the best constraints and in both cases the WBF channel is in general the most relevant.  Single Higgs production is only relevant for low-energy machines, and almost negligible once WBF $HH$ is available. For this reason, the higher is the energy, the stronger are the constraints that can be obtained in the ($\cbs$, $\cbe$) plane, both for the SM case and the BSM configurations with $\cbs^{\rm true}\ne 0$.

In both Scenario 1 and Scenario 2, although WBF $HH$ constraints alone are stronger than those for $ZHH$, the two production processes are in fact complementary and lead to improved results when they are combined. At high-energy $e^+e^-$ colliders triple Higgs production is not measurable in the SM, but its cross section strongly depends on the value of $\cbe$. In particular, at CLIC-3000, the constraint that can obtained on  $\bar c_8$ via WBF $HHH$ production are comparable to those obtainable at a future 100 TeV hadron collider.

In conclusion, we have demonstrated that the analysis of single, double and triple Higgs production at $e^+e^-$ colliders can be exploited for constraining the trilinear and quartic coupling via direct and loop-induced indirect effects. In this first sensitivity study we have assumed that BSM effects on the couplings of the Higgs boson with other particles can be neglected. This assumption has already been shown to be reasonable for the SM case in Scenario 1 in ref.~\cite{DiVita:2017vrr}, yet further studies will be necessary for the other configurations considered in this work. Also, as already mentioned, another possible sensitivity on the $\cbe$ parameter may be obtained from the high-precision measurements of single Higgs production at future $e^+ e^-$ colliders. To establish what kind of constraints could be reached on $\cbe$ in this case, a two-loop computation of $e^+ e^- \to ZH$ will be needed.

 \section*{Acknowledgements}

We acknowledge many enlightening discussions and continuous collaboration on the Higgs self-coupling determination with Giuseppe Degrassi, Christophe Grojean and Qi-Shu Yan.  This work is supported in part by the ``Fundamental interactions" convention FNRS-IISN 4.4517.08. This work has received funding from the European Union's Horizon 2020 research and innovation programme as part of the Marie Sk\l{}odowska-Curie Innovative Training Network MCnetITN3 (grant agreement no.~722104) and by  F.R.S.-FNRS under the ``Excellence of Science - EOS'' - be.h project n.~30820817. The work of D.P. is supported by the Alexander von Humboldt Foundation, in the framework of the Sofja Kovalevskaja Award Project ``Event Simulation for the Large Hadron Collider at High Precision''.

\appendix

    \section{One-loop renormalisation in double Higgs production in EFT}
    \label{appendix:ren}
In this section we provide all the ingredients that are necessary for one-loop renormalisation in double Higgs production with arbitrary $c_6$ and $c_8$ values.
 First of all, it is important to note that the only quantities that are renormalised and receive a contribution from $c_6$, $c_8$ and $c_{10}$ are 
\begin{equation}
Z_H\,, \quad \mh\,, \quad T\,, \quad c_6 \, , \label{CTcs}
\end{equation}
where $Z_H$ is the Higgs wave function and $T$ is the tadpole contribution, which we cancel via the $\delta t $ counterterm so that the physical value of $v$ does not get shifted. All the other quantities do not receive additional one-loop contributions on top of the SM ones, including $\delta v$, which is completely of SM origin. 

Thus, for our calculation the necessary ingredients for the renormalisation of the virtual corrections are:
\begin{equation}
\delta Z_H = \delta Z_H^{\rm SM} + \delta Z_H^{\rm NP}\, \, ,\label{dZHSMNP}
\end{equation}
\begin{equation}
\delta \mh^2 = (\delta \mh^2)^{\rm SM} + \delta (\mh^2)^{\rm NP} - 6  \frac{c_{6}}{\Lambda^2} v^3 \delta v\, ,
\label{newmh}
\end{equation}
\begin{equation}
\delta t = \delta t^{\rm SM} + \delta t^{\rm NP} \, ,
\label{dtsmnp}
\end{equation}
\begin{equation}
\delta c_6 = \delta c_6^{\rm NP} \, .
\label{dc6np}
\end{equation}
All the quantities with ``SM'' as apex are the SM contributions and can be found in \cite{Denner:1991kt}, those with ``NP'', which indeed stands for new physics, are the new contributions from $c_6$, $c_8$ and $c_{10}$. Besides $c_6$, which is renormalised in the $\MSbar$ scheme, all the other EW input parameters are assumed to be renormalised on-shell, with exception of fine structure $\alpha$, which we renormalise in the $G_\mu$-scheme. This is relevant for our calculation since in the SM  the renormalisation of $v$ is related to the charge renormalisation, $\delta Z_e$,  
\begin{equation}
    \frac{\delta v}{v}=\frac{\delta s_W}{s_W}-\delta Z_e +\frac{\delta \mw^2}{2 \mw^2}\,.
    \label{deltavSMexp}
\end{equation}
The appearance of the extra quantity $- 6  \frac{c_{6}}{\Lambda^2} v^3 \delta v$ in eq.~\eqref{newmh} is due to the presence of $v$ in the parametrisation of eq.~\eqref{VNP}, which as we said has an impact in the renormalisation procedure.  Before giving the explicit formulas for $ \delta Z_H^{\rm NP}, \delta (\mh^2)^{\rm NP}, \delta t^{\rm NP}$ and the counterterm for the $H^3$ vertex and $H$ propagator, we briefly discuss this technical aspect.

The explicit term $v$ used in the parametrisation of eq.~\eqref{VNP} is a subtle quantity. In a tree-level analysis it can be trivially identified with the location of the minimum of $V(\Phi)$, which defines the ground state $\ket{0}$ of the Higgs field 
\begin{equation}
|\bra{0} \Phi \ket{0} |^2=\frac{v^2}{2}\, .
\end{equation}
\begin{figure}[t]
\begin{center}\vspace*{-1.0cm}
    \begin{align}
        \vcenter{\hbox{\includegraphics[width=0.10\textwidth]{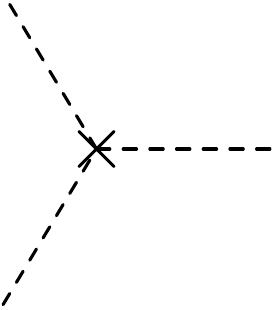}}}
=\delta \lambda_{H^3}
\end{align}
\caption{The counter term for the triple Higgs vertex.}
\label{ruletri}
\end{center}
\end{figure}
However, strictly speaking, the $v$ appearing in eq.~\eqref{VNP} is, like $\Lambda$, just a mass parameter that we chose for mapping ${c}_{2n}^{\prime}$ into $c_{2n}$.
In principle we could have chosen a generic mass $M\neq v$, but we would have not got any advantage. On the contrary, with $M= v$, SM relations such as 
\begin{equation}
 \delta \lambda =\frac{\delta \mh^2}{2v^2}
 - \frac{\mh^2 \delta v}{v^3}  + \frac{\delta t}{2 v^3}\, ,
\label{dlambdaok}
\end{equation}
\begin{eqnarray}
\delta v = \delta v^{\rm SM}~ \leftrightarrow ~ \delta v^{\rm NP} =0 \,,
\end{eqnarray}
 are preserved (see ref.~\cite{Sirlin:1985ux}); they would be different using eq.~\eqref{VNP_Stnotation}.
The crucial point is that  at one loop, or even at higher orders, $|\bra{0} \Phi \ket{0} |$ is involved in the renormalisation, while the term $v$ in  eq.~$\eqref{VNP}$ is not; as said it is just a mass parameter tuned to $v$ for our purpose. For this reason, relations among the different renormalisation constant of the SM parameters are unaltered, but in the case of $\mh$  and  $H^n$ vertexes the definitions of the renormalisation counterterms contain additional terms. 

This mechanism is at the origin of the aforementioned term in eq.~\eqref{newmh} as well as to some additional terms (second line of eq.~\eqref{dl3_cont}) that appear in the counterterm for the $H^3$ vertex in Fig.~\ref{ruletri}, where  $\delta     \lambda_{H^3}$  reads
\begin{eqnarray}
 \delta     \lambda_{H^3}&=&-i 6\Big[\frac{c_6v^3}{\Lambda^2} \left(\frac{ \delta c_{6}}{c_{6}} +  \frac{3}{2} \delta Z_H +3 \frac{\delta v}{v}  \right) \nonumber
                            + \lambda v \left(\frac{\delta \lambda}{\lambda}+   \frac{3}{2} \delta Z_H   +\frac{\delta v}{v} \right)\Big]\\
                            &\phantom{=}&-i 6\Big[\frac{4 c_{8} v^4 \delta v}{\Lambda^4}    
                            + 3 \frac{c_{6}}{\Lambda^2} v^2 \delta v \Big]\, . \label{dl3_cont}
\end{eqnarray}
Similarly, the Feynman rule for the counterterm of the Higgs propagator (see eq.~(A.4) in ref.~\cite{Denner:1991kt} ) is modified into 
\begin{equation}
    \vcenter{\hbox{\includegraphics[width=0.2\textwidth]{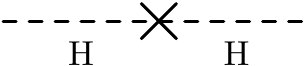}}}
~=~ -i\left(\delta \mh^2 +\mh^2 \delta Z_H -\delta Z _H k^2  + 6  \frac{c_{6}}{\Lambda^2} v^3 \delta v \right)\, , \label{CTpropdraw}
\end{equation}   
and therefore the additional term entering in eq.~\eqref{newmh} is exactly canceled.

The only missing information are the NP contributions to the counterterms in eqs.~\eqref{dZHSMNP}-\eqref{dtsmnp}, which we thus provide in the following:
\begin{eqnarray}
\delta t^{\rm NP} = -\lfact1\frac{3c_6}{\Lambda^2}v^3A_0(\mh^2)\, ,
 \label{dTNP}
\end{eqnarray}
\begin{eqnarray}
\delta Z_H^{\rm NP}
&=& \left(\frac{2c_6v^2}{\lambda\Lambda^2}+\frac{c_6^2v^4}{\lambda^2\Lambda^4}\right)\delta Z_H^{{\rm SM}, \lambda} \, , \label{dZHNP}
\end{eqnarray}
\begin{eqnarray}
\delta (\mh^2)^{\rm NP} &=& \lfact{1}\Big[\frac{c_6}{\Lambda^2}v^2  \left( 18A_0(\mh^2)+3A_0(\xi_Z\mz^2)+6A_0(\xi_W\mw^2) + 18\mh^2 B_0(\mh^2,\mh^2,\mh^2) \right)+ \nonumber  \\
                   &+& \frac{v^4}{\Lambda^{4}} \left(12c_8A_0(\mh^2) + c_6^2 18 v^2 B_0(\mh^2,\mh^2,\mh^2)\right)\Big] \, , 
\end{eqnarray}
\begin{eqnarray}
\delta c_6  &=&\divfact{}\left[ c_6\Big(54\lambda-9\frac{\mz^2+2\mw^2}{v^2}+6\frac{N_c\mt^2}{v^2}\Big)\right.\nonumber \\
    &+&\frac{c_8v^2}{\Lambda^2}\Big(64\lambda-6\frac{\mz^2+2\mw^2}{v^2}+4\frac{N_c\mt^2}{v^2}\Big)
    +\frac{45c_6^2v^2}{\Lambda^2}\nonumber \\
   &+&\frac{20c_{10}\lambda v^4}{\Lambda^4} +\frac{36c_6c_8v^4}{\Lambda^4}\Big] \, , \label{dc6}
\end{eqnarray}
where
\begin{eqnarray}
\delta Z_H^{{\rm SM}, \lambda}&=& - \lfact{9\lambda m_H^2} B_0^{\prime}(m_H^2,m_H^2,m_H^2)\label{dZHHHH}
\end{eqnarray}
is the contribution from the trilinear Higgs self-coupling to $\delta Z_H$ in the SM and $A_0$ and $B_0$ are the standard  scalar  loop integrals and $\Delta$ is the UV divergence $\Delta\equiv1/\epsilon-\gamma+\log(4\pi)$ in $D=4-2\epsilon$ dimensions. As discussed in sec.~\ref{sec:calcHH} terms up to the order $(v/\Lambda)^6$ have to be in general considered. However,
note that no terms beyond $(v/\Lambda)^2$ are present in $\delta t$, or beyond $(v/\Lambda)^4$ in $\delta Z_H$ and $\delta \mh^2$, while $c_6$ is appearing at order $(v/\Lambda)^2$, so terms up to $(v/\Lambda)^6$ are in fact present in $\delta c_6$.

We want to stress that all these contributions have to be  taken into account in order to obtain gauge invariance for the final finite result of double-Higgs production at one loop. We kept the explicit dependence on the $\xi$ parameter for a generic $R_{\xi}$-gauge in order to verify that renormalised amplitudes do not depend on $\xi$. 
 With this calculation setup, results are equivalent to those of a standard calculations based on the parameterisation of eq.~\eqref{VNP_Stnotation} and $c'_{2n}$ coefficients renormalised in the $\MSbar$ scheme.

\section{One-loop amplitudes via form factors}
\label{sec:formfactors}
\begin{figure}[t]
\begin{center}\vspace*{-1.0cm}
\includegraphics[width=0.22\textwidth]{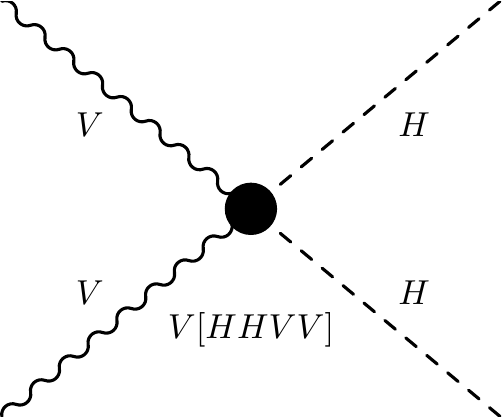}
\hspace{0.3cm}
\includegraphics[width=0.22\textwidth]{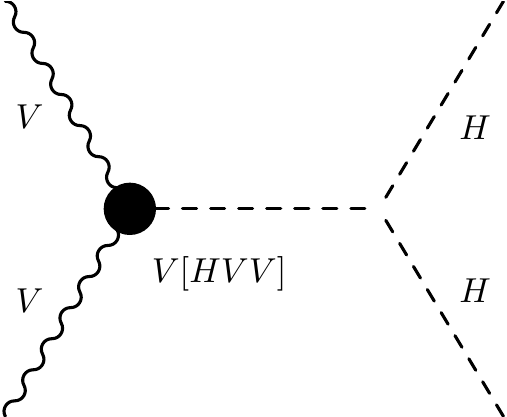}
\hspace{0.3cm}
\includegraphics[width=0.22\textwidth]{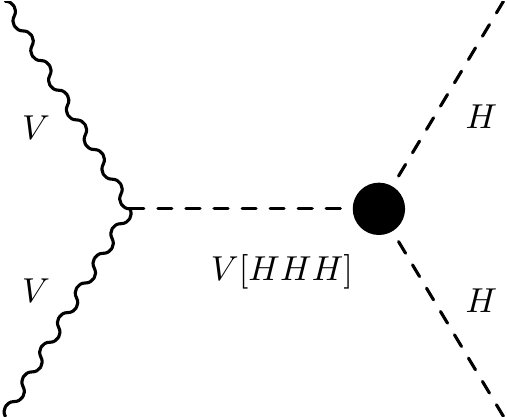}
\hspace{0.3cm}
\includegraphics[width=0.22\textwidth]{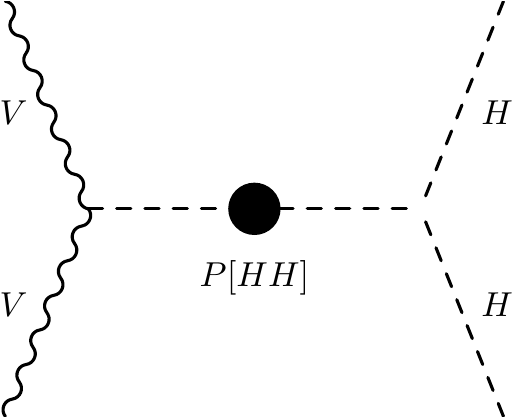}
\caption{The structure of one-loop effects in the $HHVV$ amplitude expressed via form factors.}
\label{HHVVampl}
\end{center}
\end{figure}
In this section we provide all the form factors that are necessary for the calculations of one-loop amplitudes for $ZHH$ and WBF$~HH$ production entering $\sigma^{\rm pheno}_{\rm NLO} (HH)$ in eq.~\eqref{sHHNLOpheno}. These are the form factors for the 
\begin{itemize}
\item $HVV$ vertex,
\item $HHH$ vertex,
\item $HHVV$ vertex,
\item $H$ propagator.
\end{itemize}

 We include contributions up to the order $(v/\Lambda)^6$ and therefore one-loop amplitudes  entering $\sigma^{\rm pheno}_{\rm NLO} (HH)$  can be obtained by substituting the vertexes in the corresponding tree-level amplitudes  with the aforementioned form factors. Indeed, we implemented them in a UFO~\cite{Degrande:2011ua} model file and performed the calculation within the {\sc MadGraph5\_aMC@NLO} \cite{Alwall:2014hca} framework\footnote{We have used the version 2.4.3 of this code, but we could have in principle used also a more or less recent version of it.}, as also done in ref.~\cite{Maltoni:2017ims}. We cross-checked the results via {\sc FeynArts 3.9} \cite{Hahn:2000kx} and {\sc Formcalc 9.4} \cite{Hahn:1998yk}. Loops integrals have been evaluated with {\sc LoopTools 2.13} \cite{Hahn:1998yk} and {\sc QCDLoop 2.0.3} \cite{Ellis:2007qk, Carrazza:2016gav}.

The $HVV$ form factor is the only one that is also relevant for the calculation of one-loop amplitudes entering  $\sigma^{\rm pheno}_{\rm NLO} (H)$ for single Higgs production. For this kind of processes the $(\cbs)^2$ dependence originates completely from $\delta Z_H^{\rm NP}$ in eq.~\eqref{dZHNP}, while the linear term in $\cbs$ comes from both $\delta Z_H^{\rm NP}$ and the $HVV$ form factor, which has already been calculated in ref.~\cite{Bizon:2016wgr} and induces the $C_1$ dependence on the kinematics. We repeated the calculation explicitly checking the gauge invariance, both for a generic $R_\xi$ gauge and also in unitary gauge. 

Before providing the expressions for the different form factors we want to briefly show how the calculation of the $ \sigma_{30}$,  $\sigma_{40}$, $  \sigma_{01}$, $\sigma_{11}$ and $\sigma_{21}$ terms, which are part of $\sigma^{\rm pheno}_{\rm NLO} (HH)$, can be organised. As said, the form factors provided in the following can be plugged into the tree-level diagrams in order to get the necessary one-loop amplitudes for $ZHH$ and WBF$~HH$ giving $\sigma^{\rm pheno}_{\rm NLO} (HH)$. This is schematically depicted in Fig.~\ref{HHVVampl}, where the relevant part of the $ZHH$ and WBF$~HH$ amplitudes, respectively $W^*W^*\to HH$ or $Z^*\to HHZ$,  is shown. Via the interference of tree-level amplitudes and such one-loop amplitudes obtained via  form factors, we would get a cross section that we  denote as $\sigma^{\rm FF}$, where FF stands for form-factors. It is important to note that $\sigma^{\rm FF}$ contains also $\cbs$- and $(\cbs)^2 $-dependent spurious terms, which have to be discarded. One can easily expand in powers of $\cbs^i \cbe^j$ the  $\sigma^{\rm FF}$ result and identify the different $\sigma^{\rm FF}_{ij}$ component.

We should note that for consistency we provide the form factor of $HVV$ matching the same convention used in ref.~\cite{Bizon:2016wgr}, {\it i.e.}, without including the contribution of the Higgs wave-function counterterm $\delta Z_H^{\rm NP}$. The same convention is used also for  all the other form-factors.
Note that in the case of the $H$ propagator and $HHH$ vertex, UV divergences are present in the terms relevant for our calculation and thus the other UV counterterms  have to be included in the definition of the form-factors. In conclusion the actual $\sigma_{ij}$ components can be expressed in terms of the $\sigma^{\rm FF}_{ij}$ as 
\begin{align}
    \sigma_{01,11,21}=&~\sigma_{01,11,21}^{\rm FF}\, ,\\
    \sigma_{30}=&~\sigma_{30}^{\rm FF}+2\delta Z_H^{{\rm SM}, \lambda}\sigma_{1}+4\delta Z_H^{{\rm SM}, \lambda}\sigma_{2}\, ,\\
    \sigma_{40}=&~\sigma_{40}^{\rm FF}+2\delta Z_H^{{\rm SM}, \lambda}\sigma_{2}\, ,
\end{align}
where $\sigma_{1}$ and $\sigma_{2}$ are part of $ \sigma_{\rm LO}(HH)$ in eq.~\eqref{sHHLO}. Note that $\sigma_{30}$ and $\sigma_{40}$ are written in such a form that can be easily extend to the case in which the $\delta Z_H^{\rm NP}$ contribution from external legs is resummed, as done in ref.~\cite{Degrassi:2016wml}. However, considering $|\cbs|<5$, resummation is not necessary given that $\cbs^2 \delta Z_H^{{\rm SM}, \lambda} < 4\%$.  
\subsection*{$\mathbf{HVV}$-vertex }
\begin{figure}[t]
    \begin{align*}
        \vcenter{\hbox{\includegraphics[width=0.2\textwidth]{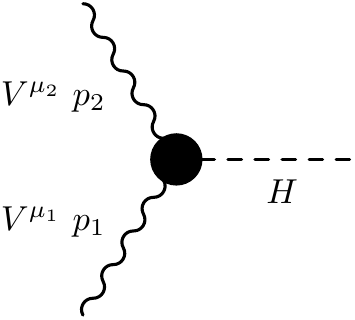}}}
        ~=
        \vcenter{\hbox{\includegraphics[width=0.2\textwidth]{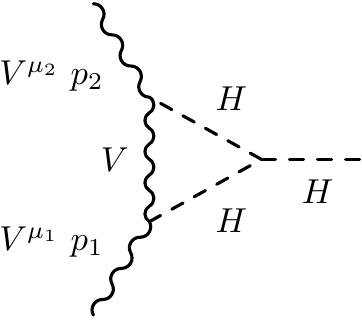}}}
        ~+
        \vcenter{\hbox{\includegraphics[width=0.2\textwidth]{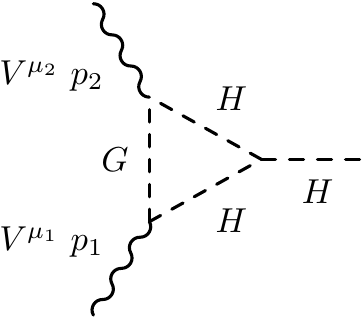}}}
        ~+
        \vcenter{\hbox{\includegraphics[width=0.2\textwidth]{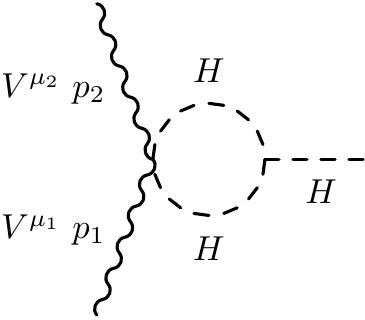}}}
    \end{align*}
    \caption{Feynman diagrams contributing to the $V[HVV]$ form factor at one loop. }
    \label{dia-hvv}
\end{figure}

The $HVV$ form factor, which will denote as $V[HVV]$, enters both the single and double Higgs production calculation and can be written as
\begin{equation}
    V^{\mu_1\mu_2}[HVV]=V^{\mu_1\mu_2}_0[HVV]+V^{\mu_1\mu_2}_1[HVV]\bar c_6   \, .
\end{equation}
For our calculation the $\cbs$-independent part can be ignored, while in a generic gauge $V_1[HVV]$ 
 is given by the three diagrams~\footnote{In the unitary gauge the second diagram does not appear}  in Fig.~\ref{dia-hvv}.  Using the convention that the corresponding Feynman rule is $iV^{\mu_1\mu_2}[HVV]$, as we will do also for the other form factors,
we can write $V^{\mu_1\mu_2}_1[HVV]$ as
\begin{equation}
    V^{\mu_1\mu_2}_1[HVV]=\frac{\lambda \mv^2}{16 \pi^2 v} T^{\mu_1\mu_2}(p_1,p_2,\mv,\mh)\, .\label{ff-hvv}
\end{equation}
In particular
\begin{equation}
    T^{\mu_1\mu_2}(p_1,p_2,\mv,\mh)=(-6B_0-24\mv^2C_0+24C_{00})g^{\mu_1\mu_2}-24p_1^{\mu_2}p_2^{\mu_1}C_{12}\,,
    \label{tensor12}
\end{equation}
where $p_1,p_2$ are the (incoming) momenta of the two vector bosons, $\mu_1,\mu_2$ are the corresponding Lorentz indices, $\mv$ with $V=W,Z$ is mass of the vector bosons, and $B_0$, $C_0$, $C_{00}$, $C_{12}$ are one-loop scalar/tensor integrals defined according to the notation used, {\it e.g.},  in ref.~\cite{Denner:1991kt} and where the following variables are understood:
\begin{align}
    B_{0}=&B_0((p_1+p_2)^2,\mh^2,\mh^2)\label{bcoef}\,,\\
    C_{0,00,12}=&C_{0,00,12}(p_1^2,(p_1+p_2)^2,p_2^2,\mv^2,\mh^2,\mh^2)\label{ccoef}\,. 
\end{align}
We remind the reader that the $ \frac{1}{2}\delta Z_H^{\rm NP}$ contribution from the external $H$ has been removed from $V[HVV]$.
\subsection*{$\mathbf{H}$ propagator }
\begin{figure}[t]
            \begin{center}
    \begin{align*}
\vcenter{\hbox{\includegraphics[width=0.22\textwidth]{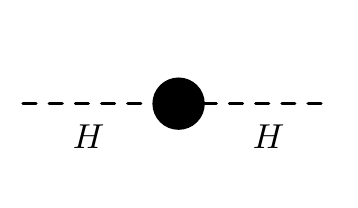}}}
        ~=
\vcenter{\hbox{\includegraphics[width=0.22\textwidth]{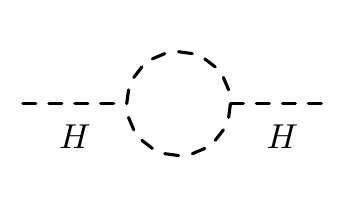}}}
        ~+
\vcenter{\hbox{\includegraphics[width=0.22\textwidth]{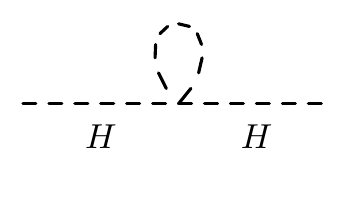}}}
        ~+
\vcenter{\hbox{\includegraphics[width=0.22\textwidth]{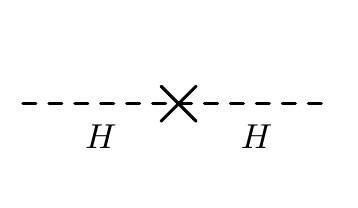}}}
    \end{align*}
                \caption{Feynman diagrams contributing to $P[HH]$.
                }
                \label{propfig}
            \end{center}
        \end{figure}

The form factors for the $HH$ two point function, which we denote as $P[HH]$, receives one-loop contributions from the diagrams in Fig.~\ref{propfig}, where the contribution of counterterm diagram is given in eq.~\eqref{CTpropdraw}. At one loop $P[HH]$ can be written as 
        \begin{equation}
            P[HH]=P_{00}[HH]+P_{10}[HH]\bar{c}_6+P_{20}[HH]\bar{c}_6^2 \, .
        \end{equation}
            In our calculation we do not include $P_{00}[HH]$ contributions, which we set to zero, while $P_{10}[HH]$ and $P_{20}[HH]$ read 
        \begin{equation}
            P_{10}[HH]=2P_{20}[HH]=\lfact{1}(6 \lambda v)^2  \big[ B_0(p^2,\mh^2,\mh^2)-B_0(\mh^2,\mh^2,\mh^2) \big]\, ,
        \end{equation}
It is important to note that $P(HH)$ does not depend on $\cbe$. Indeed, although the second diagram, the seagull, depends on $\cbe$ due to the $HHHH$ vertex,
       it is exactly cancelled by the Higgs-mass counter term.
We remind the reader that the $- \delta Z_H^{\rm NP}$ component in the counterterm  has been removed from $P[HH]$.       
\subsection*{$\mathbf{HHH}$-vertex form factor}
  The form-factor for the $HHH$ vertex,  $V[HHH]$, receives contributions from the diagrams already shown in the main text in Fig.~\ref{tril_loop} and the counter term of Fig.~\ref{ruletri}. At variance with $V[HVV]$ and $P[HH]$, $V[HHH]$  depends on both $\bar c_6$ and $\bar c_8$:
        \begin{equation}
            V[HHH]=\sum_{i+2j\le 3}V_{ij}[HHH] \bar c_6^i \bar c_8^j  \, .
        \end{equation}
For our calculation $V_{00}[HHH]$ and $V_{10}[HHH]$ can be set equal to zero, while
        \begin{align}
            V_{30}[HHH]=~&\lfact{1}(6 \lambda v)^3 
            C_0(p_1^2,p_2^2,p_3^2,\mh^2,\mh^2,\mh^2)\,,\\
            V_{20}[HHH]=~&\lfact{1}(6 \lambda v)^3 3
            C_0(p_1^2,p_2^2,p_3^2,\mh^2,\mh^2,\mh^2)\,,\\
            &+\lfact{1}108\lambda^2 v
            \left\{-\frac{1}{2}
                [B_0(\mh^2,\mh^2,\mh^2)-\Delta]+
                \sum_{i=1}^3[B_0(p_i^2,\mh^2,\mh^2)-\Delta]
            \right\}\,,\nonumber\\
            V_{01}[HHH]=~
            & \frac{\lambda v}{16 \pi^2 } \bigg\{  18 \lambda 
            \left[\sum_{i=1}^{3}[B_0(p_i^2,\mh^2,\mh^2)-\Delta]
                \right]-6\bigg[
                \frac{N_c m_t^2}{v^2}\Delta+16 \pi^2 \frac{\delta v}{v}
            \bigg]\\
            &
            +21 \frac{A_0(\mh^2)-\mh^2\Delta}{v^2}
            +3 \frac{A_0(\xi_Z\mz^2)+3\mz^2\Delta}{v^2}
            +6\frac{A_0(\xi_W \mw^2)+3\mw^2\Delta}{v^2}
        \bigg\} \,, \nonumber
        \\
        V_{11}[HHH]=~
        &\frac{\lambda v}{16 \pi^2 } 18 \lambda 
        \left[\sum_{i=1}^{3}[B_0(p_i^2,\mh^2,\mh^2)-\Delta]
        \right] \,,
    \end{align}
    where the $B_0$ and $C_0$ are the loop scalar integrals, with the dependence on external momenta and internal masses expressed with the convention in ref.~\cite{Denner:1991kt}. The term $\Delta$ is the UV divergence as defined in Appendix \ref{appendix:ren}. The $V_{30}[HHH]$ component is equivalent to the result of ref.~\cite{DiLuzio:2017tfn}, where it is assumed $V[HHH] \sim (1+\cbs)^3 V_{30}[HHH]$.   
    It should be noticed that  the contribution of $\delta v$, which as discussed in Appendix \ref{appendix:ren} is completely of SM origin, is necessary in order to obtain UV finiteness and gauge-invariance for the finite results. We kept the explicit dependence on the $\xi$ gauge parameters outside $\delta v$ precisely to make  this point manifest. In order to help the reader we report in the following the UV divergent part of $\delta v$ and the $\xi$-dependent part, which includes both finite and divergent contributions
    \begin{eqnarray}
    (\delta v)^{\rm UV}=(\delta v^{\rm SM})^{\rm UV}
    =\divfact{}\left[\frac{1}{2}\frac{(3+\xi_Z)\mz^2+2(3+\xi_W)\mw^2}{v}-\frac{N_c\mt^2}{v}\right]\, ,
\end{eqnarray}
\begin{equation}
(\delta v)_{\xi}=(\delta v^{\rm SM})_{\xi}=\lfact{1}\frac{1}{2v} \left[A_0(\mz^2\xi_Z)+2A_0(\mw^2\xi_W)\right]\, . \label{dvxi}
\end{equation}
We remind the reader that the  $\frac{3}{2} \delta Z_H^{\rm NP}$ component in the counterterm, which originates from the three $H$ external legs,  has been removed from $V[HHH]$.       
\subsection*{$\mathbf{HHVV}$-vertex form factor}
    \begin{figure}[t]
        \begin{center}
            \includegraphics[width=0.22\textwidth]{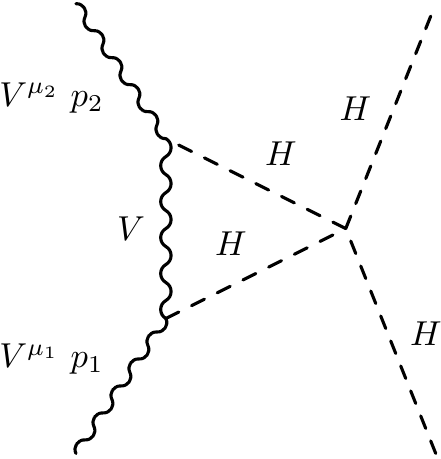}
            \hspace{0.3cm}
            \includegraphics[width=0.22\textwidth]{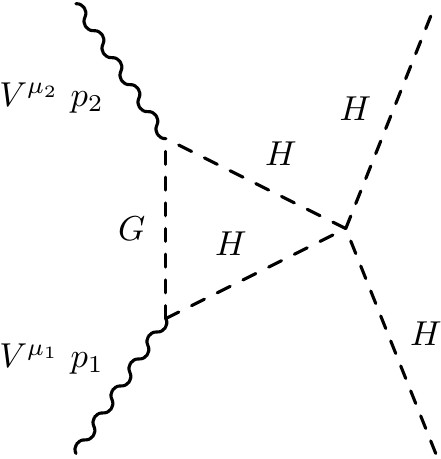}
            \hspace{0.3cm}
            \includegraphics[width=0.22\textwidth]{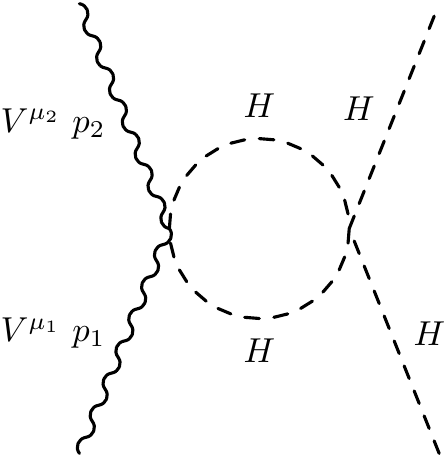}
        \end{center}
        \caption{Feynman diagrams contributing to the $V_{01}[HHVV]$ form factor.}
        \label{dia-hhvv}
    \end{figure}
    \begin{figure}[t]
        \begin{center}
            \includegraphics[width=0.22\textwidth]{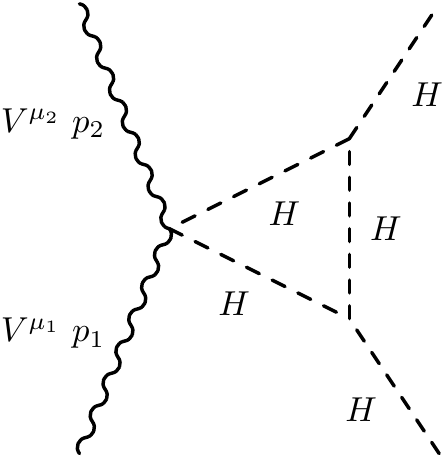}
            \hspace{0.3cm}
            \includegraphics[width=0.22\textwidth]{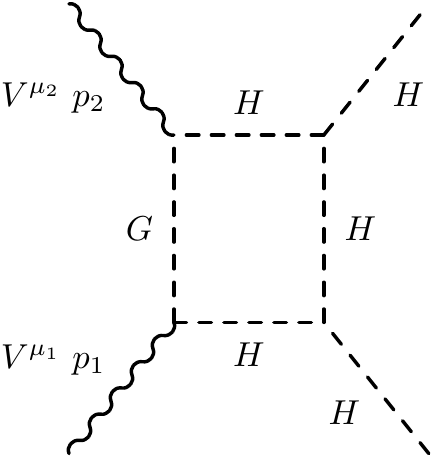}
            \hspace{0.3cm}
            \includegraphics[width=0.22\textwidth]{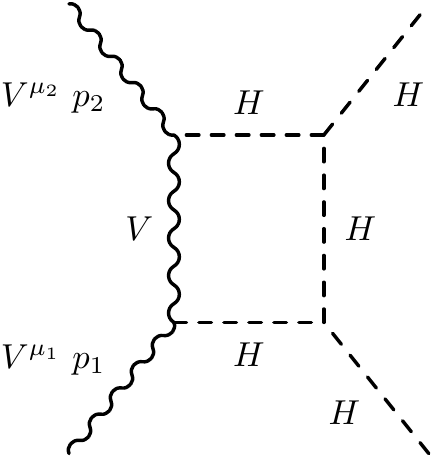}
        \end{center}
        \caption{Representative Feynman diagrams contributing to the  $V_{20}[HHVV]$ form factor.}
        \label{dia-hhvv2}
    \end{figure}
Similarly to the case of $V[HHH]$, the form factor for the $HHVV$ vertex, $V[HHVV]$, can be written as
    \begin{equation}
        V[HHVV]=\sum_{i+2j\le 2}V_{ij}[HHVV]\bar c_6^i \bar c_8^j  \, .
    \end{equation}
For our calculation only $V_{20}[HHVV]$  and $V_{01}[HHVV]$ are relevant and we set the other contributions to zero. The $V_{01}[HHVV]$ component originates from the diagrams in Fig.~\ref{dia-hhvv} and its structure is very similar to the one of
    $V_{1}[HVV]$,
      \begin{equation}
        V^{\mu_1\mu_2}_{01}[HHVV]=\frac{\lambda \mv^2}{16 \pi^2v^2}T^{\mu_1\mu_2}(p_1,p_2,\mv,\mh)\,,\label{ff-hhvv}
    \end{equation}
 where all momentum are incoming and $T^{\mu_1\mu_2}$ is given in eq.~\eqref{tensor12}.            
    The $V_{20}[HHVV]$ term instead originates from the diagrams in Fig.~\ref{dia-hhvv2}, which include boxes and thus they involve a much more complex kinematic dependence,
     \begin{equation}
        V^{\mu_1\mu_2}_{20}[HHVV]=9 \frac{ \lambda^2 \mv^2}{\pi^2} [F^{\mu_1\mu_2}(p_1,p_2,p_3,p_4,\mv,\mh)+F^{\mu_1\mu_2}(p_1,p_2,p_4,p_3,\mv,\mh)] \, ,
    \end{equation}
   where $F^{\mu_1\mu_2}$ is given by
    \begin{equation}
        \begin{split}
            F^{\mu_1\mu_2}(p_1,p_2,p_3,p_4,\mv,\mh)=& (-\frac{1}{4}C_0-\mv^2D_0+D_{00})g^{\mu_1\mu_2}+p_4^{\mu_1} p_1^{\mu_2} D_{12}\\
                                                    &+p_4^{\mu_1}(p_1+p_4)^{\mu_2} D_{22} - p_2^{\mu_1} p_1^{\mu_2} D_{13} - p_2^{\mu_1} (p_1+p_4)^{\mu_2} D_{23} \, ,
        \end{split}
    \end{equation}
    with the dependence on external momenta and internal masses of $C$ and $D$ functions as
    \begin{align}
        C_0=&C_0((p_3+p_4)^2,p_3^2,p_4^2,\mh^2,\mh^2,\mh^2)\, ,\\
        D_{i(j)}=&D_{i(j)}(p_1^2,p_4^2,p_3^2,p_2^2,(p_1+p_4)^2,(p_4+p_3)^2,\mv^2,\mh^2,\mh^2,\mh^2)\, ,
    \end{align}
according to the convention of ref.~\cite{Denner:1991kt}. Both $V_{01}[HHVV]$ and $V_{20}[HHVV]$ are UV finite and gauge-invariant.
We remind the reader that the  $ \delta Z_H^{\rm NP}$ component in the counterterm, which originates from the two $H$ external legs,  has been removed from $V[HHVV]$.       
%
%
%
%
%
\section{Perturbative limits on $\cbs$ and $\cbe$ in double Higgs production}
\label{validity}
\begin{figure}[!t]
    \begin{center}
        \includegraphics[width=0.45\textwidth]{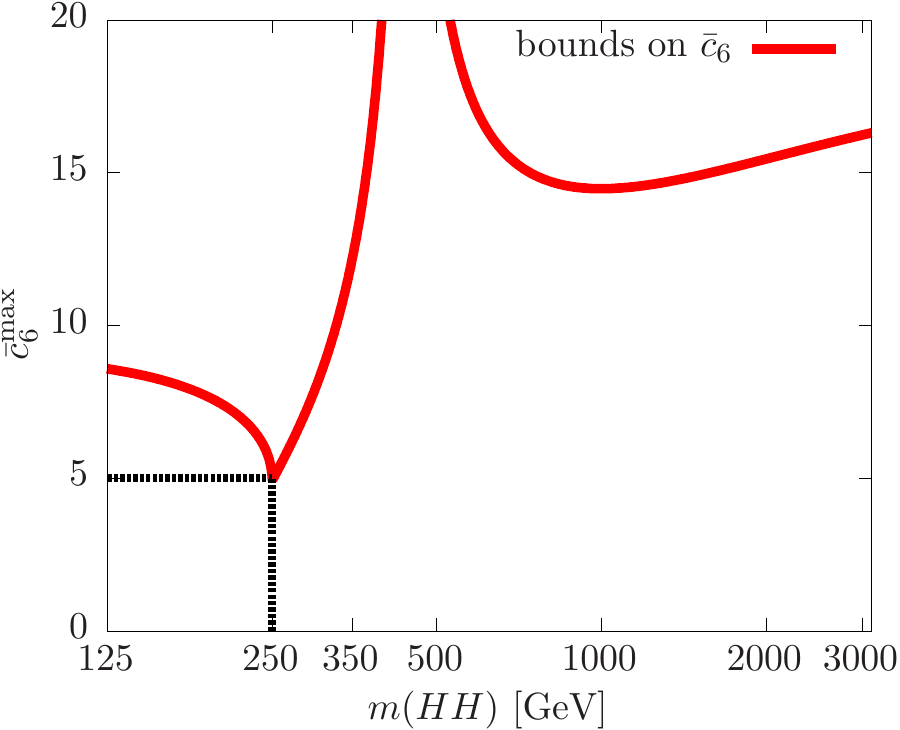}
        \hfill
        \includegraphics[width=0.45\textwidth]{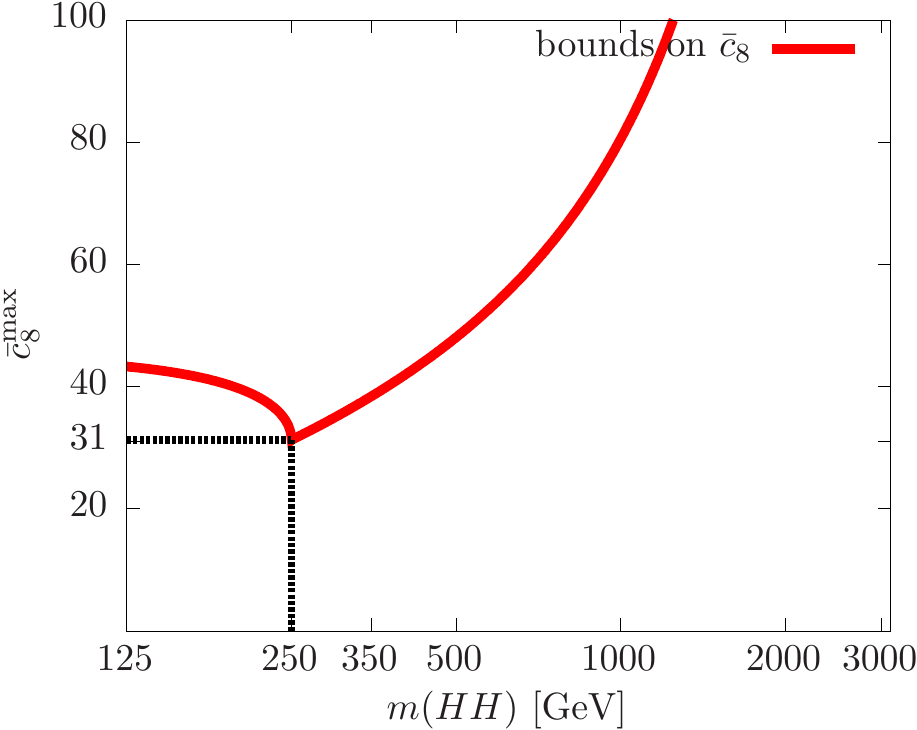}
        \caption{Maximum value of $\cbs$ (left) and $\cbe$ (right) such that the one-loop corrections to the  $HHH$ amplitude are smaller than its tree-level value. We consider two Higgs bosons on-shell and the third with virtuality equal to $m(HH)$, showing the dependence on $m(HH)$. }
        \label{plots:validity_mhh}
    \end{center}
\end{figure}
In this section we describe how we derived the range of validity of our calculation, 
\begin{equation}
|\cbs|< 5~~{\rm and }~~|\cbe|< 31\,,  
\label{limits}
\end{equation}
 which has already mentioned several times in the text.

\begin{figure}[!t]
    \begin{center}
        \includegraphics[width=0.45\textwidth]{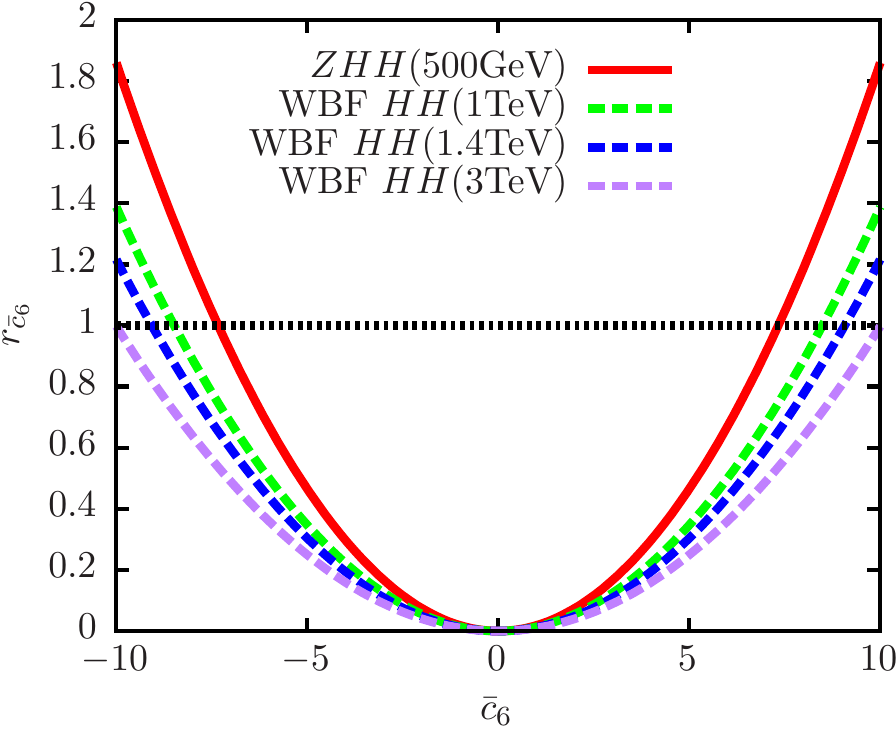}
        \hfill
        \includegraphics[width=0.45\textwidth]{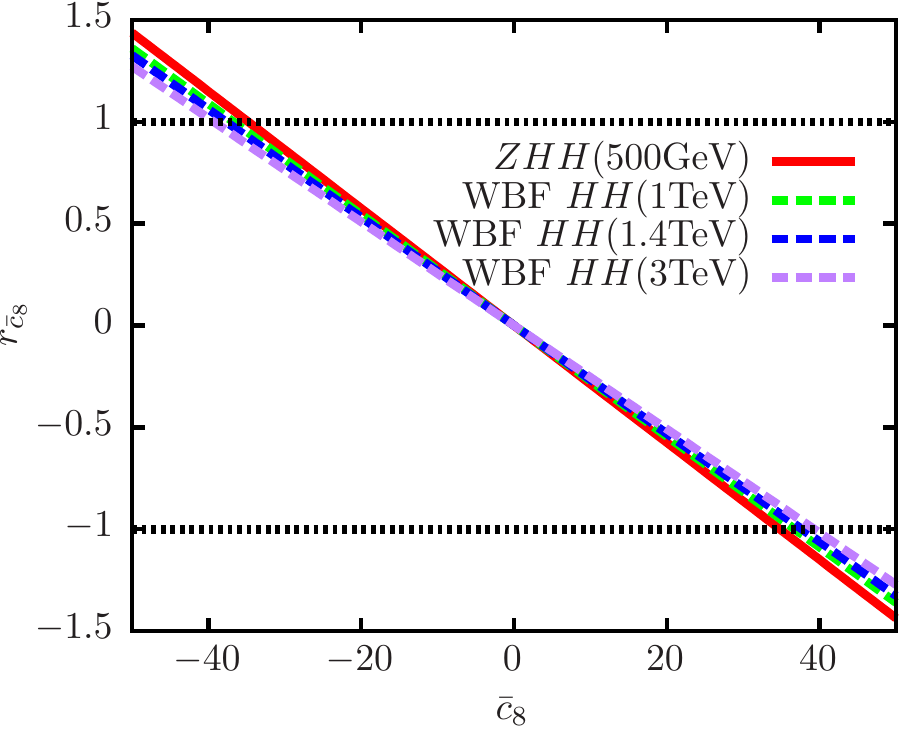}
        \caption{$\cbs$-dependence of $r_{\bar c_6}$ (left) and $\cbe$-dependence of $r_{\bar c_8}$ (right) for $ZHH$ and WBF$~HH$ at different energies.}
        \label{plots:validity}
    \end{center}
\end{figure}

First of all, we analyse the one-loop $H^*\to HH$ amplitude, the analytical expression of which can be obtained via $ \delta Z_H^{\rm NP}$ and the form factors $P[HH]$ and the $V[HHH]$ that have been provided in the previous section. We define as $\cbs^{\rm max}$($\cbe^{\rm max}$) the value of $\cbs$($\cbe$) such that the one-loop amplitude is as large as the tree-level one, {\it i.e.} the value of  of $\cbs$($\cbe$) from where perturbative convergence cannot be trusted anymore. For the estimation of $\cbs^{\rm max}$ we take into account the leading contribution from $V_{30}$ and $P_{20}$, both yielding $\cbs^3$ terms. For $\cbe^{\rm max}$ we instead consider as first step the contribution from $V_{11}$, which is the dominant term when $\cbs$ is large, and we compare it with the linearly $\cbs$ dependent part of the tree-level vertex. In such a way the value of $\cbe^{\rm max}$ is independent on $\cbs$. 

The value of $\cbs^{\rm max}$($\cbe^{\rm max}$) has a kinematic dependence. In the left plot of Fig.~\ref{plots:validity_mhh} we display the dependence of $\cbs^{\rm max}$ on $m(HH)$, ranging from 125 GeV to 3 TeV. The equivalent plot for $\cbe^{\rm max}$, taking leading term in $\cbs$, is shown on the right. Thus, we explore $m(HH)$ values both below the production threshold and in the tail of the  $m(HH)$ distributions. As can be seen in both cases, the most stringent constraints, $|\cbs|< 5$ and $|\cbe|< 31$, arise from the threshold condition $m(HH)=2 \mh$, while for different values of $m(HH)$ the bound is weaker. 

In the case of $\cbs^{\rm max}$ the constraint is independent on the value of the renormalisation scale $\mu_r$ and compatible with the result obtained in ref.~\cite{DiLuzio:2017tfn}, where the subdominant contribution of $P_{20}$ was not taken into account. Conversely, in the case of $\cbe^{\rm max}$ the constraint does depend on the value of the renormalisation scale $\mu_r$. However, we verified for $\mu_r=\mh, 4\mh$, that the most stringent $\cbe^{\rm max}$ value is anyway arising from the kinematic condition  $m(HH)=2\mh$.

The constraints of eq.~\eqref{limits} have been derived via the analysis of the $H^*\to HH$ amplitude, but also the $HVV$ and $HHVV$ vertexes contribute via loop corrections to the quantity $ \sigma^{\rm pheno}_{\rm NLO} (HH)$, eq.~\eqref{sHHNLOpheno}, that is used for our phenomenological analysis. Thus, it is important to check if they can affect the results of eq.~\eqref{limits}.  To this purpose we directly considered the quantities 
\begin{eqnarray}
    r_{\bar c_6}&\equiv&\frac{\bar c_6^4\sigma_{40}}{\bar c_6^2\sigma_2}=\frac{\sigma_{40}}{\sigma_2}\bar c_6^2\, , \nonumber \\
    r_{\bar c_8}&\equiv &\frac{\sigma_{21}\bar c_6^2 \bar c_8}{\sigma_{2}\bar c_6^2}=\frac{\sigma_{21}}{\sigma_{2}}\bar c_8\, ,
\end{eqnarray}
for $ZHHH$ and WBF$~HH$ at different energies.
The quantity $r_{\bar c_6}$ is the ratio between the term with the highest power in $\bar c_6$ from $\Delta\sigma_{\bar c_6}$ and the one with the highest power in $\bar c_6$ from $\sigma_{\rm LO}$, {\it i.e.}, the ratio of the dominant contributions at tree and one-loop level for large $\cbs$ values.  Similarly, the quantity $r_{\bar c_8}$ is the ratio between the term with the highest power in $\bar c_6$ from $\Delta\sigma_{\bar c_8}$ and  $\sigma_{\rm LO}$. Thus, both of them can be considered as a generalisation of the first step; both $HVV$ and $HHVV$ vertices are taken into account and phase-space integration is performed.

In the left plot of Fig.~\ref{plots:validity}, we show $r_{\bar c_6}$ for the case of $ZHH$ at $500~{\rm GeV}$ and of WBF at $1,1.4$ and $3~{\rm TeV}$, which are the phenomenologically relevant scenarios analysed in sec.~\ref{sec:bounds}. Requiring $ |r_{\bar c_6}| < 1$, we can get $|\bar c_6|<8$ for $ZHH$ at 500 GeV,
and $|\bar c_6|<9,10,11$ for WBF$~HH$ at 1000, 1400 and 3000 GeV, respectively. Thus, as one would expected from Fig.~\ref{plots:validity_mhh} for the $H^*\to HH$ vertex, at higher energies, far from the production threshold, limits are weaker. In the right plot we show $r_{\bar c_8}$ for the same energies an process. Also in this case the obtained limits are weaker than in eq.~\eqref{limits},  $|\bar c_8|<\sim 35-40$.

    \bibliographystyle{JHEP}
    \bibliography{article.bib}

\providecommand{\href}[2]{#2}\begingroup\raggedright\begin{thebibliography}{10}

\bibitem{Chatrchyan:2012xdj}
{\bf CMS} Collaboration, S.~Chatrchyan {\em et~al.}, {\it {Observation of a new
  boson at a mass of 125 GeV with the CMS experiment at the LHC}},  {\em Phys.
  Lett.} {\bf B716} (2012) 30--61,
  [\href{http://xxx.lanl.gov/abs/1207.7235}{{\tt 1207.7235}}].

\bibitem{Aad:2012tfa}
{\bf ATLAS} Collaboration, G.~Aad {\em et~al.}, {\it {Observation of a new
  particle in the search for the Standard Model Higgs boson with the ATLAS
  detector at the LHC}},  {\em Phys. Lett.} {\bf B716} (2012) 1--29,
  [\href{http://xxx.lanl.gov/abs/1207.7214}{{\tt 1207.7214}}].

\bibitem{Khachatryan:2016vau}
{\bf ATLAS, CMS} Collaboration, G.~Aad {\em et~al.}, {\it {Measurements of the
  Higgs boson production and decay rates and constraints on its couplings from
  a combined ATLAS and CMS analysis of the LHC pp collision data at $
  \sqrt{s}=7 $ and 8 TeV}},  {\em JHEP} {\bf 08} (2016) 045,
  [\href{http://xxx.lanl.gov/abs/1606.02266}{{\tt 1606.02266}}].

\bibitem{Baur:2003gp}
U.~Baur, T.~Plehn, and D.~L. Rainwater, {\it {Probing the Higgs selfcoupling at
  hadron colliders using rare decays}},  {\em Phys. Rev.} {\bf D69} (2004)
  053004, [\href{http://xxx.lanl.gov/abs/hep-ph/0310056}{{\tt
  hep-ph/0310056}}].

\bibitem{Dolan:2012rv}
M.~J. Dolan, C.~Englert, and M.~Spannowsky, {\it {Higgs self-coupling
  measurements at the LHC}},  {\em JHEP} {\bf 10} (2012) 112,
  [\href{http://xxx.lanl.gov/abs/1206.5001}{{\tt 1206.5001}}].

\bibitem{Papaefstathiou:2012qe}
A.~Papaefstathiou, L.~L. Yang, and J.~Zurita, {\it {Higgs boson pair production
  at the LHC in the $b \bar{b} W^+ W^-$ channel}},  {\em Phys. Rev.} {\bf D87}
  (2013), no.~1 011301, [\href{http://xxx.lanl.gov/abs/1209.1489}{{\tt
  1209.1489}}].

\bibitem{Baglio:2012np}
J.~Baglio, A.~Djouadi, R.~Gr\"{o}ber, M.~M. M\"{u}hlleitner, J.~Quevillon,
  and M.~Spira, {\it {The measurement of the Higgs self-coupling at the LHC:
  theoretical status}},  {\em JHEP} {\bf 04} (2013) 151,
  [\href{http://xxx.lanl.gov/abs/1212.5581}{{\tt 1212.5581}}].

\bibitem{Yao:2013ika}
W.~Yao, {\it {Studies of measuring Higgs self-coupling with $HH\rightarrow
  b\bar b \gamma\gamma$ at the future hadron colliders}},  in {\em
  {Proceedings, 2013 Community Summer Study on the Future of U.S. Particle
  Physics: Snowmass on the Mississippi (CSS2013): Minneapolis, MN, USA, July
  29-August 6, 2013}}, 2013.
\newblock \href{http://xxx.lanl.gov/abs/1308.6302}{{\tt 1308.6302}}.

\bibitem{Barger:2013jfa}
V.~Barger, L.~L. Everett, C.~B. Jackson, and G.~Shaughnessy, {\it {Higgs-Pair
  Production and Measurement of the Triscalar Coupling at LHC(8,14)}},  {\em
  Phys. Lett.} {\bf B728} (2014) 433--436,
  [\href{http://xxx.lanl.gov/abs/1311.2931}{{\tt 1311.2931}}].

\bibitem{deLima:2014dta}
D.~E. Ferreira~de Lima, A.~Papaefstathiou, and M.~Spannowsky, {\it {Standard
  model Higgs boson pair production in the ($ b\overline{b} $)($
  b\overline{b} $) final state}},  {\em JHEP} {\bf 08} (2014) 030,
  [\href{http://xxx.lanl.gov/abs/1404.7139}{{\tt 1404.7139}}].

\bibitem{Englert:2014uqa}
C.~Englert, F.~Krauss, M.~Spannowsky, and J.~Thompson, {\it {Di-Higgs
  phenomenology in $t\bar{t}hh$: The forgotten channel}},  {\em Phys. Lett.}
  {\bf B743} (2015) 93--97, [\href{http://xxx.lanl.gov/abs/1409.8074}{{\tt
  1409.8074}}].

\bibitem{Wardrope:2014kya}
D.~Wardrope, E.~Jansen, N.~Konstantinidis, B.~Cooper, R.~Falla, and
  N.~Norjoharuddeen, {\it {Non-resonant Higgs-pair production in the
  $b\overline{b}b\overline{b}$ final state at the LHC}},  {\em Eur. Phys.
  J.} {\bf C75} (2015), no.~5 219,
  [\href{http://xxx.lanl.gov/abs/1410.2794}{{\tt 1410.2794}}].

\bibitem{Liu:2014rva}
T.~Liu and H.~Zhang, {\it {Measuring Di-Higgs Physics via the $t \bar t hh \to
  t \bar t b \bar bb\bar b$ Channel}},
  \href{http://xxx.lanl.gov/abs/1410.1855}{{\tt 1410.1855}}.

\bibitem{Azatov:2015oxa}
A.~Azatov, R.~Contino, G.~Panico, and M.~Son, {\it {Effective field theory
  analysis of double Higgs boson production via gluon fusion}},  {\em Phys.
  Rev.} {\bf D92} (2015), no.~3 035001,
  [\href{http://xxx.lanl.gov/abs/1502.00539}{{\tt 1502.00539}}].

\bibitem{Li:2015yia}
Q.~Li, Z.~Li, Q.-S. Yan, and X.~Zhao, {\it {Probe Higgs boson pair production
  via the 3$\ell2j+$ missing $E_T$ mode}},  {\em Phys. Rev.} {\bf D92} (2015), no.~1
  014015, [\href{http://xxx.lanl.gov/abs/1503.07611}{{\tt 1503.07611}}].

\bibitem{Lu:2015jza}
C.-T. Lu, J.~Chang, K.~Cheung, and J.~S. Lee, {\it {An exploratory study of
  Higgs-boson pair production}},  {\em JHEP} {\bf 08} (2015) 133,
  [\href{http://xxx.lanl.gov/abs/1505.00957}{{\tt 1505.00957}}].

\bibitem{Cao:2015oaa}
Q.-H. Cao, B.~Yan, D.-M. Zhang, and H.~Zhang, {\it {Resolving the Degeneracy in
  Single Higgs Production with Higgs Pair Production}},  {\em Phys. Lett.} {\bf
  B752} (2016) 285--290, [\href{http://xxx.lanl.gov/abs/1508.06512}{{\tt
  1508.06512}}].

\bibitem{Cao:2015oxx}
Q.-H. Cao, Y.~Liu, and B.~Yan, {\it {Measuring trilinear Higgs coupling in WHH
  and ZHH productions at the high-luminosity LHC}},  {\em Phys. Rev.} {\bf D95}
  (2017), no.~7 073006, [\href{http://xxx.lanl.gov/abs/1511.03311}{{\tt
  1511.03311}}].

\bibitem{Behr:2015oqq}
J.~K. Behr, D.~Bortoletto, J.~A. Frost, N.~P. Hartland, C.~Issever, and
  J.~Rojo, {\it {Boosting Higgs pair production in the $b\bar{b}b\bar{b}$ final
  state with multivariate techniques}},  {\em Eur. Phys. J.} {\bf C76} (2016),
  no.~7 386, [\href{http://xxx.lanl.gov/abs/1512.08928}{{\tt 1512.08928}}].

\bibitem{Cao:2016zob}
Q.-H. Cao, G.~Li, B.~Yan, D.-M. Zhang, and H.~Zhang, {\it {Double Higgs
  production at the 14 TeV LHC and a 100 TeV $pp$ collider}},  {\em Phys. Rev.}
  {\bf D96} (2017), no.~9 095031,
  [\href{http://xxx.lanl.gov/abs/1611.09336}{{\tt 1611.09336}}].

\bibitem{Adhikary:2017jtu}
A.~Adhikary, S.~Banerjee, R.~K. Barman, B.~Bhattacherjee, and S.~Niyogi, {\it
  {Revisiting the non-resonant Higgs pair production at the HL-LHC}},
  \href{http://xxx.lanl.gov/abs/1712.05346}{{\tt 1712.05346}}.

\bibitem{CMS:2017ihs}
{\bf CMS} Collaboration, C.~Collaboration, {\it {Search for Higgs boson pair
  production in the final state containing two photons and two bottom quarks in
  proton-proton collisions at $\sqrt{s}=13~\mathrm{TeV}$}}.

\bibitem{ATL-PHYS-PUB-2014-019}
{\it {Prospects for measuring Higgs pair production in the channel
  $H(\rightarrow\gamma\gamma)H(\rightarrow b\overline{b}) $ using the ATLAS
  detector at the HL-LHC}},  Tech. Rep. ATL-PHYS-PUB-2014-019, CERN, Geneva,
  Oct, 2014.

\bibitem{Plehn:2005nk}
T.~Plehn and M.~Rauch, {\it {The quartic higgs coupling at hadron colliders}},
  {\em Phys. Rev.} {\bf D72} (2005) 053008,
  [\href{http://xxx.lanl.gov/abs/hep-ph/0507321}{{\tt hep-ph/0507321}}].

\bibitem{Binoth:2006ym}
T.~Binoth, S.~Karg, N.~Kauer, and R.~Ruckl, {\it {Multi-Higgs boson production
  in the Standard Model and beyond}},  {\em Phys. Rev.} {\bf D74} (2006)
  113008, [\href{http://xxx.lanl.gov/abs/hep-ph/0608057}{{\tt
  hep-ph/0608057}}].

\bibitem{Maltoni:2014eza}
F.~Maltoni, E.~Vryonidou, and M.~Zaro, {\it {Top-quark mass effects in double
  and triple Higgs production in gluon-gluon fusion at NLO}},  {\em JHEP} {\bf
  11} (2014) 079, [\href{http://xxx.lanl.gov/abs/1408.6542}{{\tt 1408.6542}}].

\bibitem{Baglio:2015wcg}
J.~Baglio, A.~Djouadi, and J.~Quevillon, {\it {Prospects for Higgs physics at
  energies up to 100 TeV}},  {\em Rept. Prog. Phys.} {\bf 79} (2016), no.~11
  116201, [\href{http://xxx.lanl.gov/abs/1511.07853}{{\tt 1511.07853}}].

\bibitem{Chen:2015gva}
C.-Y. Chen, Q.-S. Yan, X.~Zhao, Y.-M. Zhong, and Z.~Zhao, {\it {Probing
  triple-Higgs productions via $4b2\gamma$ decay channel at a 100 TeV hadron
  collider}},  {\em Phys. Rev.} {\bf D93} (2016), no.~1 013007,
  [\href{http://xxx.lanl.gov/abs/1510.04013}{{\tt 1510.04013}}].

\bibitem{Kilian:2017nio}
W.~Kilian, S.~Sun, Q.-S. Yan, X.~Zhao, and Z.~Zhao, {\it {New Physics in
  multi-Higgs boson final states}},  {\em JHEP} {\bf 06} (2017) 145,
  [\href{http://xxx.lanl.gov/abs/1702.03554}{{\tt 1702.03554}}].

\bibitem{Fuks:2017zkg}
B.~Fuks, J.~H. Kim, and S.~J. Lee, {\it {Scrutinizing the Higgs quartic
  coupling at a future 100 TeV proton--proton collider with taus and b-jets}},
  {\em Phys. Lett.} {\bf B771} (2017) 354--358,
  [\href{http://xxx.lanl.gov/abs/1704.04298}{{\tt 1704.04298}}].

\bibitem{McCullough:2013rea}
M.~McCullough, {\it {An Indirect Model-Dependent Probe of the Higgs
  Self-Coupling}},  {\em Phys. Rev.} {\bf D90} (2014), no.~1 015001,
  [\href{http://xxx.lanl.gov/abs/1312.3322}{{\tt 1312.3322}}]. [Erratum: Phys.
  Rev.D92,no.3,039903(2015)].

\bibitem{Ellis:1975ap}
J.~R. Ellis, M.~K. Gaillard, and D.~V. Nanopoulos, {\it {A Phenomenological
  Profile of the Higgs Boson}},  {\em Nucl. Phys.} {\bf B106} (1976) 292.

\bibitem{Lee:1977eg}
B.~W. Lee, C.~Quigg, and H.~B. Thacker, {\it {Weak Interactions at Very
  High-Energies: The Role of the Higgs Boson Mass}},  {\em Phys. Rev.} {\bf
  D16} (1977) 1519.

\bibitem{Ioffe:1976sd}
B.~L. Ioffe and V.~A. Khoze, {\it {What Can Be Expected from Experiments on
  Colliding e+ e- Beams with e Approximately Equal to 100-GeV?}},  {\em Sov. J.
  Part. Nucl.} {\bf 9} (1978) 50. [Fiz. Elem. Chast. Atom. Yadra9,118(1978)].

\bibitem{Gorbahn:2016uoy}
M.~Gorbahn and U.~Haisch, {\it {Indirect probes of the trilinear Higgs
  coupling: $gg \to h$ and $h \to \gamma \gamma$}},  {\em JHEP} {\bf 10} (2016)
  094, [\href{http://xxx.lanl.gov/abs/1607.03773}{{\tt 1607.03773}}].

\bibitem{Degrassi:2016wml}
G.~Degrassi, P.~P. Giardino, F.~Maltoni, and D.~Pagani, {\it {Probing the Higgs
  self coupling via single Higgs production at the LHC}},  {\em JHEP} {\bf 12}
  (2016) 080, [\href{http://xxx.lanl.gov/abs/1607.04251}{{\tt 1607.04251}}].

\bibitem{Bizon:2016wgr}
W.~Bizon, M.~Gorbahn, U.~Haisch, and G.~Zanderighi, {\it {Constraints on the
  trilinear Higgs coupling from vector boson fusion and associated Higgs
  production at the LHC}},  {\em JHEP} {\bf 07} (2017) 083,
  [\href{http://xxx.lanl.gov/abs/1610.05771}{{\tt 1610.05771}}].

\bibitem{Degrassi:2017ucl}
G.~Degrassi, M.~Fedele, and P.~P. Giardino, {\it {Constraints on the trilinear
  Higgs self coupling from precision observables}},  {\em JHEP} {\bf 04} (2017)
  155, [\href{http://xxx.lanl.gov/abs/1702.01737}{{\tt 1702.01737}}].

\bibitem{Kribs:2017znd}
G.~D. Kribs, A.~Maier, H.~Rzehak, M.~Spannowsky, and P.~Waite, {\it
  {Electroweak oblique parameters as a probe of the trilinear Higgs boson
  self-interaction}},  {\em Phys. Rev.} {\bf D95} (2017), no.~9 093004,
  [\href{http://xxx.lanl.gov/abs/1702.07678}{{\tt 1702.07678}}].

\bibitem{Maltoni:2017ims}
F.~Maltoni, D.~Pagani, A.~Shivaji, and X.~Zhao, {\it {Trilinear Higgs coupling
  determination via single-Higgs differential measurements at the LHC}},  {\em
  Eur. Phys. J.} {\bf C77} (2017), no.~12 887,
  [\href{http://xxx.lanl.gov/abs/1709.08649}{{\tt 1709.08649}}].

\bibitem{DiVita:2017eyz}
S.~Di~Vita, C.~Grojean, G.~Panico, M.~Riembau, and T.~Vantalon, {\it {A global
  view on the Higgs self-coupling}},  {\em JHEP} {\bf 09} (2017) 069,
  [\href{http://xxx.lanl.gov/abs/1704.01953}{{\tt 1704.01953}}].

\bibitem{Barklow:2017awn}
T.~Barklow, K.~Fujii, S.~Jung, M.~E. Peskin, and J.~Tian, {\it
  {Model-Independent Determination of the Triple Higgs Coupling at e+e-
  Colliders}},  {\em Phys. Rev.} {\bf D97} (2018), no.~5 053004,
  [\href{http://xxx.lanl.gov/abs/1708.09079}{{\tt 1708.09079}}].

\bibitem{DiVita:2017vrr}
S.~Di~Vita, G.~Durieux, C.~Grojean, J.~Gu, Z.~Liu, G.~Panico, M.~Riembau, and
  T.~Vantalon, {\it {A global view on the Higgs self-coupling at lepton
  colliders}},  {\em JHEP} {\bf 02} (2018) 178,
  [\href{http://xxx.lanl.gov/abs/1711.03978}{{\tt 1711.03978}}].

\bibitem{Jones:1979bq}
D.~R.~T. Jones and S.~T. Petcov, {\it {Heavy Higgs Bosons at LEP}},  {\em Phys.
  Lett.} {\bf 84B} (1979) 440--444.

\bibitem{CEPC-SPPCStudyGroup:2015csa}
C.-S.~S. Group, {\it {CEPC-SPPC Preliminary Conceptual Design Report. 1.
  Physics and Detector}}.

\bibitem{Gomez-Ceballos:2013zzn}
{\bf TLEP Design Study Working Group} Collaboration, M.~Bicer {\em et~al.},
  {\it {First Look at the Physics Case of TLEP}},  {\em JHEP} {\bf 01} (2014)
  164, [\href{http://xxx.lanl.gov/abs/1308.6176}{{\tt 1308.6176}}].

\bibitem{Baer:2013cma}
H.~Baer, T.~Barklow, K.~Fujii, Y.~Gao, A.~Hoang, S.~Kanemura, J.~List, H.~E.
  Logan, A.~Nomerotski, M.~Perelstein, {\em et~al.}, {\it {The International
  Linear Collider Technical Design Report - Volume 2: Physics}},
  \href{http://xxx.lanl.gov/abs/1306.6352}{{\tt 1306.6352}}.

\bibitem{CLIC:2016zwp}
{\bf CLICdp, CLIC} Collaboration, M.~J. Boland {\em et~al.}, {\it {Updated
  baseline for a staged Compact Linear Collider}},
  \href{http://xxx.lanl.gov/abs/1608.07537}{{\tt 1608.07537}}.

\bibitem{Abramowicz:2016zbo}
H.~Abramowicz {\em et~al.}, {\it {Higgs physics at the CLIC electron--positron
  linear collider}},  {\em Eur. Phys. J.} {\bf C77} (2017), no.~7 475,
  [\href{http://xxx.lanl.gov/abs/1608.07538}{{\tt 1608.07538}}].

\bibitem{Boudjema:1995cb}
F.~Boudjema and E.~Chopin, {\it {Double Higgs production at the linear
  colliders and the probing of the Higgs selfcoupling}},  {\em Z. Phys.} {\bf
  C73} (1996) 85--110, [\href{http://xxx.lanl.gov/abs/hep-ph/9507396}{{\tt
  hep-ph/9507396}}].

\bibitem{Denner:1991kt}
A.~Denner, {\it {Techniques for calculation of electroweak radiative
  corrections at the one loop level and results for W physics at LEP-200}},
  {\em Fortsch. Phys.} {\bf 41} (1993) 307--420,
  [\href{http://xxx.lanl.gov/abs/0709.1075}{{\tt 0709.1075}}].

\bibitem{Sokolov:1963zn}
A.~A. Sokolov and I.~M. Ternov, {\it {On Polarization and spin effects in the
  theory of synchrotron radiation}},  {\em Sov. Phys. Dokl.} {\bf 8} (1964)
  1203--1205. [Phys. Dokl.8,1203(1964)].

\bibitem{Patrignani:2016xqp}
{\bf Particle Data Group} Collaboration, C.~Patrignani {\em et~al.}, {\it
  {Review of Particle Physics}},  {\em Chin. Phys.} {\bf C40} (2016), no.~10
  100001.

\bibitem{Denner:2003iy}
A.~Denner, S.~Dittmaier, M.~Roth, and M.~M. Weber, {\it {Electroweak radiative
  corrections to e+ e- $\rightarrow$ nu anti-nu H}},  {\em Nucl. Phys.} {\bf B660}
  (2003) 289--321, [\href{http://xxx.lanl.gov/abs/hep-ph/0302198}{{\tt
  hep-ph/0302198}}].

\bibitem{Denner:2003yg}
A.~Denner, S.~Dittmaier, M.~Roth, and M.~M. Weber, {\it {Electroweak radiative
  corrections to single Higgs boson production in e+ e- annihilation}},  {\em
  Phys. Lett.} {\bf B560} (2003) 196--203,
  [\href{http://xxx.lanl.gov/abs/hep-ph/0301189}{{\tt hep-ph/0301189}}].

\bibitem{DiLuzio:2017tfn}
L.~Di~Luzio, R.~Gr\"{o}ber, and M.~Spannowsky, {\it {Maxi-sizing the trilinear
  Higgs self-coupling: how large could it be?}},  {\em Eur. Phys. J.} {\bf C77}
  (2017), no.~11 788, [\href{http://xxx.lanl.gov/abs/1704.02311}{{\tt
  1704.02311}}].

\bibitem{Tian:2013qmi}
J.~Tian, {\it {Study of Higgs self-coupling at the ILC based on the full
  detector simulation at $ \sqrt{s} $= 500 GeV and $ \sqrt{s} $ = 1 TeV}},  in {\em {Helmholtz
  Alliance Linear Collider Forum: Proceedings of the Workshops Hamburg, Munich,
  Hamburg 2010-2012, Germany}}, (Hamburg), pp.~224--247, DESY, DESY, 2013.

\bibitem{Belanger:2003ya}
G.~Belanger, F.~Boudjema, J.~Fujimoto, T.~Ishikawa, T.~Kaneko, Y.~Kurihara,
  K.~Kato, and Y.~Shimizu, {\it {Full 0(alpha) electroweak corrections to
  double Higgs strahlung at the linear collider}},  {\em Phys. Lett.} {\bf
  B576} (2003) 152--164, [\href{http://xxx.lanl.gov/abs/hep-ph/0309010}{{\tt
  hep-ph/0309010}}].
  
\bibitem{Barklow:2017suo}
T.~Barklow, K.~Fujii, S.~Jung, R.~Karl, J.~List, T.~Ogawa, M.~E. Peskin, and
  J.~Tian, {\it {Improved Formalism for Precision Higgs Coupling Fits}},  {\em
  Phys. Rev.} {\bf D97} (2018), no.~5 053003,
  [\href{http://xxx.lanl.gov/abs/1708.08912}{{\tt 1708.08912}}].  

\bibitem{Behnke:2013lya}
H.~Abramowicz {\em et~al.}, {\it {The International Linear Collider Technical
  Design Report - Volume 4: Detectors}},
  \href{http://xxx.lanl.gov/abs/1306.6329}{{\tt 1306.6329}}.

\bibitem{Sirlin:1985ux}
A.~Sirlin and R.~Zucchini, {\it {Dependence of the Quartic Coupling H(m) on
  M($H$) and the Possible Onset of New Physics in the Higgs Sector of the
  Standard Model}},  {\em Nucl. Phys.} {\bf B266} (1986) 389--409.

\bibitem{Degrande:2011ua}
C.~Degrande, C.~Duhr, B.~Fuks, D.~Grellscheid, O.~Mattelaer, and T.~Reiter,
  {\it {UFO - The Universal FeynRules Output}},  {\em Comput. Phys. Commun.}
  {\bf 183} (2012) 1201--1214, [\href{http://xxx.lanl.gov/abs/1108.2040}{{\tt
  1108.2040}}].

\bibitem{Alwall:2014hca}
J.~Alwall, R.~Frederix, S.~Frixione, V.~Hirschi, F.~Maltoni, O.~Mattelaer,
  H.~S. Shao, T.~Stelzer, P.~Torrielli, and M.~Zaro, {\it {The automated
  computation of tree-level and next-to-leading order differential cross
  sections, and their matching to parton shower simulations}},  {\em JHEP} {\bf
  07} (2014) 079, [\href{http://xxx.lanl.gov/abs/1405.0301}{{\tt 1405.0301}}].

\bibitem{Hahn:2000kx}
T.~Hahn, {\it {Generating Feynman diagrams and amplitudes with FeynArts 3}},
  {\em Comput. Phys. Commun.} {\bf 140} (2001) 418--431,
  [\href{http://xxx.lanl.gov/abs/hep-ph/0012260}{{\tt hep-ph/0012260}}].

\bibitem{Hahn:1998yk}
T.~Hahn and M.~Perez-Victoria, {\it {Automatized one loop calculations in
  four-dimensions and D-dimensions}},  {\em Comput. Phys. Commun.} {\bf 118}
  (1999) 153--165, [\href{http://xxx.lanl.gov/abs/hep-ph/9807565}{{\tt
  hep-ph/9807565}}].

\bibitem{Ellis:2007qk}
R.~K. Ellis and G.~Zanderighi, {\it {Scalar one-loop integrals for QCD}},  {\em
  JHEP} {\bf 02} (2008) 002, [\href{http://xxx.lanl.gov/abs/0712.1851}{{\tt
  0712.1851}}].

\bibitem{Carrazza:2016gav}
S.~Carrazza, R.~K. Ellis, and G.~Zanderighi, {\it {QCDLoop: a comprehensive
  framework for one-loop scalar integrals}},  {\em Comput. Phys. Commun.} {\bf
  209} (2016) 134--143, [\href{http://xxx.lanl.gov/abs/1605.03181}{{\tt
  1605.03181}}].

\end{thebibliography}\endgroup
    \end{document}